\documentclass[11pt]{article}
\usepackage[utf8]{inputenc}
\usepackage{amssymb,amsmath}
\usepackage{bm} %math bold symbols
\usepackage{booktabs} %some alignment stuff
\usepackage{array}
\usepackage{latexsym}
\usepackage{graphicx}
\usepackage{color}
\usepackage{datetime}
\usepackage[nosort]{cite}
\usepackage{verbatim}
\usepackage{enumerate}
\usepackage{chngpage} % allows for temporary adjustment of side margins
\usepackage{mathrsfs}
\usepackage{euscript}
\usepackage{psfrag}
\usepackage{soul}

\usepackage{datetime}

\usepackage[nosort]{cite}
\usepackage{chngpage} % allows for temporary adjustment of side margins
\usepackage{setspace}
\usepackage{tensor}
\usepackage{physics}

\usepackage{mciteplus}

\usepackage[colorlinks=true,      linkcolor=blue,      urlcolor=blue,      
            filecolor=blue,      citecolor=blue,       pdfstartview=FitH,     
						pdfpagemode=UseNone,      bookmarksopen=true]{hyperref}  % LINKS
\usepackage[all]{hypcap}     %should be loaded AFTER hyperref, which should otherwise be loaded last.

\definecolor{cardinal}{rgb}{0.6,0,0}
\definecolor{darkgreen}{rgb}{0,0.4,0}
\definecolor{golden}{rgb}{0.92, 0.7, 0}
\definecolor{midnight}{rgb}{0, 0, 0.5}
\definecolor{darkblue}{rgb}{0, 0, 0.7}
\definecolor{purple}{rgb}{0.5, 0, 0.5}

\def\sgngam{{\rm sign}(\gamma)}
\def\oneone{\rlap 1\mkern4mu{\rm l}}

\def\coeff#1#2{\relax{\textstyle \frac{#1}{#2}}\displaystyle}

\def\IR{\mathbb{R}}

\def\ZZ{\mathbb{Z}}

\def\cC{{\cal C}}

\def\cG{{\cal G}}
\def\cH{{\cal H}}

\def\cW{{\cal W}}

\def\nBPS#1{$\frac{1}{#1}$-BPS}

\numberwithin{equation}{section}

\begin{document}

\phantom{AAA}
\vspace{-10mm}

\begin{flushright}
%
%IPHT-T19/021\\
%
\end{flushright}

\vspace{1.9cm}

\begin{center}

{\huge {\bf   Effervescent Spikes in M-theory }}\\

{\huge {\bf \vspace*{.25cm}  }}

\vspace{1cm}

{\large{\bf { Iosif Bena$^{1}$, Rapha\"el Dulac$^{1}$, Dimitrios Toulikas$^{1,2}$\\ and  Nicholas P. Warner$^{1,3,4}$}}}
 
\vspace{1cm}

$^1$Institut de Physique Th\'eorique, \\
Universit\'e Paris Saclay, CEA, CNRS,\\
Orme des Merisiers, Gif sur Yvette, 91191 CEDEX, France \\[12pt]

$^2$Department of Physics, \\ 
Ben-Gurion University of the Negev, \\
Beer Sheva 84105, Israel \\ [12pt]

\centerline{$^3$Department of Physics and Astronomy}
\centerline{and $^4$Department of Mathematics,}
\centerline{University of Southern California,} 
\centerline{Los Angeles, CA 90089, USA}

\vspace{10mm} 
{\footnotesize\upshape\ttfamily iosif.bena @ ipht.fr,  raphael.dulac @ ipht.fr,   toulikas @ post.bgu.ac.il, warner @ usc.edu} \\

\vspace{2.2cm}
 
\textsc{Abstract}
\end{center}

\noindent AdS$_3 \times$ S$^3 \times$ S$^3$ solutions warped over a Riemann surface, $\Sigma$, are indexed by a parameter, $\gamma$, that defines the superconformal algebra, $D(2,1; \gamma) \oplus D(2,1; \gamma)$ they preserve. We show that  these solutions come from multiple back-reacted M2-M5 spikes, and that different values of $\gamma$ correspond to different scaling limits of the same M2-M5 solutions. We find that when $\gamma$ switches from positive to negative, the infrared region of the AdS$_3$ switches from the tip of spikes, far from the M5 branes, to the bottom of the spikes, far from the M2 branes. We also explain how the bubbling negative-$\gamma$ solutions emerge from the geometric transition of multiple M2-M5 spikes.

\begin{adjustwidth}{3mm}{3mm} % to adjust the L and R margins

\vspace{-1.2mm}
\noindent

\end{adjustwidth}

%\end{titlepage}
\thispagestyle{empty}
\newpage

%%%%%%%%%%%%%%%%%%%%%%%%%%%%%%%%%%%%%

\tableofcontents

%%%%%%%%%%%%%%%%%%%%%%%%%%%%%%%%%%%%%
\section{Introduction}
\label{sec:Intro}
%%%%%%%%%%%%%%%%%%%%%%%%%%%%%%%%%%%%%

Supersymmetric AdS$_3 \times$ S$^3 \times$ S$^3$ solutions warped over a Riemann surface, $\Sigma$, \cite{Bachas:2013vza} are widely used both as holographic duals of CFT defects and as holographic duals of boundary conformal field theories\footnote{See \cite{Okazaki:2025rjl,Harvey:2025ttz, Harvey:2023pdv, Drukker:2022pxk} for interesting recent progress.}. However, the relation between these solutions and the branes whose back-reaction gives rise to them is not well understood. This relation is necessary, both for the correct application of holography, and for the correct identification of the CFT's dual to  AdS$_3 \times$ S$^3 \times$ S$^3$ backgrounds \cite{Witten:2024yod}. More broadly, brane intersections create spikes \cite{Callan:1997kz, Constable:1999ac}, as one brane pulls on another, and these spikes can exhibit layered self-similar structures, {\it mohawks}  \cite{Bena:2023rzm},  that promote the common brane directions to an AdS space, with its attendant conformal symmetry.

The M-theory solutions are based in the intersections of M2,  M5 and M5' branes along a common $\IR^{1,1}$   \cite{Lunin:2007mj}.  As one approaches the intersection along the three radial directions inside each species of brane, but transverse to the  $\IR^{1,1}$, one  finds a scaling behavior that  promotes the $\IR^{1,1}$ to a Poincar\'e AdS$_3$ \cite{Bena:2023rzm}.  The Riemann surface  then parametrizes  the scale-invariant, or projective, combinations of these three radial directions,  and the S$^3$'s  are  the spheres in the M5 and M5' (transverse to the  $\IR^{1,1}$). All the solutions have some Yang-tableau descriptions that define how the stack of M2's is partitioned to end on stacks of M5's. 

The original construction of these AdS$_3 \times$ S$^3 \times$ S$^3 \times \Sigma$  solutions was also   characterized by a parameter, $\gamma$,   whose physics was somewhat mysterious.  It is one of the purposes of this paper to elucidate the physical meaning of this parameter in terms of the  scaling on the branes, and show that solutions with the same charges but different values of $\gamma$ represent near-brane limits of {\em the same} system of intersecting M2 and M5 branes in flat space, but scaled in a different manner. In particular, we will show that the asymptotic brane configurations define a {\it universal space of Poincar\'e supersymmetries} that are independent of $\gamma$; the $\gamma$-dependence emerges in the superconformal extension of the Poincar\'e algebra and so only emerges in the scaling limit.   We will also show that the sign of $\gamma$ not only determines which directions are scaled up and which ones are being scaled down as one ``zooms in'' on  the brane intersection, but  also exposes, or collapses,  the  geometric transitions of the brane system. 

Mathematically, the parameter, $\gamma$, defines the precise superconformal algebra, $D(2,1; \gamma) \oplus D(2,1; \gamma)$, of the near-intersection limit.  This algebra has a symmetry, $\gamma \to \gamma^{-1}$, that corresponds to interchanging the two  $S^3$'s in the geometry.  One can therefore assume that $-1 \le \gamma \le 1$.  The special values of  $\gamma$ are very succinctly described in \cite{Bachas:2013vza}, and there are two that of particular importance here:  $\gamma= -1/2$ and $\gamma= 1$ .  Mathematically, the real forms of the corresponding superalgebras  are:
\begin{equation} 
D(2,1;\gamma)  ~=~ 
\begin{cases}
\  OSp(4^*|2)  \qquad   \ \ \text{for}  \quad \gamma= -1/2\,, \\ 
\ OSp(4|2,\mathbb{R})  \qquad  \text{for}  \quad \gamma = 1 \,,
\end{cases}
\label{algebra1}
\end{equation} 
and one has
\begin{equation*} 
\begin{aligned} 
D(2,1;\gamma) \oplus D(2,1;\gamma) \subset OSp(8^*|4) ~~ \, & \qquad \ \text{for}   \quad \gamma = -1/2 \,,
\nonumber \\
D(2,1;\gamma) \oplus D(2,1;\gamma) \subset OSp(8|4,\mathbb{R}) & \qquad  \  \text{for} \quad  \gamma = 1\,,
\end{aligned} 
%\label{algebra2}
\end{equation*} 
where the right-hand superalgebras are those associated with ${\rm AdS}_7 \times {\rm S}^4$ and ${\rm AdS}_4 \times {\rm S}^7$  respectively.  

Indeed, in addition to the fact that $\gamma >0$ solutions allow asymptotic regions that are maximally symmetric AdS$_4 \times $S$^7$, while $\gamma <0$ solutions allow  maximally symmetric asymptotic AdS$_7 \times$ S$^4$ regions, it was observed in \cite{Bachas:2013vza} that the $\gamma <0$ and the $\gamma >0$ solutions have drastically different source structure: The $\gamma > 0$ solutions cannot have non-trivial topology and fluxes and, in general, have singular M5 sources, while the solutions with  $\gamma < 0$ are smooth, with no singular sources, having only non-trivial cycles threaded  by cohomological fluxes.  In this paper we will explain in detail how the bubbling solutions come from the geometric transition of multiple M2-M5 spikes, followed by taking a scaling limit that reveals the AdS$_3 \times$ S$^3 \times$ S$^3$ structure.

In \cite{Bena:2024dre}, the solutions with $\gamma = 1$, and a particular choice of Riemann surface, were related   to scaling limits of the supergravity solutions sourced by certain eight-supercharge intersections of M2 and M5 branes in flat space \cite{Lunin:2007mj}.  The spikes created by M2 branes pulling on the M5 brane world-volumes were shown to create a {\it mohawk} structure:  The number of M2 branes ending on a group of M5 branes determines the ``steepness'' of a spike; furthermore, any junction of a large number of M2's terminating on M5 branes will be resolved into multiple spikes nested, one ``inside'' the other,  layered and separated according to their steepness. The radial direction of AdS$_3$ is the direction along these spikes, and every partition of the numbers of M2 branes terminating on M5 branes corresponds to one solution. It is this partitioning that is encoded in the corresponding Young Tableaux.

However, it was not clear if the mohawk is an artifact of the choice of $\gamma$ or the simplicity of the Riemann surface, or is a general feature of all AdS$_3 \times$ S$^3 \times$ S$^3 \times \Sigma$ solutions.  For example, {\it a priori}, it could have been plausible that solutions with general $\gamma$ could be the IR solutions corresponding to a system of bent branes, or to branes placed in non-trivial transverse fluxes, which might still flow in the infrared to conformal theories.

We will show that this is not so: For all values of $\gamma$ the  AdS$_3 \times$ S$^3 \times$ S$^3 \times \Sigma$ solutions are different scaling limits of the same system of 8-supercharge semi-infinite M2 branes ending on multiple M5 branes. Thus, the coincidence of the Young-diagram numerology for positive and negative $\gamma$ observed in \cite{Bachas:2013vza} is no coincidence at all. All these solutions are limits of {\em the same} system of M2 branes ending on M5 and M5' branes.

To understand the physics of $\gamma$ we start by  introducing radial coordinates, $(u,v,z)$, inside the M5, M5' and M2 but transverse to the common $\IR^{1,1}$. Thus $u$ and $v$ are coordinates in the $\IR^4$ inside the  world-volumes of the M5 and M5' branes and $z$ is the remaining ``radial''  coordinate along the M2 brane world-volume.   We will show that these scale as:
\begin{equation}
u  ~\sim~   \lambda^\frac{\gamma  }{(\gamma +1)} \, u  \,,   \qquad  v  ~\sim~   \lambda^\frac{1}{(\gamma +1)}  \, v  \,,   \qquad z  ~\sim~   \lambda^{- \frac{2\gamma  }{(\gamma +1)} }\,   z  \,,
\label{scaling0}
\end{equation}
and, in particular, $u^2 z$ and $ u \,v^{-\gamma}$  are the scale invariant, projective coordinates that define the Riemann surface, $\Sigma$.  The universal scale invariance of  $u^2 z$ is the result of M2 branes ending on and pulling on the M5 branes: the end of the M2 branes is a point in the $\IR^4$ of the M5, whose radial coordinate is  $u$, so the profile of all the M2-M5 ``harmonic spikes'' is $z \sim 1/u^2$. Note also that scaling $\lambda \to 0$ means that one is scaling $v \to 0$ for any value of $\gamma$ (with $-1 < \gamma <1$), but the  scaling behavior of $u$ and $z$ critically depends on the sign of $\gamma$.  For $\gamma > 0$ one has $u \to 0$, $z \to \infty$, which means that one is zooming out along the M2 branes while zooming in on the M5 radial direction.  For $\gamma < 0$, the zooming in $(u,z)$ is exactly the reverse: one zooms out along the M5 branes while zooming in on the M2 ``radial'' direction.  This difference explains precisely why the sign of $\gamma$ favors one or other of  ${\rm AdS}_7 \times {\rm S}^4$ or ${\rm AdS}_4 \times {\rm S}^7$ asymptotics.

From this perspective,  so long as one does not change the sign of $\gamma$, changing its value is a smooth deformation that slices the underlying brane intersection differently.  This suggests that $\gamma$ is a {\it real} parameter, as opposed to a rational parameter,  as concluded in \cite{Bachas:2013vza}.  However, the Riemann surface in \cite{Bachas:2013vza} was chosen to be compact and have a particular form, and the rationality of $\gamma$ was a consequence of choosing an integer basis of cycles on $\Sigma$ and using the quantizaton of the M5  charges.  Based on (\ref{scaling0}), the surface, $\Sigma$, emerges projectively and may involve some singular points if $\gamma$ is not rational.  The underlying brane intersection only cares about the sign of $\gamma$, while its value tells us how we can slice it differently to make different conformal field theories.   

This is a little reminiscent of spectral flow in the holographic duals of the D1-D5 CFT (see, for example, \cite{Bena:2008wt,Giusto:2012yz}): as far as the algebra is concerned, a spectral flow parameter can be any real number, but if the algebra emerges from a compact circle, the parameter must be half-integer, or, at least, rational for an orbifold. It must also be rational if the algebra is to respect the topology of the dual geometry up to orbifolds. 

In Section \ref{sec:GenRes}  we summarize some of the results in  \cite{Lunin:2007mj, Bena:2023rzm}.  The solutions based on AdS$_3 \times$ S$^3 \times$ S$^3 \times \Sigma$   \cite{Lunin:2007ab,DHoker:2008lup,DHoker:2008rje,DHoker:2008wvd,DHoker:2009lky, DHoker:2009wlx,Bobev:2013yra,Bachas:2013vza} are reviewed in  Section \ref{sec:AdSlmits}, and we then show how these solutions emerge as a scaling limit of the generic brane intersections of  \cite{Lunin:2007mj, Bena:2023rzm}.  Remarkably enough, there is a one-parameter family of such scaling limits and this parameter is, of course, $\gamma$.  Since various AdS limits are going to be important to our discussion, we make a brief excursion,  in Section \ref{sec:AdS}, into various coordinatizations of global and Poincar\'e AdS.   In Section \ref{sec:solutions}, we consider  the solutions in \cite{Bachas:2013vza} with $\gamma <0$, and we discuss the M2 charges of these solutions in Section \ref{sec:M2chg-spikes}.  Section \ref{sec:probes} contains an analysis of brane probes in these solutions, which allows us to explain, in Section \ref{sec:Geometric}, how the back-reacted solutions come from the geometric transition of an M2-M5 mohawk, and how this this is visible in $\gamma<0$ solutions. We finish with some concluding remarks in Section \ref{sec:Conclusions}.

%%%%%%%%%%%%%%%%%%%%%%%%%%%%%%%%%%%%%
\section{\nBPS{4} M2-M5-M5' intersections}
\label{sec:GenRes}
%%%%%%%%%%%%%%%%%%%%%%%%%%%%%%%%%%%%%

%%%%%%%%%%%%%%%%%%%
\subsection{The metric and supersymmetries}
\label{ss:metric}
%%%%%%%%%%%%%%%%%%% 

We  construct supersymmetric solutions that have charges corresponding to M2 branes along the directions $012$, M5 branes along the directions $013456$, and a second set of M5 branes, called M5' branes, along the directions $01789\,10$.  We parametrize these directions via $(x^0, x^1,x^2) = (t,x,y)$ and we denote $(x^3, \dots,x^6)$  by the coordinates $\vec u \in \IR^4$, and  the remaining coordinates, $(x^7, \dots,x^{10})$,  by  $\vec v \in {\IR}^4$.  The eleven-dimensional metric has the form:
\begin{equation}
\begin{aligned}
ds_{11}^2 ~=~  e^{2  A_0}\, \bigg[ -dt^2 &~+~ dy^2 ~+~ e^{-3  A_0} \, (-\partial_z w )^{-\frac{1}{2}}\, d \vec u \cdot d \vec u  ~+~ e^{-3  A_0} \, (-\partial_z w )^{\frac{1}{2}}\, d \vec v \cdot d \vec v \,    \\
  &  ~+~  (-\partial_z w ) \, \Big( dz ~+~(\partial_z w )^{-1}\,  \big (\vec \nabla_{\vec u} \, w \big)  \cdot  d \vec u \Big)^2  \bigg]\,,
\end{aligned}
 \label{11metric}
\end{equation}
which is  conformally flat along $(t,y) \in \IR^{(1,1)}$,  $\vec u \in \IR^4$ and  $\vec v \in \IR^4$.  The metric involves a non-trivial fibration of the ``M2 direction,'' $z$, over the   $\IR^4$ of the  M5 brane,   parametrized by coordinates, $\vec u$.  The functions $A_0(\vec u, \vec v, z)$ and $w(\vec u, \vec v, z)$ will be defined below, and, for obvious reasons we require $\partial_z w <0$.  

As noted in \cite{Lunin:2007mj, Bena:2023rzm}, there is actually a democracy between the $\vec u$ and $\vec v$ directions: one can recast  (\ref{11metric}) as a fibration over the $\vec v$-plane by interchanging the role of $z$ and $w$, making  $w$ the coordinate along the M2 direction, and taking $z(\vec u\,, \vec u\,, w)$ to be the unknown function.    

We will use the set of frames:
\begin{equation}
\begin{aligned}
e^0 ~=~  & e^{A_0}\, dt \,, \qquad e^1~=~  e^{A_0}\, dy \,,  \qquad e^2 ~=~    (-\partial_z w )^{\frac{1}{2}} \, \Big( dz ~+~(\partial_z w )^{-1}\,  \big (\vec \nabla_{\vec u} \, w \big)  \cdot  d \vec u \Big) \,, \\
e^{i+2}~=~  & e^{- \frac{1}{2} A_0} \, (-\partial_z w )^{-\frac{1}{4}}\,  du_i \,, \qquad e^{i+6} ~=~  e^{- \frac{1}{2} A_0} \, (-\partial_z w )^{\frac{1}{4}}\,  dv_i\,,   \qquad {i = 1,2,3,4}   \,.
\end{aligned}
 \label{11frames}
\end{equation}

The three-form vector potential is given by:
\begin{equation}
C^{(3)} ~=~   - e^0 \wedge e^1 \wedge e^2 ~+~ \frac{1}{3!}\, \epsilon_{ijk\ell} \,  \Big((\partial_z w )^{-1}\, (\partial_{u_\ell} w) \,  du^i \wedge du^j \wedge du^k ~-~ (\partial_{v_\ell} w)  \, dv^i \wedge dv^j \wedge dv^k  \Big)  \,.
 \label{C3gen}
\end{equation}
where $\epsilon_{ijk\ell}$ is the $\epsilon$-symbol on $\IR^4$.

The supersymmetries of this system will be defined in terms of the frame components along the M2 and M5 directions: 
\begin{equation}
 \Gamma^{012} \, \varepsilon  ~=~ - \varepsilon \,,   \qquad  \Gamma^{013456} \, \varepsilon  ~=~\varepsilon \,. 
 \label{projs1}
\end{equation}
This defines the eight supersymmetries of the M2-M5 system. These projectors define the universal Poincar\'e supersymmetries that are independent of $\gamma$.  (The full specification of these projectors requires a specification of frames and this is given below in (\ref{firstframes}).) 

Recalling that 
\begin{equation}
 \Gamma^{0123456789\,10}    ~=~ \oneone \,,  
 \label{prodgammas}
\end{equation}
one sees that (\ref{projs1})  implies
\begin{equation}
 \Gamma^{01789\,10} \, \varepsilon  ~=~-\varepsilon \,,
 \label{projs2}
\end{equation}
and hence adding the M5' branes along the directions $01789\,10$ does not break the supersymmetry any further.   Thus, despite representing three sets of intersecting branes, the system has eight supercharges (so is \nBPS{4}). 

The goal is then to solve the gravitino equation	
\begin{equation}
\delta \psi_\mu ~\equiv~ \nabla_\mu \, \epsilon ~+~ \coeff{1}{288}\,
\Big({\Gamma_\mu}^{\nu \rho \lambda \sigma} ~-~ 8\, \delta_\mu^\nu  \, 
\Gamma^{\rho \lambda \sigma} \Big)\, F_{\nu \rho \lambda \sigma} ~=~ 0 \,,
\label{11dgravvar}
\end{equation}
subject to the foregoing projection conditions.

%%%%%%%%%%%%%%%%%%%
\subsection{The master function}
\label{ss:master}
%%%%%%%%%%%%%%%%%%% 

Denote the Laplacians on each $\IR^4$ via:
\begin{equation}
{\cal L}_{u} ~\equiv~  \nabla_{\vec u} \cdot \nabla_{\vec u} \,, \qquad {\cal L}_{v} ~\equiv~  \nabla_{\vec v} \cdot \nabla_{\vec v} \,,
 \label{Laps}
\end{equation}
and suppose that $G_0(\vec u, \vec v, z)$ is a solution to the ``master equation:''
\begin{equation}
 {\cal L}_{v} G_0 ~=~  ({\cal L}_{u} G_0)\,(\partial_z \partial_z G_0) ~-~ ( \nabla_{\vec u} \partial_z G_0)\cdot  (  \nabla_{\vec u} \partial_z G_0)\,.
 \label{master}
\end{equation}
One then finds that there are eight Killing spinors solving the BPS equations, (\ref{11dgravvar}), provided one determines $w$ and $A_0$ via:
\begin{equation}
w ~=~ \partial_z G_0  \,, \qquad  e^{-3  A_0} \, (-\partial_z w )^{\frac{1}{2}} ~=~ {\cal L}_{v}  G_0 \,.
 \label{solfns}
\end{equation}
One can also verify that these equations, along with (\ref{master}), imply
\begin{equation}
 e^{-3  A_0} \, (\partial_z w )^{-\frac{1}{2}} ~-~ (\partial_z w )^{-1} \, (\nabla_{\vec u} \,w )\cdot  (\nabla_{\vec u} \, w ) ~=~  - {\cal L}_{u}  G_0 \,.
 \label{reln0}
\end{equation}

The differential equation (\ref{master}) has a very interesting form but it is non-linear  and cannot be explicitly solved   in general.  It also has variant, but very similar, forms for all manners of \nBPS{4} brane intersections \cite{Lunin:2007ab,Lunin:2007mj,Lunin:2008tf}.   Despite the non-linearity, it was argued in \cite{Lunin:2007ab,Lunin:2007mj,Lunin:2008tf}, using perturbative methods, that given a brane distribution specified by boundary conditions and sources, the ``brane-intersection'' equations, like   (\ref{master}), will have a unique solution and so  (\ref{master})  does indeed completely determine the M2-M5-M5' intersections of interest to us.  

If one uses the reformulation described above in which $w$ becomes a coordinate and $z(\vec u\,, \vec u\,, w)$ becomes the unknown function, then the equations governing the solution are precisely those above  but with $\vec u$ interchaged with  $\vec v$ and $z$ interchanged  with $w$ throughout the discussion.

%%%%%%%%%%%%%%%%%%%
\subsection{Imposing spherical symmetry}
\label{ss:sphsymm}
%%%%%%%%%%%%%%%%%%% 

One can easily impose spherical symmetry in the two $\IR^4$'s to arrive at the metric:
\begin{equation}
\begin{aligned}
ds_{11}^2 ~=~ e^{2  A_0}\, \bigg[&  - dt^2 ~+~ dy^2  ~+~  (-\partial_z w ) \, \Big( dz ~+~(\partial_z w )^{-1}\,  \big (\partial_u w \big)  \, d u \Big)^2   \\
& ~+~ e^{-3  A_0} \, (-\partial_z w )^{-\frac{1}{2}}\, \big( du^2  ~+~ u^2 \, d\Omega_3^2 \big)   ~+~ e^{-3  A_0} \, (-\partial_z w )^{\frac{1}{2}}\, \big( dv^2  ~+~ v^2 \, d{\Omega'}_3^2 \big)  
  \bigg]\,,
\end{aligned}
 \label{11metric-symm}
\end{equation}
where $u = |\vec u|$, $v = |\vec v|$ and $d\Omega_3^2$, $d{\Omega'}_3^2$ are the metrics of unit three-spheres in each $\IR^4$ factor. 
The obvious choice for a set of frames is then:
\begin{equation}
\begin{aligned}
e^0 ~=~  &  e^{A_0}\, dt \,, \qquad e^1~=~  e^{A_0}\,  dy \,,  \qquad e^2 ~=~  e^{A_0}\,   (-\partial_z w )^{\frac{1}{2}} \,  \Big( dz ~+~(\partial_z w )^{-1}\,  \big (\partial_u w \big)  \, d u \Big)  \,, \\  e^3 ~=~ &   e^{- \frac{1}{2} A_0} \, (-\partial_z w )^{-\frac{1}{4}}\, du \,,\qquad e^4~=~  e^{- \frac{1}{2} A_0} \, (-\partial_z w )^{\frac{1}{4}}\,  \, dv  \,,\\
e^{i+4}~=~& e^{- \frac{1}{2} A_0} \, (-\partial_z w )^{-\frac{1}{4}}\,    \sigma_i  \,, \qquad e^{i+7}  ~=~ e^{- \frac{1}{2} A_0} \, (-\partial_z w )^{\frac{1}{4}}\, \tilde \sigma_i \,,   \qquad {i = 1,2,3}   \,,
\end{aligned}
 \label{firstframes}
\end{equation}
where $\sigma_i$ and $\tilde \sigma_i$ are left-invariant one-forms on the unit three-spheres.

Similarly one has:
\begin{equation}
C^{(3)} ~=~   - e^0 \wedge e^1 \wedge e^2 ~+~  (\partial_z w )^{-1} \, \big (u^3 \partial_u w \big)  \, {\rm Vol}({S^3}) ~+~  \, \big(v^3 \partial_v w \big) \,  {\rm Vol}({{S'}^3})    \,,
 \label{C3symm}
\end{equation}
where ${\rm Vol}({S^3})$ and ${\rm Vol}({{S'}^3})$ are the volume forms of the unit three-spheres.  Note there is a sign-flip in the flux along the ${S'}^3$ compared to  (\ref{C3gen}).  This is because of the orientation change in  (\ref{firstframes})  compared to (\ref{11frames}) where the $e^4$ is now the radial $v$-direction.

%%%%%%%%%%%%%%%%%%%%%%%%%%%%%%%%%%%%%
\section{Near-brane M5-M2 intersections}
\label{sec:AdSlmits}
%%%%%%%%%%%%%%%%%%%%%%%%%%%%%%%%%%%%%

The  \nBPS{4} geometry  created by intersecting M2, M5 and M5' branes can have a  near-brane limit that includes an AdS$_3$  factor \cite{Bena:2024dre}. One   find a family of such solutions by searching for solutions with an $SO(2,2) \times SO(4) \times SO(4)$ isometry and whose geometry contains factors of AdS$_3$ $\times S^3 \times S^3$. The most general such geometry can therefore depend on the remaining two spatial directions, whose coordinates will be denoted by $(\xi,\rho)$.  Such  solutions have been extensively studied in  \cite{deBoer:1999gea,Lunin:2007ab,DHoker:2008lup,DHoker:2008rje,DHoker:2008wvd,DHoker:2009lky, DHoker:2009wlx,Bobev:2013yra,Bachas:2013vza}.

%%%%%%%%%%%%%%%%%%%
\subsection{The  AdS$_3$ $\times S^3 \times S^3 \times \Sigma_2$ solutions}
\label{ss:AdS3sols}
%%%%%%%%%%%%%%%%%%%

The Ansatz \cite{Bachas:2013vza} makes complete use of the isometries: 
\begin{equation}
\begin{aligned}
ds_{11}^2 &~=~   e^{2A} \, \big( \, \hat f_1^2 \, ds_{AdS_3}^2 ~+~ \hat f_2^2 \, ds_{S^3}^2 ~+~ \hat f_3^2 \, ds_{{S'}^3}^2 ~+~ h_{ij} d x ^i  d x^j \,   \big)  \,, \\
C^{(3)}  &~=~  b_1\, \hat e^{012} ~+~ b_2\, \hat e^{345} ~+~ b_3\, \hat e^{678}  \,,
\end{aligned}
 \label{DHokerAnsatz}
\end{equation}
where the metrics $ds_{AdS_3}^2$, $ds_{S^3}^2$  and $ds_{{S'}^3}^2$ are the metrics of unit radius on  AdS
and  the three-spheres and $\hat e^{012}$, $\hat e^{345}$   and $\hat e^{678}$ are the corresponding unit volume forms. In \cite{Bachas:2013vza} there was no specified form of the AdS$_3$, but here we will take it to be Poincar\'e AdS because we wish to connect the solution to the near-brane limit of flat branes.

The functions, $e^{2A}$, $\hat f_j$, $b_j$, and the two-dimensional metric, $h_{ij}$, are, {\it a priori}, arbitrary functions of $(x^1 = \xi, x^2 = \rho)$ (and the $e^{2A}$ factor is redundant, but has been introduced for later convenience).  However, the final result in  \cite{Bachas:2013vza} is to pin down all these functions and express them in terms of a complex function, $G$, and a real function, $h$, that satisfy some linear equations.  The challenge is then to find the solutions for $G$ and $h$ that lead to sensible supergravity backgrounds.

First, the two dimensional metric must be that of a Riemann surface with K\"ahler potential, $\log(h)$:
\begin{equation}
h_{ij} d x ^i  d x^j ~=~ \frac{\partial_\zeta h \partial_{\bar \zeta}   h }{  h^2}  \, |d\zeta|^2    \,,
 \label{KahlerMet}
\end{equation}
where $\zeta$ is a complex coordinate and  $h$ is required to be harmonic:
\begin{equation}
 \partial_\zeta  \partial_{\bar \zeta} h  ~=~ 0   \,.
 \label{harmonic}
\end{equation}
We will define real and imaginary parts of $w$ via:
\begin{equation}
\zeta ~=~  \xi ~+~ i\, \rho \quad \Rightarrow \quad \partial_\zeta ~=~  \coeff{1}{2}\, \big( \partial_\xi ~-~ i\, \partial_\rho \big) \,, \qquad \partial_{\bar{\zeta}}~=~  \coeff{1}{2} \,\big( \partial_\xi ~+~ i\, \partial_\rho \big)  \,.
 \label{zetaparts}
\end{equation}
It is also convenient to introduce the harmonic conjugate, $\tilde h$, of $h$, defined by requiring that  $-\tilde h + i h$ be holomorphic:
\begin{equation}
 \partial_{\bar \zeta}  (- \tilde h + i h)  ~=~   0 \,.
 \label{harmconj}
\end{equation}
Since $-\tilde h + i h$ is holomorphic we can use $(\tilde{h},h)$   as local coordinates on the Riemann surface, or, equivalently we can take 
\begin{equation}
- \tilde h ~+~ i h   ~=~   \beta \, \zeta ~=~   \beta \, ( \xi ~+~ i\, \rho )\,,
 \label{simph}
\end{equation}
where $\beta$ is a constant parameter introduced for later convenience.   Since we will ultimately be interested in the Poincar\'e half plane, we will impose 
\begin{equation}
h\,, \beta \,, \rho   ~>~  0   \,.
 \label{posithbr}
\end{equation}

Thus  we may (locally) take  the Riemann surface metric to be a multiple of that of the Poincar\'e upper half-plane:  
\begin{equation}
h_{ij} d x ^i  d x^j ~=~ \frac{ d\xi^2 ~+~ d\rho^2}{4 \, \rho^2}  \,,
 \label{PoincareMet}
\end{equation}
where the factor of $4$ comes from the factors of $\frac{1}{2}$ in partial derivatives (\ref{zetaparts}). 

The complex function, $G$, is required to satisfy the (linear) equation: 
\begin{equation}
 \partial_\zeta   \, G ~=~  \coeff{1}{2}\, (\, G  +  \overline G\,) \,  \partial_\zeta \log(h) \,. 
 \label{Geqn}
\end{equation}
If one writes $G$ in terms of real and imaginary parts, $G = g_1 + i g_2$, and uses the local coordinates (\ref{simph}), then one has:
\begin{equation}
 \partial_\xi  g_1 ~+~  \partial_\rho    g_2 ~=~ 0 \,,  \qquad  \partial_\xi  g_2 ~-~   \partial_\rho    g_1 ~=~ - \frac{1}{\rho} \, g_1\,. 
 \label{gjeqns1}
\end{equation}

It is convenient to introduce potentials, $\Phi$, and $\tilde \Phi$, associated with $G$.  First, one defines $\Phi$ via:
\begin{equation}
 \partial_\zeta   \, \Phi  ~=~  \overline G \,  \partial_\zeta  h  \qquad \Leftrightarrow \qquad \partial_\xi \Phi ~=~ - \beta \, g_2  \,, \quad  \partial_\rho \Phi ~=~ \beta \, g_1 \,. 
 \label{Phidefn}
\end{equation}
The existence of such a $\Phi$ is guaranteed by the first equation in (\ref{gjeqns1}). The second equation in (\ref{gjeqns1}) implies that $\Phi$ must satisfy
\begin{equation}
\Big(\partial_\xi^2  ~+~ \partial_\rho^2 ~-~ \frac{1}{\rho} \,  \partial_\rho \Big)   \, \Phi ~=~  0 \,. 
 \label{Phieqn}
\end{equation}

Similary, the  second equation in (\ref{gjeqns1}) implies that there is a conjugate potential, $\tilde \Phi$, defined by:
\begin{equation}
\partial_\xi \tilde \Phi ~=~ - \frac{\beta}{\rho} \, g_1 ~=~ - \frac{1}{\rho} \, \partial_\rho \Phi     \,, \qquad  \partial_\rho \tilde \Phi ~=~  - \frac{\beta}{\rho} \, g_2 ~=~  \frac{1}{\rho} \, \partial_\xi \Phi  \,. 
 \label{Phidefns}
\end{equation}
The first equation in (\ref{gjeqns1}) then implies that $\tilde \Phi$ must satisfy
\begin{equation}
\partial_\xi^2 \tilde \Phi  ~+~ \frac{1}{\rho} \, \partial_\rho \big( \, \rho\,  \partial_\rho \tilde \Phi  \, \big)   \, ~=~  0 \,. 
 \label{tPhieqn}
\end{equation}

Finally, define the functions:
\begin{equation}
W_\pm  ~\equiv~  | G ~\pm~ i |^2 ~+~ \gamma^{\pm 1} \, ( G    \overline G~-~1)   \,, 
 \label{Wpmdefn}
\end{equation}
where $-1 \le \gamma \le 1$ is the deformation parameter that appears in the relevant exceptional superalgebra $D(2,1;\gamma)  \oplus D(2,1;\gamma)$ \cite{Bachas:2013vza}.

The representation structure is  invariant under $\gamma \to \gamma^{-1}$.  Moreover, in the supergravity solution this merely exchanges the two $S^3$'s. We can therefore restrict $ -1 \le \gamma \le 1$, and because the AdS$_3$ factor reduces to $\mathbb{R}^{1,2}$  at $\gamma=-1$  \cite{Bachas:2013vza},  we will take:
\begin{equation}
-1 ~<~ \gamma ~\le~  1  \,.
 \label{gam-rng}
\end{equation}
 We also recall from \cite{Bachas:2013vza} that there is a discontinuity at $\gamma=0$ (and at $\gamma=\infty$) because one of the $S^3$'s decompactfies to an $\IR^3$.  

The sign of $\gamma$ is determined by its relation to the magnitude of $G$ via:
\begin{equation}
 \gamma \, ( G    \overline G~-~1) ~\ge~ 0  \,.
 \label{gammacond}
\end{equation}
The metric functions in  (\ref{DHokerAnsatz})  are given by: 
\begin{equation}
 \hat f_1^{-2}   ~=~   \gamma^{-1} \, (\gamma +1)^2 \,  ( G    \overline G~-~1)  \,, \qquad  \hat f_2^{-2}    ~=~  W_+ \,, \qquad \hat f_3^{-2}   ~=~  W_-  \,,   
 \label{f-functions}
\end{equation}
and
\begin{equation}
\begin{aligned}
  e^{6A}   ~=~   {\rm sign}(\gamma)\, h^2\, ( G    \overline G~-~1)  \, W_+\, W_-   ~=~ | \gamma | \, (\gamma +1)^{-2}  \,   h^2\,  \hat f_1^{-2}   \,  \hat f_2^{-2}  \,  \hat f_3^{-2}      \,.
\end{aligned}
 \label{eA-function}
\end{equation}

The flux functions, $b_i$, are given by:
\begin{equation}
\begin{aligned}
b _1  &~=~ \frac{\nu _1}{c_1^3} \, \hat b_1 ~=~ \frac{\nu _1}{c_1^3}\, \bigg[\,    \frac{ h \, (G + \overline G) }{(G    \overline G~-~1) } ~+~ \gamma^{-1} \, (\gamma +1)^2 \, \Phi ~-~ (\gamma - \gamma^{-1})\, \tilde h  ~+~ b _1^0\, \bigg]\,, \\
b _2  &~=~ \frac{\nu _2}{c_2^3} \, \hat b_2 ~=~ \frac{  \nu _2}{c_2^3}\, \bigg[ -\gamma \, \frac{ h \, (G + \overline G) }{W_+ } ~+~ \gamma (\Phi - \tilde h ) ~+~  b _2^0\, \bigg]\,,  \\ 
b _3 & ~=~ \frac{\nu _3}{c_3^3} \, \hat b_3 ~=~   \frac{\nu _3}{c_3^3}\, \bigg[\,   \frac{  1}{\gamma }\, \frac{ h \, (G + \overline G) }{  W_-} ~-~ \gamma ^{-1}  (\Phi +\tilde h ) ~+~ b _3^0\,\bigg]\,,
\end{aligned}
\label{bfunctions}
\end{equation}
where the $b_j^0$ are arbitrary constants, representing a choice of  gauge that will be discussed later.
The coefficients, $c_j$ are determined \cite{Bachas:2013vza}  in terms of $\gamma$ via:
\begin{equation}
c_1  = {\rm sign}(\gamma)\, |\gamma|^{1/ 2} + |\gamma|^{-1/ 2}    \,,   \qquad c_2 =  -  {\rm sign}(\gamma)\,  |\gamma|^{1/ 2}    \,, \qquad c_3 = -|\gamma|^{-1/ 2}    \,.
 \label{cparams}
\end{equation}
and the $\nu_i $ are simply signs, with $|\nu_i| =1$. Supersymmetry constrains these signs so that 
\begin{equation}
\nu_1\nu_2\nu_3= - \tilde \sigma\,, \label{sigsusy}
\end{equation}
where  $\tilde \sigma$ is set by choosing the sign of the square-root of (\ref{eA-function}): 
\begin{equation}
\begin{aligned}
  e^{3A} \, c_1 \, c_2\,c_3 \,  \hat f_1   \,  \hat f_2   \,  \hat f_3  ~=~ \tilde \sigma  \,   h \,      \,,
\end{aligned}
 \label{sigdefn}
\end{equation}
and requiring $h >0$.

%%%%%%%%%%%%%%%%%%%
\subsection{Mapping the AdS$_3$ solutions to M2-M5-M5' intersections}
\label{ss:mapping}
%%%%%%%%%%%%%%%%%%%

Our goal here is to show how to map the AdS$_3$ solutions of Section \ref{ss:AdS3sols} into the spherically-symmetric M2-M5-M5' brane-intersection solutions of Section \ref{ss:sphsymm}.  

The first step is to remember that the AdS$_3$ comes from a scaling invariance in the  near-brane limit, and that the Riemann surface coordinates, $(\rho, \xi)$, must emerge projectively from the ``radial variables,'' $(u,v,z)$.

We therefore use a scale variable, $\mu$, and  write the AdS$_3$ metric as a Poincar\'e section:
\begin{equation}
 ds_{AdS_3}^2  ~=~   \frac{d \mu^2}{\mu^2} ~+~ \mu^2 \, \big(-dt^2 ~+~ dy^2 \big)\,,
 \label{PoinAdS}
\end{equation}
where the Poincar\'e $\IR^{1,1}$ factor represents the common directions of the brane intersection, and is to be identified with the same factor in  (\ref{11metric-symm}).

A direct comparison of  (\ref{11metric-symm}) and  (\ref{DHokerAnsatz}), using (\ref{PoinAdS}) and  (\ref{PoincareMet}), leads to:
\begin{equation}
 e^{2A}\, \hat f_1^2\, \mu^2  ~=~ e^{2A_0}  \,, \qquad   e^{2A}\, \hat f_2^2   ~=~ e^{- A_0}  \,(-\partial_z w)^{-\frac{1}{2}} \,  u^2  \,, \qquad   e^{2A}\, \hat f_3^2   ~=~ e^{- A_0}  \,(-\partial_z w)^{\frac{1}{2}} \,  v^2   \,,
 \label{identities1}
\end{equation}
along with 
\begin{equation}
\begin{aligned}
e^{2A} \, \bigg( \, \hat f_1^2 \,  \frac{d \mu^2}{\mu^2}  ~+~ \frac{ d\xi^2 ~+~ d\rho^2}{4 \, \rho^2}   \bigg) ~=~  & e^{- A_0}  \Big( (-\partial_z w)^{-\frac{1}{2}} \, du^2~+~  (-\partial_z w)^{\frac{1}{2}} \, dv^2  \Big)  \\
& +~  e^{2 A_0} \,  (-\partial_z w ) \, \Big( dz ~+~(\partial_z w )^{-1}\,  \big (\partial_u w \big)  \, d u \Big)^2 \,.
\end{aligned}
 \label{identities2}
\end{equation}
Using   (\ref{simph}), (\ref{f-functions}),  (\ref{eA-function}) and  (\ref{identities1}) in (\ref{identities2}) one finds that one must have:
\begin{equation}
\begin{aligned}
   \frac{\gamma}{( 1+\gamma)^2 }  & \,    \frac{1}{( G    \overline G -1) }   \,  \frac{d \mu^2}{\mu^2}  ~+~ \frac{ d\xi^2 ~+~ d\rho^2}{4 \, \rho^2} \\
~=~  &   \frac{1}{ W_+}   \, \frac{du^2}{u^2} ~+~    \frac{1}{ W_-}  \,  \frac{dv^2}{v^2}   ~+~ \frac{{\rm sign}(\gamma)}{\beta^2\,  \rho^2\,  ( G    \overline G -1) }  \frac{W_+}{ W_- }   \, \bigg( u^2 dz ~+~ \tilde b_2  \, \frac{d u}{u} \bigg)^2 \,,
\end{aligned}
 \label{identities3}
\end{equation}
where
\begin{equation}
\tilde b_2 ~\equiv~  (\partial_z w )^{-1}\,  \big (u^3 \partial_u w \big) \,.
 \label{tb2defn}
\end{equation}
Note that $\tilde b_2$ matches part of the gauge potential (\ref{C3symm}) and should therefore match the gauge potential term, $b_2$, in (\ref{DHokerAnsatz}), which is given by (\ref{bfunctions}).

One can also manipulate  (\ref{identities1}), using (\ref{eA-function}) and (\ref{simph}),  to obtain:
\begin{equation}
u^2 v^2  ~=~ \frac{\beta^2 \, |\gamma|}{(\gamma +1)^2} \,\,\mu^2  \, \rho^2    \,, \qquad    (-\partial_z w)\, \frac{v^2}{u^2} ~=~ \frac{W_+}{ W_-}   \,,\qquad   e^{A_0}~=~  \frac{ \beta  \sqrt{|\gamma|} }{(\gamma +1)} \, \mu\, \rho\,e^{-2 A}  (W_+ W_-)^{\frac{1}{2}} \,.
\label{identities4}
\end{equation}

The  metric  (\ref{DHokerAnsatz})  is scale invariant under:
\begin{equation}
\mu   ~\to~  \lambda \,\mu  \,, \qquad (t,y) ~\to~  \lambda^{-1} (t,y) \,,
 \label{scaling1}
\end{equation}
with all other parts of the metric, and the other coordinates, independent of this scaling.   We  therefore  impose this on  (\ref{11metric-symm})  by requiring the radial coordinates and metric functions, each have a fixed scaling dimension: 
\begin{equation}
(u \,, v \,, z \,,  w)  ~\to~   (\lambda^{\alpha_1}\, u \,,    \lambda^{\alpha_2}\,v \,, \lambda^{\alpha_3}\,z \,,     \lambda^{\alpha_4}\,w)   \,, \qquad   e^{A_0} ~\to~ \lambda^{\alpha_5} \, e^{A_0}   \,,
 \label{scaling3}
\end{equation}
for some constants, $\alpha_j$.  

Using this in  (\ref{identities1}) and (\ref{identities2}) leaves one with a single scaling parameter:
\begin{equation}
(u \,, v \,, z \,,  w)  ~\to~   (\lambda^{\alpha}\, u \,, \lambda^{1- \alpha}\,v \,, \lambda^{-2 \alpha}\,z \,,  \lambda^{-2(1-\alpha)}\,w)   \,, \qquad   e^{A_0} ~\to~ \lambda\, e^{A_0} \,.
 \label{scaling4}
\end{equation}
This leads to the Ansatz:
\begin{equation}
\begin{aligned}
u  &~=~ \mu^{\alpha}\, m_1 (\xi,\rho)  \,,   \qquad  v  ~=~ \mu^{1-\alpha}\, m_2 (\xi,\rho)  \,,   \qquad  z  ~=~\mu^{-2\alpha}\, m_2 (\xi,\rho)  \,,   \\
  w  &~=~ \mu^{-2(1-\alpha)}\, m_4  (\xi,\rho) \,,   \qquad  e^{A_0}  ~=~\mu  \, m_5 (\xi,\rho)  \,,
\end{aligned}
 \label{variables1}
\end{equation}
for some functions, $m_j$.

The first identity in  (\ref{identities4}), and the form of the fibration on the right-hand side of  (\ref{identities3}), leads to a slightly more refined change of variables:
\begin{equation}
u  ~=~  a\, \mu^{\alpha}\,    e^{\,p_1 (\xi,\rho)}  \,,   \qquad  v  ~=~  a\, \mu^{1-\alpha}\,  \rho\,  e^{-p_1 (\xi,\rho)}     \,,   \qquad  z  ~=~a^{-2} \, \mu^{-2 \alpha} \,e^{-2 p_1 (\xi,\rho)}  \,  p_2 (\xi,\rho) \,,
 \label{variables2}
\end{equation}
where the $p_j$ are arbitrary functions and 
\begin{equation}
a ~\equiv~  \bigg( \frac{\beta \, \sqrt{|\gamma|}}{(\gamma +1)}\bigg)^{1/2} \,.
\label{adefn}
 \end{equation}

Substituting the change of variable (\ref{variables2}) into  (\ref{identities3})  generates an overdetermined system of equations for $b_2$ and the derivatives of the $p_j$.    {\it A priori},  this system is very complicated, involving square-roots of a quadratic in $W_\pm$. However, the system dramatically simplifies when one takes:
\begin{equation}
\alpha ~=~ \frac{\gamma  }{(\gamma +1)}  \qquad \Rightarrow  \qquad 1- \alpha ~=~ \frac{ 1}{(\gamma +1)}\,.
\label{alphadefn}
 \end{equation}
Indeed, the system collapses to:
\begin{equation}
\begin{aligned}
  \partial_\xi p_1 &~=~ -\frac{1}{2 \rho} \, g_1 \,,  \quad \partial_\rho p_1 ~=~ \frac{1}{2 \rho} \, (g_2 +1) \,, \qquad \tilde b_2 ~=~ 2 \, p_2 ~+~   \sgngam \, \varepsilon_1 \, \frac{2\,\beta\, \rho\, g_1}{\sqrt{|\gamma |} \,W_+} \,,   \\
   \partial_\xi p_2 &~=~   \sgngam \, \varepsilon_1 \,\frac{\beta}{2\, \sqrt{|\gamma |} } \, (g_2 -1) \,,  \quad \partial_\rho p_2 ~=~   - \sgngam \, \varepsilon_1 \,\frac{\beta}{2\, \sqrt{|\gamma |} } \,g_1 \,,
\end{aligned}
 \label{diffconstraints}
\end{equation}
where $g_1$ and  $g_2$  are the real and imaginary parts of $G$, and $\varepsilon_1 =\pm 1$.  We have fixed  some of the signs in our analysis so that  (\ref{gjeqns1})  provides the integrability conditions for the $p_j$.

One can integrate this system to arrive at
\begin{equation}
p_1 ~=~ \frac{1}{2}\, \log(\rho)~-~   \frac{1}{2\, \beta} \,  \tilde \Phi \,, \qquad 
p_2~=~ -   \frac{\sgngam\, \varepsilon_1 }{2\, \sqrt{|\gamma |}} \,  \big(  \Phi ~+~ \beta \, \xi \big)\,.
\label{pjres}
 \end{equation}
Using  (\ref{diffconstraints}) one finds that $\tilde b_2$ must have the form:
\begin{equation}
\tilde b_2 ~=~     \frac{\sgngam\, \varepsilon_1 }{ \sqrt{|\gamma |}} \,  \bigg( \frac{ 2\, \beta \, \rho\, g_1}{W_+} ~-~ \big(  \Phi + \beta \, \xi \big)  \bigg)~=~     \frac{\sgngam\, \varepsilon_1 }{ \sqrt{|\gamma |}} \,  \bigg(  \frac{ h \,(G + \overline G)}{W_+} ~-~ \big(  \Phi - \tilde h \big)  \bigg) \,.
\label{b2deduced}
 \end{equation}
 From (\ref{cparams}) one sees that  $c_2^3 =- \gamma |\gamma |^{1/2} $   and hence 
\begin{equation}
b_2 ~=~  \frac{  \nu _2}{\sqrt{|\gamma |}}\, \bigg[ \frac{ h \, (G + \overline G) }{W_+ } ~-~ (\Phi - \tilde h )  \bigg]\,,
\label{b2DHoker}
 \end{equation}
where we have dropped the constants of integration. There is thus a perfect match between (\ref{b2deduced}) and (\ref{b2DHoker}) if $\nu_2 =\sgngam\,  \varepsilon_1$.

To summarize, one finds a perfect match between the results of Section \ref{ss:AdS3sols} and the solutions describing spherically-symmetric M2-M5-M5' intersections of Section \ref{ss:sphsymm} if one takes:
\begin{equation}
u  =  a\, \mu^\alpha \, \rho^{\frac{1}{2}}   \, e^{- \frac{1}{2\, \beta} \,  \tilde \Phi}  \,,   \qquad  v  =  a\, \mu^{1-\alpha} \, \rho^{\frac{1}{2}}   \, e^{+ \frac{1}{2\, \beta} \,  \tilde \Phi}   \,,   \qquad  z  =  - \frac{\varepsilon_1\, (1+\gamma) }{2\, \beta\, \gamma} \,  \mu^{-2\,\alpha} \, \rho^{-1} \,   e^{\frac{1}{\beta} \,  \tilde \Phi }  \,    \big(  \Phi +  \beta \, \xi \big) \,,
 \label{variables3}
\end{equation}
where 
\begin{equation}
\alpha ~=~ \frac{\gamma  }{(\gamma +1)}  \,, \qquad a ~\equiv~  \bigg( \frac{\beta \, \sqrt{|\gamma|}}{(\gamma +1)}\bigg)^{1/2} \,.
\label{aparams}
 \end{equation}
This implies the  scaling noted in (\ref{scaling0}).  Indeed, we have:
\begin{equation}
u  ~\sim~   \mu^\frac{\gamma  }{(\gamma +1)}   \,,   \qquad  v  ~\sim~   \mu^\frac{1}{(\gamma +1)}    \,,   \qquad z  ~\sim~   \mu^{- \frac{2\gamma  }{(\gamma +1)} }   \,.
 \label{scaling2}
\end{equation}

The function,  $w$,  is given by:
\begin{equation}
    w~=~ \frac{\varepsilon_1~ (1+\gamma)}{2~\beta~\rho~\mu^{2(1-\alpha)}}e^{-\frac{1}{\beta}\tilde{\Phi}}\left(\Phi-\beta \xi\right) \,.
\label{wsimp}
\end{equation}
Note, in particular, that we have the following scale invariant combinations:
\begin{equation}
u^2 z ~=~  -\sgngam\,  \frac{\varepsilon_1}{2\, \sqrt{|\gamma|}} \,     \big(  \Phi +  \beta \, \xi \big)   \,, \qquad v^2 w ~=~  \frac{\varepsilon_1~ \sqrt{|\gamma|}}{2} \,\big(\Phi-\beta \xi\big)  \,.
\label{scalinv1}
 \end{equation}
%

%%%%%%%%%%%%%%%%%%%%%%
\subsection{Some comments on scaling}
\label{ss:scalingcomments}
%%%%%%%%%%%%%%%%%%%%%%

In the Poincar\'e metric  (\ref{PoinAdS}),  one approaches the infrared as $\mu \to 0$.  If we restrict to regions in which $\rho$, $\Phi$ and $\tilde \Phi$ are finite, then it is relatively straightforward to identify the branes that dominate the AdS$_3 $ infrared.

From  (\ref{scaling2}) and  (\ref{gam-rng})  it is evident that $\mu \to 0$ means  $v \to 0$, and hence we are going to the origin in the $\IR^4$ spanned by the M5' branes, and thus focusing on the intersection locus of the M2 branes with the M5 branes.   If $\gamma >0$, then we also have $u \to 0$ and $z \to \infty$, which means we are zooming into the branes located at the origin of the $\IR^4$ occupied by the M5 branes but zooming out along the M2 direction, far up on the M2 spike. This limit is dominated by the M2 branes.  If $\gamma  < 0$, then we  have $z \to 0$ and $u \to \infty$, which means we are zooming in on a point\footnote{Rather than a point we mean, of course, the particular $\IR^{1,1}$ defined by $(t,y)$ at $z=0$.} on the M2 branes, while zooming out  along the $\IR^4$  spanned by the M5 branes.  This limit is dominated by the M5 branes.   

As discussed extensively in  \cite{Bachas:2013vza,Bena:2023rzm}, the singularities in  $\Phi$ and $\tilde \Phi$ represent singular brane sources, typically with very distorted (AdS$_3 \times S^3$) world-volumes.

We also note the symmetry $\gamma \to \gamma^{-1}$ is very much a feature of the brane scaling.  Indeed, one sees from  (\ref{Wpmdefn}) that one has invariance under:
\begin{equation}
\gamma ~\to~   \gamma^{-1} \,,  \qquad  W_\pm  ~\to~  W_\mp \,, \qquad  G  ~\to~  -G \,,
\label{invol1}
 \end{equation}
which implies, from (\ref{Phidefns}) and (\ref{bfunctions}), that 
\begin{equation}
\Phi  ~\to~ - \Phi  \,,  \qquad  \tilde \Phi  ~\to~ - \tilde \Phi  \,, \qquad  b_1  ~\to~  -b_1 \,, \qquad  b_2 ~\leftrightarrow~  b_3 \,,
\label{invol2}
 \end{equation}
which makes it evident that the  role of the two $S^3$'s is being flipped.  It also follows from   (\ref{variables3}) and (\ref{wsimp})  that 
\begin{equation}
  u ~\leftrightarrow~  v \,, \qquad  z ~\leftrightarrow~   w \,,
\label{invol3}
 \end{equation}
underlining the democracy between the M5 and M5' branes.  We have broken that democracy by taking (\ref{gam-rng}) and using the $z$-fibration ansatz of Section \ref{ss:sphsymm}.

It is also important to note that one can relate the Poincar\'e supersymmetries defined by (\ref{firstframes}) and (\ref{projs1}) to the supersymmetry analysis of  \cite{Bachas:2013vza}, and to the natural system of frames for the metric  (\ref{DHokerAnsatz}), by substituting the coordinate change  (\ref{variables3}) into (\ref{firstframes}).  This gives the decomposition of $e^2$, $e^3$ and $e^4$ into the coordinates $(\mu, \rho, \xi)$, and hence to a standard system of frames for the metric (\ref{DHokerAnsatz}).   This provides the local frame rotation that maps the supersymmetry analysis presented here to that of \cite{Bachas:2013vza}.  (An explicit example of this was discussed in detail in \cite{Bena:2023rzm}.)  One should note that, because $(u,v,z)$ depend on different powers of $\mu$, the frame rotation depends explicitly on $\gamma$, and so this determines how the embedding of the Poincar\'e supersymmetries into the full superconformal structure depends on $\gamma$.   Thus the Poincar\'e supersymmetries are universally defined by  (\ref{firstframes}) and (\ref{projs1}), but their relation to the superconformal algebra depends upon the scaling of the flat-brane coordinates, and hence upon $\gamma$.

 %%%%%%%%%%%%%%%%%%%%%%%%%%%%%%%%%%%%%
\section{A brief summary of AdS patches}
\label{sec:AdS}
%%%%%%%%%%%%%%%%%%%%%%%%%%%%%%%%%%%%%

%%%%%%%%%%
\def\adsvar{\sigma}
%%%%%%%%%%

We start by considering a unit hyperbolic surface in flat $\IR^{p+q, 2}$:
\begin{equation}
 X_0^2 ~+~ X_{p+q+1}^2 ~-~ \sum_{j=1}^{p+q} \,  X_{j}^2 ~=~   1 \,.
\label{hyperbola1}
 \end{equation}
Now write the  coordinates $X_{p}, \dots, X_{p+q}$ in terms of an $S^{q}$-sphere of radius $ \sinh \adsvar$:
\begin{equation}
  \sum_{j=p}^{p+q} \,  X_{j}^2 ~=~   \sinh^2 \adsvar \,
\label{sphere1}
 \end{equation}
It follows that the remaining coordinates define a  $(p,1)$-dimensional hyperbolic surface, $\cH^{p,1}$, of radius $ \cosh \adsvar$:
\begin{equation}
 X_0^2 ~+~ X_{p+q+1}^2 ~-~   \sum_{j=1}^{p-1} \,  X_{j}^2 ~=~   \cosh^2 \adsvar \,.
\label{hyperbola2}
 \end{equation}

Using the standard route to obtaining the global AdS metric  from these hyperbolic surfaces, one finds that 
\begin{equation}
ds_{AdS_{p+q+1}}^{2 \, \text{global}}  ~=~   d\adsvar^2 ~+~ \cosh^2 \adsvar \, ds_{AdS_{p}}^{2 \, \text{global}} ~+~  \sinh^2 \adsvar \, ds_{S^{q}}^2   \,,
\label{AdSmet1}
 \end{equation}
where all the $ds^2$'s are unit metrics.

The range of $\adsvar$ has one minor subtlety.  For $ q >0$, the sphere metric defined by (\ref{sphere1}) means $0 \le \adsvar < \infty$.  However, for $q=0$, the zero-sphere is the two points, $\{-1 \,,+1\}$,  or, equivalently, one must replace  (\ref{sphere1}) by:
\begin{equation}
X_{p} ~=~   \sinh \adsvar \,
\label{Xpform}
 \end{equation}
 and therefore take $-\infty < \adsvar < \infty$.
 
 Thus, for $q=0$ one has:
\begin{equation}
ds_{AdS_{p+1}}^{2 \, \text{global}}  ~=~   d\adsvar^2 ~+~ \cosh^2 \adsvar \, ds_{AdS_{p}}^{2 \, \text{global}}    \,,
\label{AdSmet2}
 \end{equation}
where $-\infty < \adsvar < \infty$.

The Poincar\'e patch of the global AdS$_{p+1}$ can be defined by the light cone:
\begin{equation}
r  ~\equiv~ X_0 ~-~ X_{p-1}   ~\ge~   0  \,,
\label{lightcone1}
 \end{equation}
which also defines a Poincar\'e patch of the original global AdS$_{p+q+1}$.  So we have
\begin{equation}
ds_{AdS_{p+q+1}}^{2 \, \text{Poincar\'e}}  ~=~   d\adsvar^2 ~+~ \cosh^2 \adsvar \, ds_{AdS_{p}}^{2 \, \text{Poincar\'e}} ~+~  \sinh^2 \adsvar \, ds_{S^{q}}^2   \,,
\label{AdSmet3}
 \end{equation}
with  $-\infty < \adsvar < \infty$ for $q=0$ and $0 \le  \adsvar < \infty$ for $q>0$.

To be more explicit, it is convenient to introduce  $\vec x \in \IR^{q+1}$ with   $\vec x~\sim~ (X_{p}, \dots, X_{p+q})$  and  $(t, \vec y ) \in \IR^{1,p-2}$  to parametrize the Poincar\'e slices of AdS$_{p}$.  Then the Poincar\'e patch of the original global AdS$_{p+q+1}$ can be obtained from:
\begin{equation}
\begin{aligned}
X_0 & ~=~ \frac{1}{2\,r} \, \bigg[ 1 + r^2 \, \big(1 ~+~ |\vec x|^2~+~ |\vec y|^2 ~-~  t^2 \big)    \bigg]  \,, \qquad X_{p+q+1}~=~   r \, t\,, \\
X_{p-1} & ~=~    \frac{1}{2\,r} \, \bigg[ 1 - r^2 \, \big(1 ~-~ |\vec x|^2~-~ |\vec y|^2 ~+~  t^2 \big)    \bigg]    \,, \\
  X_{j} & ~=~   r \, y_j \,, \quad j =1,\dots, p-2\,, \qquad \qquad X_{p+j-1}  ~=~   r \, x_j \,, \quad j =1,\dots, q+1 \,, \\
\end{aligned}
\label{Pslice}
 \end{equation}
and then the metric may be written:
\begin{equation}
ds^2 ~=~ \frac{dr^2}{r^2} ~+~r^2\,\Big(  -  dt^2 \ ~+~ |d\vec x|^2~+~ |d\vec y|^2 \Big)\,.
\label{Pmet1}
 \end{equation}

Now observe that
\begin{equation}
x^2   ~=~ \frac{1}{r^2} \sum_{j=p}^{p+q} \,  X_{j}^2 ~=~ \frac{\sinh^2 \adsvar}{r^2}   \,,  \qquad |d\vec x|^2 ~=~ dx^2 ~+~ x^2 d\Omega_{q}^2 \,,
\label{reln1}
 \end{equation}
where $x \equiv |\vec x|$.  Define a new coordinate, $\nu$, to replace $r$ via:
\begin{equation}
r ~=~e^ \nu \,\cosh  \adsvar  \,,  
\label{nudefn}
 \end{equation}
and note that  $-\infty < \nu < \infty$.   Then the Poincar\'e metric on AdS$_{p+q+1}$, (\ref{Pmet1}), can be written:
\begin{equation}
ds^2 ~=~d\adsvar^2 ~+~ \cosh^2  \adsvar\,  \Big[ \, d\nu^2 ~+~e^{2 \nu}  \big(- dt^2   ~+~ |d\vec y|^2 \big)\, \Big]~+~ \sinh^2 \adsvar \,  d\Omega_{q}^2  \,,
\label{Pmet2}
 \end{equation}
which explicitly exhibits the Poincar\'e patch of $AdS_{p}$.

Now recall that for $q=0$ we have $-\infty < \adsvar < \infty$ and so the  Poincar\'e patch of AdS$_{p+1}$ is given by:
\begin{equation}
ds^2 ~=~d\adsvar^2 ~+~ \cosh^2  \adsvar\,  \Big[ \, d\nu^2 ~+~e^{2 \nu}  \big(- dt^2   ~+~ |d\vec y|^2 \big)\, \Big]  \,.
\label{Pmet3}
 \end{equation}
for  $-\infty < \adsvar, \nu < \infty$.  

Finally, observe that the metrics (\ref{AdSmet3}) and (\ref{Pmet3}) are invariant under $\adsvar \to -\adsvar$, and so one can  quotient by this $\ZZ_2$ symmetry.  We will refer to this space (either in global or Poincar\'e form) as AdS$_{p+1}/\ZZ_2$,  defining it using the coordinates introduced above  but with  $0 \le  \adsvar < \infty$.

It also instructive  (\ref{variables3})

%%%%%%%%%%%%%%%%%%%%%%%%%%%%%%%%%%%%%
\section{Solutions with $\gamma <0$}
\label{sec:solutions}
%%%%%%%%%%%%%%%%%%%%%%%%%%%%%%%%%%%%%

In \cite{Bena:2023rzm}, we discussed a simple solution from among those of \cite{Bachas:2013vza} coming from the scaling limit of a simple M2-M5 intersection.  We now make a repeat performance for $\gamma < 0$.  Following  \cite{Bena:2023rzm} we are also going to take the Riemann surface to be the Poincar\'e upper half-plane with $\beta =2$: 
\begin{equation}
\zeta ~=~  \xi ~+~ i\, \rho  \,,\qquad - \tilde h ~+~ i h   ~=~   2 \, \zeta ~=~   2 \, ( \xi ~+~ i\, \rho ) \,, \qquad h  ~=~  -i (\zeta-\bar{\zeta}) \,,
 \label{PoincareChoice}
\end{equation}
and the metric given globally by (\ref{PoincareMet}).

%%%%%%%%%%%%%%%%%%%%%%
\subsection{Families of smooth geometries}
\label{ss:Families}
%%%%%%%%%%%%%%%%%%%%%%

The first, and perhaps most dramatic, difference in passing to $\gamma < 0$ is  that the constraint (\ref{gammacond}) means that we must have
\begin{equation}
| G | ~\le~ 1  \,, 
 \label{Gcond}
\end{equation}
everywhere. In particular, $G$ cannot have poles.  Moreover, it was shown in   \cite{Bachas:2013vza} that one can only have $|G| =1$ on  boundaries of the Riemann  surface, where, in fact, one must have $G = \pm i$, except at singularities.  Indeed, it is was argued that, for $\gamma < 0$, the most general possible two choices for $G$ involve only ``flip'' singularities:
\begin{align}
G~=~ -i\left(1+\sum_{j=1}^{2n+2}(-1)^j\frac{\zeta-\xi_j}{\vert \zeta-\xi_j \vert} \right)  \qquad \text{or} \qquad G=-i \sum_{j=1}^{2n+1}(-1)^j\frac{\zeta-\xi_j}{\vert \zeta-\xi_j \vert} \,.
\label{Gforms1}
\end{align}
We will assume, without loss of generality, that 
\begin{align}
\xi_i  ~<~ \xi_j   \quad \text {for} \quad i < j \,.
\label{xijorder}
\end{align}

We will refer to the solutions defined by the first function as {\it even flip solutions} and the solutions defined by the second function as {\it odd flip solutions}.  This is, of course something of a misnomer as  there are always an even number of flips in $G$:  the second function also has a flip at infinity, while the first function does not.  So the {\it even} and {\it odd} designations refer to flips at {\it finite} $\zeta$.  

The important difference between these two classes of solution lies in the asymptotics at infinity on the half-plane.  For even flip solutions, $G$ does not have a flip at infinity. Combined with the fact that $h$ has a pole at infinity, this implies  \cite{Bachas:2013vza} that the solution goes to an AdS$_7'\times$S$^4$ limit, where the  AdS$_7'$ reflects the presence of a  highly deformed object with M5 brane charges, that carries a large M2 flux. Solutions that have an odd number of flips on the real axis also have  a flip at infinity. This implies that the solution approaches (AdS$_4/\ZZ_2) \times$S$^7$  in that region (the $\ZZ_2$ quotient was defined in Section \ref{sec:AdS}), and hence it is asymptotic to an M2 brane metric.  

For future reference we also recall from \cite{Bachas:2013vza} that as one approaches a flip point at finite distance,  $\xi_j $, (where there is no pole in $h$,  (\ref{simph}))  the metric limits to that of AdS$_3 \times \IR^8$, which means this region is smooth, empty space.

%%%%%%%%%%%%%%%%%%%%%%
\subsection{The associated functions}
\label{ss:Phifns}
%%%%%%%%%%%%%%%%%%%%%%

For the first choice of $G$ in (\ref{Gforms1}), we have:
\begin{equation}
\begin{aligned}
\Phi~=~& 2\,  \Bigg[  \xi ~+~ \sum_{j=1}^{2n+2} \,(-1)^j\, \sqrt{(\xi-\xi_j)^2+\rho^2} \,  \Bigg]~=~ 2\,\Bigg[\, \xi ~+~  \sum_{j=1}^{2n+2} \,(-1)^j\, r_j \, \Bigg] \,,\\
\tilde{\Phi} ~=~&   2\, \Bigg[ \log ( \rho) ~-~ \sum_{j=1}^{2n+2} \,(-1)^j\, \log \bigg( \xi-\xi_j +\sqrt{(\xi-\xi_j )^2+\rho^2}\bigg)  \Bigg] \\
~=~&   2\,  \log ( \rho) ~+~ \sum_{j=1}^{2n+2} \,(-1)^j\, \log \bigg(\frac{\sqrt{(\xi-\xi_j )^2+\rho^2}-(\xi-\xi_j )}{\sqrt{(\xi-\xi_j )^2+\rho^2}+(\xi-\xi_j )}   \bigg)   \\
~=~&   2\,   \Bigg[ \log ( \rho) ~+~ \sum_{j=1}^{2n+2} \,(-1)^j\, \log \bigg(\tan \frac{\theta_j}{2}   \bigg) \Bigg]\,,
\end{aligned}
\label{phitildephi1}
\end{equation}
while for the second, we have
\begin{equation}
\begin{aligned}
\Phi~=~& 2\, \sum_{j=1}^{2n+1} \,(-1)^j\, \sqrt{(\xi-\xi_j)^2+\rho^2}~=~ 2\, \sum_{j=1}^{2n+1} \,(-1)^j\, r_j \,,\\
\tilde{\Phi} ~=~&   2\, \sum_{j=1}^{2n+1} \,(-1)^j\, \log \bigg(\frac{\sqrt{(\xi-\xi_j )^2+\rho^2}-(\xi-\xi_j )}{\sqrt{(\xi-\xi_j )^2+\rho^2}+(\xi-\xi_j )}   \bigg)   \\
~=~&   2\,    \sum_{j=1}^{2n+1} \,(-1)^j\, \log \bigg(\tan \frac{\theta_j}{2}   \bigg) \,,
\end{aligned}
\label{phitildephi2}
\end{equation}
where  $r_j$ and $\theta_j$ are defined by:
\begin{align}
r_j ~\equiv~\sqrt{(\xi-\xi_j)^2+\rho^2}  \,, \qquad \cos \theta_j ~\equiv~\frac{(\xi-\xi_j)}{\sqrt{(\xi-\xi_j)^2+\rho^2}} \,, \qquad \sin \theta_j ~\equiv~\frac{\rho}{\sqrt{(\xi-\xi_j)^2+\rho^2}} \,.
\label{rjtjdefns}
\end{align}

Note that as $\rho \to 0$, one has:
\begin{align}
\tilde{\Phi}  ~\sim~ \pm 2 \, \log (\rho) 
\label{tPhibdry}
\end{align}
where the sign depends on where $\xi$ lies relative to the flip points, $ \xi_j$.   Recalling  (\ref{variables3}), with $\beta =2$, we have:
\begin{equation}
u  ~\sim~   a\, \mu^\alpha \, \rho^{\frac{1}{2}( 1 \mp 1)}   \,,   \qquad  v  ~\sim~   a\, \mu^{1-\alpha} \, \rho^{\frac{1}{2}( 1 \pm 1)}    \,,
 \label{variables4}
\end{equation}
which means that along the $\xi$-axis either $u \to 0$ or $v \to 0$,  depending on the  position of $\xi$  relative to the flip points, $ \xi_j$.

%%%%%%%%%%%%%%%%%%%%%%
\subsection{Topology, fluxes and smoothness}
\label{ss:Topology}
%%%%%%%%%%%%%%%%%%%%%%

One should first observe that at $\rho =0$ one has:
\begin{equation}
G~=~ -i\left(1+\sum_{j=1}^{2n+2}(-1)^j \, \text{sign}(\xi-\xi_j)  \right)  \qquad \text{or} \qquad G=-i \sum_{j=1}^{2n+1}(-1)^j\, \text{sign}(\xi-\xi_j)   \,,
\label{Gforms-axis}
\end{equation}
which means that $G=\pm i$ on the $\xi$-axis. 

\begin{figure}[h]
    \centering
    \includegraphics[width=.8 \textwidth]{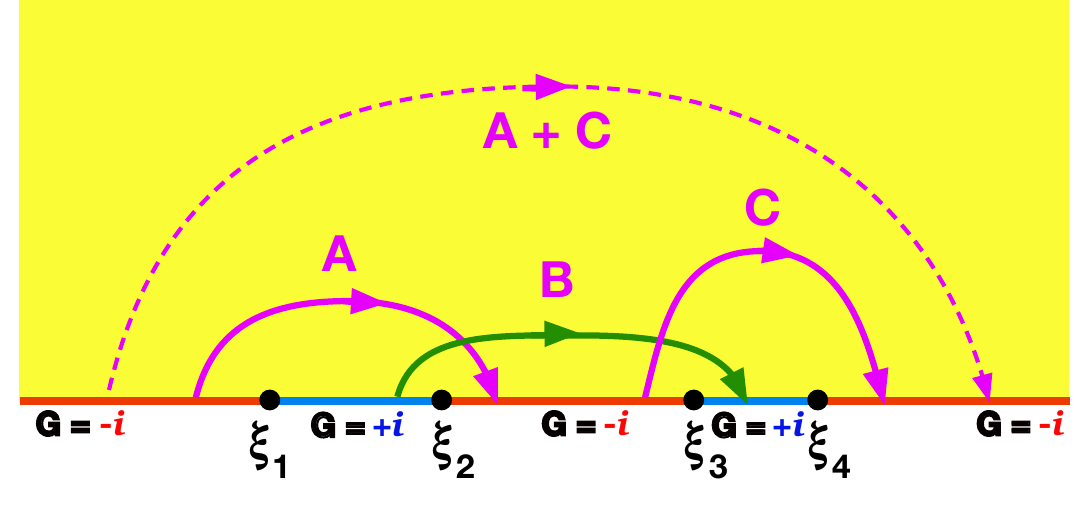}  
    \caption{The Poincar\'e upper half-plane, showing the boundary values of $G$ with an {\it even number} of flip points at finite $\xi_j$ (and hence no flip at infinity).   Path $B$ defines a homology  sphere, ${S}^4$, while paths $A$ and $C$ define homology  spheres, ${S'}^4$.  There is a net M5 charge at infinity and so the cycle $A+C$ is not contractible, and the cycles $A$ and $C$ carry independent 5-brane charges.}
    \label{fig:Topology1}
\end{figure}

\begin{figure}[h]
    \centering
    \includegraphics[width=.8 \textwidth]{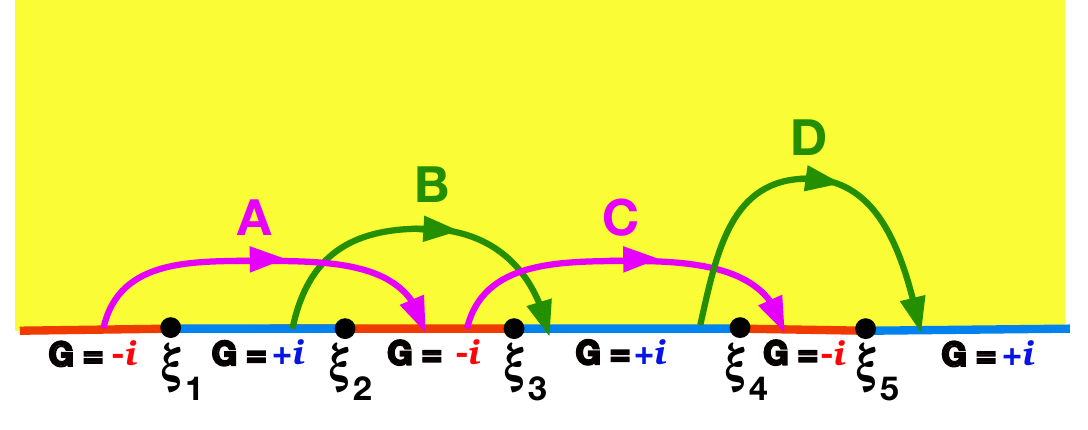}  
    \caption{The Poincar\'e upper half-plane, showing the boundary values of $G$ with an {\it odd number} of flip points at finite $\xi_j$ (and a flip at infinity).  The paths $A$ and $C$ define homology  spheres, ${S'}^4$, while $B$ and $D$ define homology  spheres, ${S}^4$.  At infinity there is  an (AdS$_4/\ZZ_2) \times$S$^7$, and the M2 charge is equal to the integral of $F_7$ on this S$^7$ (see Section \ref{sec:M2chg-spikes}). }
    \label{fig:Topology2}
\end{figure}

 A careful analysis of the metric  (\ref{DHokerAnsatz}) using the metric functions  (\ref{f-functions})  and  (\ref{eA-function}) shows that when  $G \to -i$ the warp factor for $S^3$ remains finite, while the warp factor for ${S'}^3$ vanishes, which means that this ${S'}^3$  pinches off.  Conversely, when  $G \to + i$ the   ${S'}^3$ remains finite, and the ${S}^3$  pinches off.  This means that if one follows a path that runs inside the Riemann surface,  terminating on two consecutive regions on the $\xi$-axis where $G = -i$, then ${S'}^3$ sweeps out an ${S'}^4$ while the $S^3$ remains finite. See Figs.~\ref{fig:Topology1} and~\ref{fig:Topology2}.   Similarly,   following a path that runs inside the Riemann surface,  terminating on two adjacent regions on the $\xi$-axis where $G = i$, then $S^3$ sweeps out an $S^4$ while the ${S'}^3$ remains finite. 
 
One can verify that the entire eleven-dimensional metric is smooth and that the singularities in $G$ (and $\tilde{\Phi}$) simply correspond to the North and South poles of the $S^4$ and  ${S'}^4$  homology spheres.  It is these spheres that carry the cohomological fluxes that source M5 and M5' charge, and thus M2 charge through the Chern-Simons interaction.

%%%%%%%%%%%%%%%%%%%%%%%%%%%%%%%%
%%%%%%%%%%%%%%%%%%%%%%%%%%%%%%%%

%%%%%%%%%%%%%%%%%%%%%%%%%%%%%%%%
%%%%%%%%%%%%%%%%%%%%%%%%%%%%%%%%

It was shown in \cite{Bachas:2013vza} that the fluxes on these cycles are given by:
\begin{equation}
Q_{5\,, j} ~=~  \frac{4\nu_3}{\gamma c_3^3} \,(\xi_{2j} - \xi_{2j-1})    \qquad \text{and}\qquad 
Q'_{5\,, j}  ~=~  \frac{4\nu_2\gamma}{ c_2^3}\, (\xi_{2j+1} - \xi_{2j}) \,,
 \label{M5charges}
\end{equation}
where the $c_i$ are defined in (\ref{cparams}). Note that in \eqref{M5charges} we have divided by the volume of a unit-radius three-sphere.

\begin{figure}[h]
    \centering
    \includegraphics[width=.60 \textwidth]{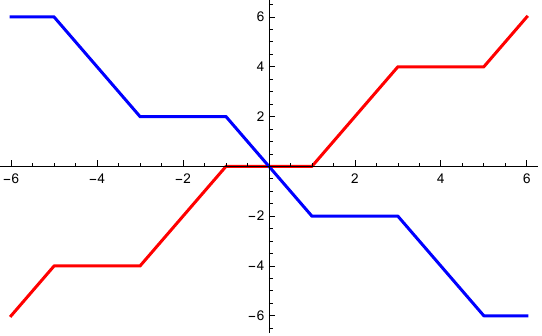}  
    \caption{ A plot of the steepness functions, $u^2 z$ (in red, bottom-left to top-right)  and $v^2 w$ (in blue, top-left to bottom-right) for $\xi_j =(-5, -3,-1,1,3,5)$.  The function, $\Phi$, involves a constant of integration and so the actual steepness of the brane spikes can involve a vertical translation of this figure.  Note that the plateaus and linear behavior alternate between the two functions. }
    \label{fig:steepness}
\end{figure}

We also note  that for the first choice of $G$ in (\ref{Gforms1}), the invariant coordinates (\ref{scalinv1}) at $\rho =0$  become
\begin{equation}
u^2 z ~=~   \frac{\varepsilon_1}{\sqrt{|\gamma|}} \,     \Bigg[\, 2\, \xi ~+~  \sum_{j=1}^{2n+2} \,(-1)^j\, \big|  \xi - \xi_j \big| \, \Bigg]   \,, \qquad v^2 w ~=~ \varepsilon_1~ \sqrt{|\gamma|}\, \sum_{j=1}^{2n+2} \,(-1)^j\, \big|  \xi - \xi_j \big|  \,.
\label{scalinv2}
 \end{equation}
These functions alternate between locally constant plateaus and linear behavior, as seen in Fig.~\ref{fig:steepness}.  In particular, $u^2 z$ is constant over the intervals $(\xi_{2j-1},\xi_{2j})$, where $G=+i$,  while $v^2 w$ is constant over the intervals $(\xi_{2j},\xi_{2j+1})$, where $G= - i$.  This means that $u^2 z$ is constant where $S^3$ pinches off and  $v^2 w$ is constant where ${S'}^3$ pinches off.

%%%%%%%%%%%%%%%%%%%%%%%%%%%%%%%%
%%%%%%%%%%%%%%%%%%%%%%%%%%%%%%%%

%%%%%%%%%%%%%%%%%%%%%%
\subsection{A simple example}
\label{ss:PureM5}
%%%%%%%%%%%%%%%%%%%%%%

%%%%%%%%%%
\def\apoint{c}
%%%%%%%%%%

%%%%%%%%%%%
\subsubsection{The functions}
\label{ss:simpfns}
%%%%%%%%%%%

The simplest example involves two flips taken at finite points which we take to be at $\zeta = \pm \apoint $, with $\apoint >0$:
\begin{align}
G~=~ -i \bigg(1~-~ \frac{\zeta+ \apoint}{| \zeta+\apoint |}~+~ \frac{\zeta- \apoint}{| \zeta-\apoint |}  \bigg)  \,.
\label{Gsimp1}
\end{align}
This means that
\begin{equation}
\begin{aligned}
\Phi~=~& 2\,  \Bigg[  \xi ~-~  \sqrt{(\xi+\apoint)^2+\rho^2} ~+~  \sqrt{(\xi-\apoint)^2+\rho^2}\,  \Bigg]\,,\\
\tilde{\Phi} ~=~&
 2\,  \log ( \rho) ~+~  \log \bigg(\frac{\sqrt{(\xi+\apoint)^2+\rho^2}-(\xi+\apoint)}{\sqrt{(\xi+\apoint)^2+\rho^2}+(\xi+\apoint)}   \bigg)~-~  \log \bigg(\frac{\sqrt{(\xi-\apoint)^2+\rho^2}-(\xi-\apoint)}{\sqrt{(\xi-\apoint)^2+\rho^2}+(\xi-\apoint)}   \bigg) \,.
\end{aligned}
\label{Phissimp1}
\end{equation}

It is convenient to use a conformal mapping: 
\begin{equation}
\begin{aligned}
\zeta ~=~ & \apoint \, \cosh (2 \,\kappa) \,, \qquad  \kappa ~\equiv~  \sigma + i \chi   \\
&\Rightarrow   \quad \xi ~=~    \apoint \, \cosh ( 2 \, \sigma) \, \cos (2 \, \chi) \,, \quad \rho ~=~  \apoint \, \sinh ( 2 \, \sigma) \, \sin (2 \, \chi)  \,,
 \end{aligned}
\label{confmap}
\end{equation}
where the upper half-plane ($\rho >0$) is covered by taking:
\begin{align}
0  ~\le~ \sigma ~<~ \infty  \,, \qquad  0  ~\le~ \chi ~\le~  \frac{\pi}{2}   \,.
\label{sigchi-rng}
\end{align}

One then finds 
\begin{equation}
\begin{aligned}
1- G \bar G ~=~ &  \frac{8}{\Delta} \,\sinh^2 ( \sigma) \, \sin^2 (2 \, \chi)  \,,  \\
W_+ ~=~ &  \frac{4}{\Delta} \, \Big[ 1 ~-~ 2\, \gamma \,\sinh^2 \sigma\ \Big] \, \sin^2 (2 \, \chi)   \,, \\
W_- ~=~ &  \frac{16}{\Delta} \, \bigg[\cosh^2\sigma\,  ~-~ \frac{1}{2}\, \big(2 + \gamma^{-1}\big)\, \sin^2 (2 \, \chi) \bigg]   \,\sinh^2 ( \sigma) \,.
 \end{aligned}
\label{metfuns}
\end{equation}
where
\begin{align}
\Delta~\equiv~\big| \sinh(2 \, \kappa) \big|^2 ~=~ \sinh^2 ( 2\, \sigma) ~+~ \sin^2 (2 \, \chi)  \,.
\label{Deltadefn}
\end{align}
%

%%%%%%%%%%%
\subsubsection{The metric}
\label{ss:simpmet}
%%%%%%%%%%%

The Riemann-surface metric  (\ref{PoincareMet}) becomes
\begin{equation}
ds_2^2 ~=~ \frac{ d\xi^2 +  d\rho^2}{4 \, \rho^2}  ~=~ \frac{\Delta}{ \sinh^2 ( 2\, \sigma) \, \sin^2 (2 \, \chi)} \, \big( d\sigma^2 + d \chi^2 \big) \,,
 \label{RSmet1}
\end{equation}
and the full eleven-dimensional metric may be written:
\begin{equation}
\begin{aligned}
ds_{11}^2  ~=~ &  \bigg[ \, \frac{2\,(1 - 2\, \gamma \,\sinh^2 \sigma)}{ \cosh^2  \sigma }    \bigg]^{1/3} \,  \bigg[1  ~-~ \frac{1}{2}\, \big(2 + \gamma^{-1}\big)\, \frac{\sin^2 (2 \, \chi)}{\cosh^2  \sigma}      \bigg]^{1/3}  \\
 &\times   \Bigg[ 4 \,  \bigg(\, d\sigma^2  ~+~ \frac{ (-\gamma)}{2 \,(\gamma +1)^2} \, \cosh^2  \sigma \, ds^2_{\text{AdS}_3} ~+~ \frac{\sinh^2  \sigma \, \cosh^2  \sigma }{(1 - 2\, \gamma \,\sinh^2 \sigma)}  \,  ds^2_{S^3} \,\bigg)  \\ 
& \qquad ~+~4 \,  d \chi^2~+~   \frac{\sin^2 (2 \, \chi) }{\Big(1  ~-~ \frac{1}{2}\, \big(2 + \gamma^{-1}\big)\, \frac{\sin^2 (2 \, \chi)}{\cosh^2  \sigma}   \Big)  }\,   ds^2_{{S'}^3} \,  \Bigg] \,.
 \end{aligned}
\label{simp11met}
\end{equation}
For $\gamma = -1/2$ this becomes: 
\begin{equation}
ds_{11}^2  ~=~  2^{1/3} \,   \Big[ \,4 \,  \big(\, d\sigma^2  ~+~ \cosh^2  \sigma \,ds^2_{\text{AdS}_3} ~+~\sinh^2  \sigma \,  ds^2_{S^3} \,\big) 
 ~+~ d \theta^2~+~   \sin^2\theta  \, ds^2_{{S'}^3} \,  \Big] \,,
\label{vsimp11met}
\end{equation}
where
\begin{equation}
\theta  ~\equiv~  2 \,\chi \,.
 \label{thetadefn1}
\end{equation}
From the discussion in Section \ref{sec:AdS} and the coordinate ranges (\ref{sigchi-rng}), one sees that this is precisely the metric on AdS$_7 \times \text{S}^4$.

%%%%%%%%%%%
\subsubsection{The flat-brane system}
\label{ss:simpCart}
%%%%%%%%%%%

Using (\ref{confmap}) in (\ref{Phissimp1}), one obtains:
\begin{equation}
\Phi~=~ 2\,\apoint \,  \big[ \, \cosh ( 2 \, \sigma)  ~-~2 \, \big]\, \cos (2 \, \chi)  \,, \qquad 
\tilde{\Phi} ~=~
 2\,  \log \bigg[ \frac{2\, \apoint \cosh^3 \sigma\, \sin (2 \, \chi) }{\sinh \sigma}   \bigg] \,.
\label{trfPhi}
\end{equation}
From (\ref{variables3}) and (\ref{wsimp}),  with $\beta = 2$, one then obtains
\begin{equation}
\begin{aligned}
u  ~=~&  a\, \mu^ \frac{\gamma  }{(\gamma +1)} \, \rho^{\frac{1}{2}}   \, e^{- \frac{1}{4} \,  \tilde \Phi}  ~=~ a\, \mu^ \frac{\gamma  }{(\gamma +1)} \,\tanh \sigma \,, \\
  v~=~&    a\, \mu^ \frac{1 }{(\gamma +1)} \, \rho^{\frac{1}{2}}   \, e^{+ \frac{1}{4} \,  \tilde \Phi} ~=~2\, \apoint\, a\, \mu^ \frac{1 }{(\gamma +1)} \, \cosh^2 \sigma\, \sin \theta  \,, \\
z  ~=~&   - \frac{\varepsilon_1\, (1+\gamma) }{4\, \gamma\, \rho} \,  \mu^{-\frac{2\, \gamma  }{(\gamma +1)}} \,   e^{\frac{1}{2} \,  \tilde \Phi }  \,    \big(  \Phi +  2 \, \xi \big) ~=~   - 2\, \apoint\, \frac{\varepsilon_1\, (1+\gamma) }{ \gamma} \,\mu^ {- \, \frac{2\,\gamma  }{(\gamma +1)}} \,  \cosh^2 \sigma\, \cos \theta \,,\\
w  ~=~&    \frac{\varepsilon_1\, (1+\gamma) }{4\,\rho} \,  \mu^{-\frac{2 }{(\gamma +1)}} \,   e^{-\frac{1}{2} \,  \tilde \Phi }  \,    \big(  \Phi - 2 \, \xi \big) ~=~   -  \frac{\varepsilon_1}{ 4\, \apoint}\, (1+\gamma)  \,\mu^ {-\frac{2   }{(\gamma +1)} }\,   \frac{\cos\theta}{\cosh^4 \sigma \,\sin^2 \theta}   \,,
\end{aligned}
 \label{simpuvz}
\end{equation}
where 
\begin{equation}
 a ~\equiv~  \bigg( \frac{2 \, \sqrt{|\gamma|}}{(\gamma +1)}\bigg)^{1/2} \,.
\label{aparams2}
 \end{equation}
For $\gamma =-1/2$, this becomes:
\begin{equation}
\begin{aligned}
u  & ~=~ \sqrt{2} \, \mu^{-1} \,\tanh \sigma \,, \qquad
  v~=~    2\,  \sqrt{2} \,\apoint\,  \mu^2 \, \cosh^2 \sigma\, \sin \theta  \,,  \qquad
  z  ~=~     2\, \apoint\,  \varepsilon_1 \,\mu^2\, \cosh^2 \sigma\, \cos \theta \\
w & ~=~   -  \frac{\varepsilon_1}{ 8 \, \apoint}  \,  \frac{\cos\theta}{\mu^ {4 }\,  \cosh^4 \sigma \,\sin^2 \theta}  ~=~   -     \frac{ \sqrt{2}\, \apoint    \, z }{v^ {2 }\,   \sqrt{v^2 + 2\, z^2}}  \,,
  \end{aligned}
 \label{vsimpuvz}
\end{equation}
and rescaling leads to:
\begin{equation}
\begin{aligned}
\tilde u  & ~=~ \mu^{-1} \,\tanh \sigma \,, \qquad
 \tilde  v~=~    \mu^2 \, \cosh^2 \sigma\, \sin \theta  \,,  \qquad
\tilde  z  ~=~     \mu^2\, \cosh^2 \sigma\, \cos \theta \,, \\
w & ~=~   -  \frac{\varepsilon_1}{ 8 \, \apoint}  \,  \frac{\cos\theta}{\mu^ {4 }\,  \cosh^4 \sigma \,\sin^2 \theta}  ~=~    -  \frac{\varepsilon_1}{ 8 \, \apoint}  \,   \frac{ \tilde z }{\tilde v^ {2 }\,   \sqrt{\tilde v^2 + \tilde z^2}} \,.
 \end{aligned}
 \label{vsimpuvz2}
\end{equation}
In particular, one has
\begin{equation}
 \mu^{-2}  ~=~  \tilde u^2 ~+~  \frac{1}{\sqrt{\tilde v^2 +  \tilde z^2}}\,,  \qquad \sinh \sigma  ~=~ \tilde u \, (\tilde v^2 +  \tilde z^2)^{1/4}  \,, \qquad \tan \theta ~=~ \frac{ \tilde  v}{ \tilde  z}
 \label{relns1}
\end{equation}
and 
\begin{equation}
r   ~=~\mu \, \cosh \sigma ~=~(\tilde v^2 +  \tilde z^2)^{1/4}  \,.
 \label{relns2}
\end{equation}
%

%%%%%%%%%%%
\subsubsection{The M5-brane metrics}
\label{ss:M5mets}
%%%%%%%%%%%

Recall the standard metric for a stack of M5 branes:
\begin{equation}
ds_{11}^2   ~=~H(\tilde r)^{-\frac{1}{3}}  \, \eta_{\mu \nu}\, d x^\mu  dx^\nu ~+~ H(\tilde r)^{\frac{2}{3}}  \, \big( \, d\tilde r^2 ~+~\tilde r^2  \, d \Omega_4^2 \, \big)     \,, \qquad H(\tilde r) = 1 ~+~ \frac{Q}{\tilde r^3} \,,
 \label{M5stack1}
\end{equation}
where $\tilde r$ is the radial coordinate transverse to the branes, which, in the flat brane coordinates, means
\begin{equation}
\tilde r ~=~ \sqrt{\tilde v^2 +  \tilde z^2} \,.
 \label{tr-defn}
\end{equation}
In the near-brane limit, the metric becomes: 
\begin{equation}
ds_{11}^2   ~=~Q^\frac{2}{3} \, \bigg[ \, \frac{d \tilde r^2 }{\tilde r^2} ~+~ Q^{-1}  \,\tilde r \,  \eta_{\mu \nu}\, d x^\mu  dx^\nu ~+~    \, d \Omega_4^2 \, \bigg]   \,.
 \label{M5stack2}
\end{equation}
One gets the canonical form of Poincar\'e AdS by changing variables to $\tilde r = r^2$ and rescaling the $x^\mu$:
\begin{equation}
ds_{11}^2   ~=~Q^\frac{2}{3} \, \bigg[ \, 4\, \bigg(\,\frac{d  r^2 }{ r^2} ~+~   r^2 \,  \eta_{\mu \nu}\, d x^\mu  dx^\nu\, \bigg) ~+~    \, d \Omega_4^2 \, \bigg]   \,,
 \label{M5stack3}
\end{equation}
which precisely matches (\ref{vsimp11met}).  Note also that $r = \sqrt{\tilde r} =  (\tilde v^2 +  \tilde z^2)^{1/4}$, which means that  (\ref{relns2}) is precisely the correct relation between the Poincar\'e coordinate, $r$, and the Cartesian coordinates transverse to the branes.

While we have focussed on the Poicar\'e metrics here, we note that if one uses global AdS$_3$ in (\ref{vsimp11met}), then one obtains the metric on global AdS$_7 \times \text{S}^4$, which does not come from the backreaction of flat M5 branes.

%%%%%%%%%%%%%%%%%%%%%%%%%%%%%%%%%%%%%
\section{M2 charges and spikes}
\label{sec:M2chg-spikes}
%%%%%%%%%%%%%%%%%%%%%%%%%%%%%%%%%%%%%

%%%%%%%%%%%%%%%%%%%%%%
\subsection{M2 charges}
\label{ss:M2charges}
%%%%%%%%%%%%%%%%%%%%%%

To compute the M2 charge of the solution, we need to find the  6-form potential for the  {\it conserved} dual of the Maxwell field obtained from the 3-form potential of \eqref{DHokerAnsatz}. Because of the Chern-Simons interaction, we have the following equation of motion for $C^{(3)}$:
\begin{equation}
	d \star F^{(4)}=-\frac{1}{2}F^{(4)}\wedge F^{(4)} \,,
\end{equation}
%Therefore, to define a conserved charge we introduce the following 7-form:
The 6-form potential  is then defined from the conserved $7$-form:
\begin{equation}
	dC^{(6)}=\star F^{(4)}+\frac{1}{2}C^{(3)}\wedge F^{(4)} + \text{exact} \,.
\label{dC6a}
\end{equation} 
Indeed, the $7$-form on the right-hand side is the flux  that defines the M2 Page charge, which is conserved but gauge dependent. 
It is also convenient to recall that the {\it Maxwell charge} of these branes is defined by using $\star F^{(4)}$ alone, and that this charge is gauge invariant but not conserved. The fact that it is not conserved means that it depends on the details of the Gaussian surface and so,  in practice, the Maxwell charge is typically only useful in characterizing   charges at infinity.  Indeed, when the gauge fields fall off sufficiently fast at infinity, the Maxwell and Page charges coincide. 

For the solutions of Section \ref{sec:AdSlmits}, $dC^{(6)}$ can be written as:
\begin{equation}
\label{dC6b}
	dC^{(6)}=-d\Omega_1 \hat{e}^{345678}+d\Omega_2 \hat{e}^{678012}+d\Omega_3 \hat{e}^{345012} \,,
\end{equation}
where the $d\Omega_i$ are one-forms on the Riemann surface, $\Sigma$, and the six-forms, $\hat{e}^{abcdef}$ are wedge products of the unit volume forms introduced in \eqref{DHokerAnsatz}. 

The non-trivial 7-cycles over which we integrate (\ref{dC6a}) are either S$^7$ or S$^4 \times$ S$^3$. The S$^7$ cycle is only present in solutions that have an odd number of flip singularities in $G$ at finite points on $\Sigma$, and is constructed by the two S$^3$'s fibered along a curve in $\Sigma$ and extending between the two regions in $\partial \Sigma$ around the flip point at infinity. The cycles with topology S$^4 \times $ S$^3$ are defined over the blue and red intervals of Fig.~\ref{fig:Topology1} and are constructed by the product of the S$^4$ homology spheres that were defined in Section \ref{ss:Topology} times the S$^3$ that remains finite along the interval in question.

The second and third terms in \eqref{dC6b} involve the unit AdS$_3$ volume form, $\hat{e}^{012}$, and are thus electric parts of the $C^6$ flux, coming from the M5 and M5' branes. The M2 charge comes from the magnetic term in (\ref{dC6b}).  We therefore only need the first term in (\ref{dC6b}) and its form means that the integral   over $S^3 \times {S'}^3$ is elementary, with each sphere contributing $2 \pi^2$. Thus, the  calculation of the M2 charges  reduces to a line integral of $d\Omega_1$  along the curve in $\Sigma$. Since $d\Omega_1$ is exact on this interval, there is a differentiable function, $\Omega_1$, from which the M2 charge can be easily read off:
\begin{equation}
	\frac{1}{4\pi^4}\int_{X_7} \,   d  C^{(6)}  ~=~  -\Omega_1\, \Big |_{\rho=0,\, \xi_j}^{\rho=0,\,\xi_{j+1}}   \,,
\label{dC6c}
\end{equation}
where we have divided by the product of the volumes of the two unit radius S$^3$'s.

Using \eqref{dC6a}, \eqref{dC6b} and the expressions for the flux functions in \eqref{f-functions}, one finds the following expression for $\partial_\zeta \Omega_1$: 
\begin{equation}
\partial_{\zeta} \Omega_1=\partial_{\zeta} \left( \frac{\nu_1 \tilde \sigma}{c_2^3 c_3^3}\,  \widehat \Omega_1 \right) ~=~ -\bigg[\frac{ i\, h  \, (G   \overline G-1)^2 }{W_+ W_- }  \, \partial_{\zeta} \hat b_1 ~-~ \frac{1}{2}\, \big( \hat b_2\, \partial_{\zeta} \hat b_3 -  \hat b_3 \,\partial_{\zeta} \hat b_2 \big) ~-~ \frac{\epsilon}{2}\partial_{\zeta} (\hat b_2 \hat b_3) \bigg] \,, 
 \label{Omega1eqn}
\end{equation}
where the $ \hat b_i $ are defined in (\ref{bfunctions}),  we have set  $\tilde \sigma=-1$ (see (\ref{sigsusy}), (\ref{sigdefn})) for a negative $\gamma$ solution, and we chose the following values for the signs $\nu_i$: $\nu_1=-1, \nu_2=-1, \nu_3=1$.   Note that, as in \cite{Bena:2024dre},  we have also introduced a normalized function, $\widehat \Omega_1$, and using  (\ref{cparams})  for our choice of parameters with   $\gamma < 0$ , one finds:
\begin{equation}
\widehat \Omega_1   ~=~  -\Omega \,.
 \label{Omegareln1}
\end{equation}

The last term in \eqref{Omega1eqn} represents an exact piece, as in \eqref{dC6a}.  This  needs to be chosen in such a way as to make $d  C^{(6)}$ non-singular.  At the boundary of $\Sigma$, the flux functions $\hat b_i$ are generally finite and hence there is a risk that the first term in $d  C^{(6)}$ is singular because \eqref{Omega1eqn} is non-vanishing, while a three-sphere is pinching off.   For the $S^7$ this is easily arranged by choosing the constants of integration, $b_i^0$ in (\ref{bfunctions}) so that the appropriate $\hat b_i$ vanishes at each end of the interval.  For the  S$^4 \times$ S$^3$ cycles it is a little more subtle. 

The important point is that $d \hat b_2 \wedge \hat e^{345}$ and $d \hat b_3   \wedge \hat  e^{678}$ are smooth $4$-forms that can be integrated over $S^4$ and ${S'}^4$.  If it is $S^3$ (and not ${S'}^3$) that is pinching off to create ${S}^4$ then we want to ensure that only $d \hat b_2$, and not $\hat b_2$ appears in  $d  C^{(6)}$.  (The integral of  $\hat b_3  \hat  e^{678}$ over ${S'}^3$ is then non-singular because ${S'}^3$ is {\it not} pinching off.)  This means one must choose $\epsilon = -1$ when the interval runs between points with $G=i$, creating  the cycle  ${S}^4 \times {S'}^3$.  Similarly, one  must choose $\epsilon = 1$ when the interval runs between points with  $G=-i$, creating  the cycle  ${S}^3 \times {S'}^4$. 

Focusing from now on  $\widehat{\Omega}_1$, \eqref{Omega1eqn} has the following solution \cite{Bachas:2013vza}:
\begin{equation}
\begin{aligned}
 \widehat \Omega_1 ~=~ & \frac{h}{2\,W_+} \Big[ \gamma \, h  \, (G   \overline G-1)  ~+~  (\Phi + \tilde h)(G  +  \overline G)\Big]~-~ \frac{h}{2\,W_-} \Big[ \frac{1}{\gamma}\, h  \, (G   \overline G-1)  ~+~  (\Phi - \tilde h)(G  +  \overline G)\Big]  \\ 
 & ~-~ \frac{1}{2\gamma}\,\hat b_2^0 \, \bigg[ \frac{ h  \, (G    + \overline G)}{  W_-} - (\Phi + \tilde h  )  \bigg]~-~ \frac{\gamma}{2}\,\hat b_3^0 \, \bigg[ \frac{ h  \, (G    + \overline G)}{W_+ } -  (\Phi - \tilde h) )  \bigg] \\
 &~-~\tilde h \, \Phi ~+~\Lambda~-~ \frac{1}{2}\, \epsilon \, \hat b_2\, \hat b_3\,, 
\end{aligned}
 \label{Omega1sol} 
\end{equation}
where $\Lambda$ satisfies:
\begin{equation}
\partial_{\zeta} \Lambda  ~=~ i \, h\, \partial_{\zeta} \Phi ~-~ 2i \, \Phi\, \partial_{\zeta}  h   \,. 
 \label{Lambda-defn} 
\end{equation}
As noted in \cite{Bena:2024dre},  the integrability condition for the equation for $\Lambda$ follows from the equation (\ref{Phieqn}) for $\Phi$.

Using \eqref{bfunctions} we find 
\begin{equation}
\begin{aligned}
 - \frac{1}{2}\, \epsilon \, \hat b_2\, \hat b_3  ~=~ &  \frac{1}{2}\, \epsilon \,   \frac{ h^2  \, (G    + \overline G)^2}{W_+W_- }   ~+~  \frac{1}{2}\, \epsilon  \,  \bigg( \gamma \big(\Phi - \tilde h \big) + \hat b_2^0 \bigg) \bigg(\frac{1}{\gamma} \big(\Phi + \tilde h \big)-\hat b_3^0\bigg) \\ ~-~   &\epsilon \,   \bigg[\gamma \frac{h}{2\,W_+}\bigg(\frac{1}{\gamma} \big(\Phi + \tilde h \big)-\hat b_3^0\bigg)(G  +  \overline G) ~+~ \frac{1}{\gamma}\frac{h}{2\,W_-} \bigg(\gamma \big(\Phi - \tilde h \big) + \hat b_2^0 \bigg)(G  +  \overline G)   \bigg] 
\end{aligned}
 \label{b2b3} 
\end{equation}
and hence 
\begin{equation}
\begin{aligned}
 \widehat \Omega_1 ~=~ &  \frac{1}{2}\, \epsilon \,   \frac{ h^2  \, (G    + \overline G)^2}{W_+W_- }  ~+~ \frac{h^2 \, (G  \overline G -1)}{2} \bigg[ \frac{\gamma}{W_+}-\frac{1}{\gamma}\frac{1}{W_-} \bigg]   \\  
 &~-~  \frac{\gamma}{2}\,(1- \epsilon)  \, \big( \hat b_3^0 - \frac{1}{\gamma}(\Phi + \tilde h)\big) \,  \bigg[\frac{h}{W_+} \,(G  +  \overline G) -  \big(\Phi - \tilde h\big) \bigg] \\
 & ~-~  \frac{1}{2\gamma}\,(1+ \epsilon ) \,  \big( \hat b_2^0+\gamma (\Phi - \tilde h)\big) \, \bigg[ \frac{h}{W_-} \,(G  +  \overline G)   -  \big(\Phi + \tilde h\big)\bigg] \\ 
 & ~-~  \frac{1}{2}\, \epsilon  \,  \big(\Phi^2  - \tilde h^2  \big)  ~-~\tilde h \, \Phi ~+~\Lambda ~-~  \frac{1}{2}\, \epsilon  \,   \hat b_2^0 \, \hat b_3^0 \,.  
\end{aligned}
 \label{Omega1sol-full} 
\end{equation}

First recall (\ref{simph}) that we are taking  $h =   \beta  \rho,  \tilde h = -   \beta  \xi$.
We are interested in the limit of $\widehat{\Omega}_1$ as $\rho \rightarrow 0$. In this limit, $G \rightarrow \mp i $, which means $W_{\pm} \sim {\cal O}(\rho^2)$ and $W_{\mp}\rightarrow 4$. It follows that the first line in \eqref{Omega1sol-full} vanishes. Moreover, if we are interested in probing the $G\rightarrow -i$ region, we must use the gauge $\epsilon=+1$ for $\Omega_1$ to be well-defined. This leaves: 
\begin{equation}
\begin{aligned}
\widehat \Omega_1\big|_{\rho =0} ~=~ &\Big( \Lambda~-~   \tilde h \, \Phi ~+~  \frac{1}{2}\,   \big(\Phi^2 - \tilde h^2)   ~+~ \frac{1}{\gamma}\, \hat b_2^0  \,  \big(\Phi +  \tilde h) \Big. \\
\Big.  &~-~  \frac{1}{2}\, \hat b_2^0 \, \hat b_3^0 ~-~  \big( \hat b_2^0+ \gamma(\Phi - \tilde h)\big)\frac{1}{\gamma} \frac{h}{W_-}\, (G+ \overline G) \Big)\Big|_{\rho =0}     \\
~=~&\Big( \Lambda~-~   \tilde h \, \Phi ~+~  \frac{1}{2}\,   \big(\Phi^2 - \tilde h^2)   ~+~ \frac{1}{\gamma}\, \hat b_2^0  \,  \big(\Phi +  \tilde h) ~-~  \frac{1}{2}\, \hat b_2^0 \, \hat b_3^0 \Big)\Big|_{\rho =0} \,,
\end{aligned}
 \label{Omega1bdry1} 
\end{equation}
where we used the fact that the last term of the second line vanishes in a $G=-i$ region. 
Conversely, in the $G\rightarrow +i$ regions, we must use the gauge $\epsilon=-1$, and one is left with:
\begin{equation}
\begin{aligned}
\widehat \Omega_1\big|_{\rho =0} ~=~ &\Big( \Lambda~-~   \tilde h \, \Phi ~-~  \frac{1}{2}\,   \big(\Phi^2 - \tilde h^2)   ~+~  \gamma \, \hat b_3^0  \,  \big(\Phi -  \tilde h) ~+~  \frac{1}{2}\, \hat b_2^0 \, \hat b_3^0 \Big)\Big|_{\rho =0}     \,.
\end{aligned}
 \label{Omega1bdry2} 
\end{equation}
As for $\hat b_2^0$ and $\hat b_3^0$, we will follow \cite{Bachas:2013vza} and choose a gauge such that $b_3=0$ on the boundary interval $(-\infty, \xi_1]$, and $b_2=0$ either on the boundary interval $[\xi_{2n+1},\xi_{2n+2}]$ for the even-flip solutions, or on the boundary internal $[\xi_{2n+1},+\infty)$ for the odd-flip solutions. This means that 
\begin{equation}
\begin{aligned}
	\hat b_3^0&=\frac{1}{\gamma}(\Phi+\tilde{h})|_{\rho=0, (-\infty,\xi_1]}=\frac{2}{\gamma}\sum_{j=1}^{2n+2}(-1)^j\xi_j \,, \,\,\text{even number of G-flips} \,,\\
	\hat b_3^0&=\frac{1}{\gamma}(\Phi+\tilde{h})|_{\rho=0, (-\infty,\xi_1]}=\frac{2}{\gamma}\sum_{j=1}^{2n+1}(-1)^j\xi_j \,, \,\,\text{odd number of G-flips} \,,
\end{aligned}
\label{b30 gauge}
\end{equation}
and 
\begin{equation}
\begin{aligned}
	\hat b_2^0&=-\gamma (\Phi-\tilde h)_{\rho=0, [\xi_{2n+1},\xi_{2n+2}]}=2\gamma \left(\sum_{j=1}^{2n+1}(-1)^j\xi_j-\xi_{2n+2} \right) \,, \,\,\text{even number of G-flips} \,,\\
	\hat b_2^0&=-\gamma (\Phi-\tilde h)_{\rho=0, [\xi_{2n+1},+\infty]}=2\gamma \sum_{j=1}^{2n+1}(-1)^j\xi_j \,, \,\, \text{odd number of G-flips} \,.
\end{aligned}
\label{b20 gauge}
\end{equation}

For the solutions of \eqref{Gforms1} we find
\begin{equation}
\begin{aligned}
	\Lambda&=-4\xi^2-2\rho^2-4\sum_{j=1}^{2n+2}(-1)^j(\xi-\xi_j)\sqrt{(\xi-\xi_j)^2+\rho^2} \,,	\\
	\Lambda&=-4\sum_{j=1}^{2n+1}(-1)^j(\xi-\xi_j)\sqrt{(\xi-\xi_j)^2+\rho^2} \,.
\end{aligned}
\end{equation}
Taking the $\rho \rightarrow 0$ limit of the above expressions and of the $\Phi$'s in \eqref{phitildephi1} and \eqref{phitildephi2} we  find
\begin{equation}
\begin{aligned}
	\widehat{\Omega}_1|_{\rho=0}&=4\sum_{j=1}^{2n+2}(-1)^j(\xi+\xi_j)|\xi-\xi_j|+2 \left(\sum_{j=1}^{2n+2}(-1)^j|\xi-\xi_j|\right)^2+\frac{2}{\gamma}\hat{b}_2^0\sum_{j=1}^{2n+2}(-1)^j|\xi-\xi_j|-\frac{1}{2}\hat{b}_2^0\hat{b}_3^0 \,, \\
	\widehat{\Omega}_1|_{\rho=0}&=4\sum_{j=1}^{2n+1}(-1)^j\xi_j|\xi-\xi_j|+2 \left(\sum_{j=1}^{2n+1}(-1)^j|\xi-\xi_j|\right)^2\\
	&-2\xi^2+\frac{2}{\gamma}\hat{b}_2^0\left(-\xi+\sum_{j=1}^{2n+1}(-1)^j|\xi-\xi_j|\right)-\frac{1}{2}\hat{b}_2^0\hat{b}_3^0 \,.
\end{aligned}
\end{equation} 
One can then associate an M2 Page charge to a blue interval $[\xi_{2m-1},\xi_{2m}]$ (which corresponds to a four-cycle that carries an M5 flux), where $m=1,\dots,n+1$ for an even-flip solution and $m=1,\dots,n$ for an odd-flip one. For both types of solution this is given by:
\begin{equation}
\label{M2chargef}
	Q_{M2,m}\equiv  \widehat{\Omega}_1|_{\rho=0,\xi=\xi_{2m}+\varepsilon}-\widehat{\Omega}_1|_{\rho=0,\xi=\xi_{2m-1}-\varepsilon}  = 16\,(\xi_{2m}-\xi_{2m-1})\sum_{j=2m}^{2n+1}(-1)^{j+1}\xi_j \,,
\end{equation}
where we  have used \eqref{dC6c} and (\ref{Omegareln1}). 
Using \eqref{M5charges} we can express this relation in terms of 5-brane fluxes:
\begin{equation}
\label{M2chargefM5}
	\frac{Q_{M2,m}}{Q_{M5\,, m}} ~=~  Q'_{M5'\,, m}+Q'_{M5'\,, m+1}+\dots + Q'_{M5'\,, n}  \,,
\end{equation}
which shows directly how the M2 charge comes from the smooth cohomological M5 and M5' fluxes via the Chern-Simons interaction.

We can also associate an M2 Page charge to red intervals $[\xi_{2m},\xi_{2m+1}]$ (which correspond to four cycles that carry an M5' flux), with $m=1,\dots,n\,$, using now \eqref{Omega1bdry2} for $\widehat{\Omega}_1$ at the boundary. Repeating the above procedure we find for the two kinds of solutions
\begin{equation}
\begin{aligned}
	\widehat{\Omega}_1|_{\rho=0}&=4\sum_{j=1}^{2n+2}(-1)^j(\xi_j-\xi)|\xi-\xi_j|-2 \left(\sum_{j=1}^{2n+2}(-1)^j|\xi-\xi_j|\right)^2  + 2\gamma\hat{b}_3^0\sum_{j=1}^{2n+2}(-1)^j|\xi-\xi_j| \\ 
& \qquad \qquad +4\xi \gamma \hat{b}_3^0+\frac{1}{2}\hat{b}_2^0\hat{b}_3^0 \,,
\end{aligned}
\end{equation}
\begin{equation}
\begin{aligned}
	\widehat{\Omega}_1|_{\rho=0}&=4\sum_{j=1}^{2n+1}(-1)^j\xi_j|\xi-\xi_j|-2 \left(\sum_{j=1}^{2n+1}(-1)^j|\xi-\xi_j|\right)^2+2\xi^2+2\gamma\hat{b}_3^0\left(\xi+\sum_{j=1}^{2n+1}(-1)^j|\xi-\xi_j|\right)\\ 
& \qquad \qquad +\frac{1}{2}\hat{b}_2^0\hat{b}_3^0 \,,
\end{aligned}
\end{equation}
and the M2 Page charge is given for both solutions by: 
\begin{equation}
\label{M2chargefp}
	Q'_{M2,m}\equiv \widehat{\Omega}_1|_{\rho=0,\xi=\xi_{2m+1}+\varepsilon}-\widehat{\Omega}_1|_{\rho=0,\xi=\xi_{2m}-\varepsilon}=16\,(\xi_{2m+1}-\xi_{2m})\sum_{j=1}^{2m}(-1)^{j}\xi_j \,.
\end{equation}
Using again \eqref{M5charges}, the above equation can be written as:
\begin{equation}
\label{M2chargefpM5}
	\frac{Q'_{M2,m}}{Q'_{M5',m}} ~=~ Q_{M5,1}+Q_{M5,2}+\dots +Q_{M5,m}   \,.
\end{equation}

Now, we would like to express \eqref{M2chargefM5} and \eqref{M2chargefpM5} in terms of integer brane charges. The M5- and M2-brane charges are quantized in units of the M5- and M2-brane tensions:
\begin{equation}
\begin{aligned}
	Q_{M5}&=\frac{1}{2\pi^2}2\kappa_{11}^2T_5 \, N_{M5} \,, \\
	Q_{M2}&=\frac{1}{(2\pi^2)^2}2\kappa_{11}^2T_2 \,N_{M2} \,,
\end{aligned}
\label{QuantizedCharges}
\end{equation}
where $2\kappa_{11}^2=(2\pi)^8\ell_{11}^9$, $T_5=1/ \left((2\pi)^5\ell_{11}^6 \right)$, $T_2=1/ \left((2\pi)^2\ell^3_{11}\right)$, and $\ell_{11}$ is the eleven-dimensional Planck length. Using these, we can write \eqref{M2chargefM5} as 
\begin{equation}
	N_{M2,m}=N_{M5,m}\left( N'_{M5'\,, m}+N'_{M5'\,, m+1}+\dots + N'_{M5'\,, n} \right) \,.
\label{NM2m}
\end{equation}
%where for simplicity we chose the signs $\nu_i$ so that all charges are positive.
The total quantized M2 Page charge on all the (blue) intervals with M5 flux is now given by:
\begin{equation}
	N_{M2}^{\text{blue}}=\sum_{m=1}^{M}N_{M5,m}\left( N'_{M5'\,, m}+N'_{M5'\,, m+1}+\dots + N'_{M5'\,, n} \right) \,, 
\label{totalM2}
\end{equation}
where $M$ is equal to $n$ or $n+1$ depending on whether we have an odd- or even-flip solution.

Similarly, for the quantized M2 Page charge associated with (red) intervals with M5' flux we obtain
\begin{equation}
	N'_{M2,m} ~=~ N'_{M5',m}\left(N_{M5,1}+N_{M5,2}+\dots +N_{M5,m}  \right) \,,
\label{NpM2m}
\end{equation} 
and the total $N_{M2}^{'\,\text{red}}$ Page charge is 
\begin{equation}
\label{totalM2p}
	N_{M2}^{'\,\text{red}}=\sum_{m=1}^n N'_{M5',m} \left( N_{M5,1}+N_{M5,2}+\dots + N_{M5,m} \right)\,.
\end{equation}

This charge structure  can be encoded via a Young Tableau \cite{Bachas:2013vza}.  The red and blue intervals correspond to (bubbled) stacks of M5 and M5' charges.  A  stack of $N_{M5,m}$  M5 branes is represented in the tableau by $N_{M5,m}$ rows that all have the same length, $N_{M2,m}/N_{M5,m}$ (see  (\ref{NM2m})), so that the area of this block  of rows in the tableau is  $N_{M2,m}$.  Similarly, each stack of $N'_{M5',m}$  M5' branes corresponds to a set  columns that all have the length $N'_{M2,m}/N'_{M5',m}$  (see  (\ref{NpM2m})) so that the area of this block  of columns in the tableau is  $N'_{M2,m}$.  The shape of the tableau means that   (\ref{totalM2}) is simply the total number of boxes in the tableau obtained by summing the number of boxes row by row, while (\ref{totalM2p}) is the total number of  boxes in the tableau obtained by summing the number of boxes column  by column. 
Thus total M2 Page charge of the blue intervals, $N_{M2}^{\text{blue}}$, and the total M2 Page charge of the red intervals, $ N_{M2}^{'\,\text{red}}$, must be the same. 

Since the M2 Page charge is conserved, this relationship should be visible through deformation of contours. One does, however, have to make a consistent and correct gauge choice so that the one does not inadvertently change  the gauge and hence the integrand.
It is not hard to see that for even-flip solutions, the total M2 charge is either  $N_{M2}^{\text{blue}}$ or  $N_{M2}^{'\,\text{red}}$, depending on the color at infinity. 

\begin{figure}[h]
    \centering
    \includegraphics[width=.8 \textwidth]{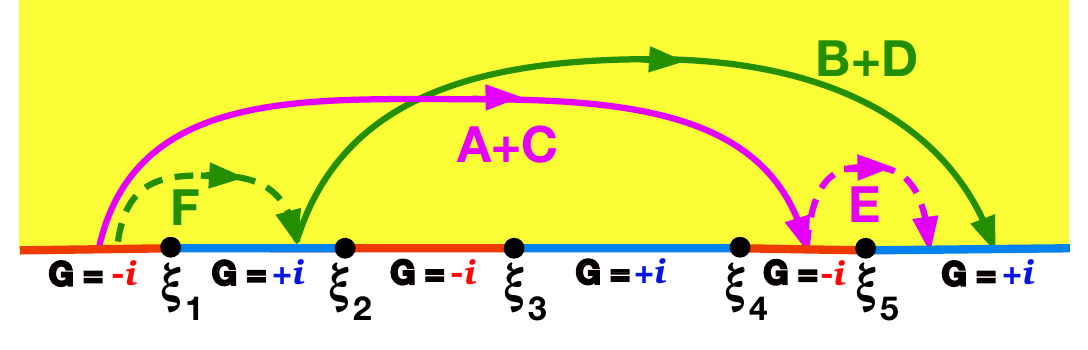}  
    \caption{Deforming the cycles of Fig.~\ref{fig:Topology2}, and introducing cycles $E$ and $F$ around $\xi_5$ and $\xi_1$, respectively. The cycles $A,B,C,D$ are now to be viewed as $7$-cycles with topology $S^4 \times S^3$, while $E$ and $F$ are to be viewed as $7$-cycles with topology $S^7$. The cycles $A+C+E$ and $B+D+F$ are homologous and can be deformed to wrap the $S^7$ at infinity. }
    \label{fig:TopologyS7}
\end{figure}

 For the odd-flip solutions one must proceed more carefully. {\it A priori} one chooses the $b_j^0$ to make the fluxes smooth on the $S^7$ at infinity.  One also has the choice of taking $\epsilon = \pm 1$, thereby making the fluxes smooth on the all the cycles surrounding either the red or blue intervals.   For simplicity, we will consider the five-flip solution depicted in Fig.~\ref{fig:Topology2}.  The fundamental cycles are then $A,B,C,D$, and these have the topology of $S^4 \times S^3$ and each can contribute to the M2 page charge.   These cycles can be added to give the cycles $A+C$ and $B+D$ depicted in Fig.~\ref{fig:TopologyS7}.  Now add the $7$-cycles, $E$ and $F$, defined by fibering $S^3 \times S^3$ along arcs around  $\xi_5$ and $\xi_1$, respectively. These define $S^7$'s, and because $h$ does not have a pole at $\xi_1$ and $\xi_5$, each of these arcs is topologically trivial (because the spatial sections are simply $\IR^8$ near these points \cite{Bachas:2013vza}). 

The important point is that $A+C+E$ and $B+D+F$ are both homologous to the contour around infinity and so both of them measure the M2 Page charge at  infinity. Since the contours $E$ and $F$ are trivial, they cannot contribute to this charge and so the Page charge can be computed from either $A+C$ or $B+D$, which means that:
\begin{equation}
	N_{M2}^{\text{total}}= N_{M2}^{'\,\text{red}}=N_{M2}^{\text{blue}}	\,.
\label{Pagetotal}
\end{equation}

There is, however, one detail to be checked. While $E$ and $F$ are topologically trivial, they could still give non-trivial contributions to the Page charge if the flux integrand is singular at $\xi_5$ or $\xi_1$.  However, for the cycles $A$ and $C$ we must take $\epsilon = +1$ so that that the integrated flux is regular when the ${S'}^3$ collapses at $G=-i$. This choice also ensures regularity of the flux on the left-hand side of the contour $E$. On the right-hand side of the contour $E$, we have $G=i$ and the ${S}^3$ collapses, which means that we must have $b_2 =0$, but we set the boundary conditions at infinity (see (\ref{b20 gauge})) so that $b_2=0$ on the  interval $[\xi_5,+\infty)$.  Thus the $\epsilon = +1$ flux is smooth everywhere along $A+C+E$.   Regularity   for $B+D+F$ follows similarly because of the gauge choice  (\ref{b30 gauge})  and we take the flux with $\epsilon = -1$. Thus one gets the  total M2 charge, measured at infinity, by  decomposing into different choices of cycles and with integrands in different gauges that have been fixed by regularity.

%%%%%%%%%%%%%%%%

%%%%%%%%%%%%%%%%%%%%%%%%%%%%%%%%
%%%%%%%%%%%%%%%%%%%%%%%%%%%%%%%%

More broadly, we also want to note that the total M2 charge comes purely from the cohomological M5 and M5' fluxes threading topologically-non-trivial four-cycles, and there are no singular sources.  Moreover, the M2 charge comes about from the intersection  cohomology, much as it does in five dimensions \cite{Gibbons:2013tqa}.  

The fact that, when determining the total M2 Page charge of odd-flip solutions, one should consider only either $N_{M2}^{'\,\text{red}}$ or $N_{M2}^{\text{blue}}$ can also be seen by computing the gauge-invariant M2 Maxwell charge at the flip point at infinity. This charge can also be extracted from the radius of the asymptotic $\text{AdS}_4/\mathbb{Z}_2 \times S^7$ metric.   Indeed, it is straightforward to compute this metric. 

Introduce polar coordinates on the Riemann surface:
\begin{equation}
	\xi=\lambda^2 \cos \theta\,, \quad \quad \quad \rho=\lambda^2 \sin \theta \,,
\end{equation}
where $0 \leq \theta \leq \pi$. Using these and expanding \eqref{DHokerAnsatz} at large $\lambda$ we find: 
\begin{equation}
	ds_{11}^2=\left[ A_1 \lambda^4 ds^2_{AdS_3}+A_2 \frac{d \lambda^2}{\lambda^2} \right] +A_2 \left[d\left(\frac{\theta}{2} \right)^2+\sin^2\left(\frac{\theta}{2} \right)ds^2_{S^3}+\cos^2 \left(\frac{\theta}{2} \right) ds^2_{S'^3} \right] \,, 
\label{asymAdS4}
\end{equation}
where 
\begin{equation}
	A_1=\frac{-\gamma}{(1+\gamma)^2}\frac{16}{A_2^2}=\frac{-\gamma}{(1+\gamma)^2}\frac{8\times 2^{1/3}}{\left(Q_{M2}^{\rm Maxwell}\right)^{2/3}} 
\label{A1A2}
\end{equation}
and $Q_{M2}^{\rm Maxwell}$ is the sum over $m$ of either \eqref{M2chargef} or \eqref{M2chargefp} for an odd-flip solution. In \eqref{asymAdS4} the second factor is an $S^7$ metric, while the first factor is a section of an AdS$_4$ metric. This can be seen by suitably comparing with \eqref{AdSmet3}. 

By taking the near horizon limit of the standard M2-brane metric, it is not hard to see that the $S^7$ radius, $A_2$, is equal to 
\begin{equation}
	A_2=\left(2^5\pi^2 N_{M2}^{\rm Maxwell} \right)^{1/3} \ell^2 \,,
\end{equation}
where $N$ is the number of M2-branes. Using now \eqref{A1A2} and \eqref{QuantizedCharges} we finally find that
\begin{equation}
N_{M2}^{\rm Maxwell}= N_{M2}^{'\,\text{red}}=N_{M2}^{\text{blue}} \,.
\end{equation}
Hence, as we have already noted before \eqref{Pagetotal}, the M2  Maxwell charge evaluated at the flip point at infinity equals the total Page charge, which can either be computed in a gauge suited for the blue intervals or in a gauge suited for red intervals.

%%%%%%%%%%%%%%%%%%%%%%%%%%%%%%%%%%%%%
\subsection{Scale-invariant coordinates}
\label{sec:scale-invariant}
%%%%%%%%%%%%%%%%%%%%%%%%%%%%%%%%%%%%%
As we mentioned in Section \ref{ss:mapping}, there are two scale-invariant combinations of coordinates, involving the ``radial" coordinates, $(u,v,z,w)$:
\begin{equation}
\hat z ~\equiv~ u^2 z ~=~  -\sgngam\,  \frac{\varepsilon_1}{2\, \sqrt{|\gamma|}} \,     \big(  \Phi +  \beta \, \xi \big)   \,, \qquad \hat w ~\equiv~ v^2 w ~=~  \frac{\varepsilon_1~ \sqrt{|\gamma|}}{2} \,\big(\Phi-\beta \xi\big)  \,,	
\end{equation}
and it is useful to study their behavior at the boundary of $\Sigma$. 

Starting with $\hat z$ and looking at the solution with an even number of $G$-flips (the first choice of $G$ in \eqref{Gforms1}) we obtain
\begin{equation}
	\lim_{\rho \rightarrow 0} \hat z= \frac{1}{2\sqrt{|\gamma|}} \left(4\xi +2\sum_{j=1}^{2n+2} (-1)^j|\xi-\xi_j| \right) \,,
\label{zhat1}
\end{equation}
where, as usual, we took $\beta=2$, and we set $\varepsilon_1=1$, since $\nu_2=\text{sign}(\gamma)\varepsilon_1$ and $\nu_2=-1$. 

When evaluating the above sum at the red interval $[\xi_{2m},\xi_{2m+1}]$, there is an even number of $|\xi-\xi_j|<0$ terms and an even number of $|\xi-\xi_j|>0$ terms. Therefore, all $\xi$'s coming from the sum cancel and we are left with the following linear function in $\xi$:
\begin{equation}
	\lim_{\rho \rightarrow 0} \hat z\big|_{\xi \in [\xi_{2m},\xi_{2m+1}]} =  \frac{1}{\sqrt{|\gamma|}} \left(2\xi - \sum_{j=1}^{2m}(-1)^j\xi_j+\sum_{j=2m+1}^{2n+2}(-1)^j\xi_j \right) \,.
\label{zhat2}
\end{equation}
On the other hand, at the blue interval $[\xi_{2m-1},\xi_{2m}]$ the sum contributes a $-2\xi$ term and we are left with the following constant expression:
\begin{equation}
	\lim_{\rho \rightarrow 0} \hat z\big|_{\xi \in [\xi_{2m-1},\xi_{2m}]} = -\frac{1}{\sqrt{|\gamma|}} \left( \sum_{j=1}^{2m-1}(-1)^j\xi_j-\sum_{j=2m}^{2n+2}(-1)^j\xi_j \right) \,,
\label{zhat3}
\end{equation}
which is equal to \eqref{zhat2} evaluated at $\xi=\xi_{2m-1}$. 

From the foregoing discussion it should be clear that as we move along the $\rho=0$ axis, $\hat z$ is a continuous monotonically increasing function that is linear in $\xi$ in the red intervals, where $G=-i$, and constant along the blue intervals, where $G=i$. Moreover, the values of $\hat z$ in two subsequent red intervals differ from each other by a jump which is proportional to the M5 charge associated with the blue interval that separates them.

To make contact with the previous section, it is actually more useful to compute the following quantity:
\begin{equation}
	\lim_{\rho \rightarrow 0} \left( \hat{z}+ \frac{1}{2}\frac{\nu_2}{c_2^3}\, \hat b_2^0 \right)\Big|_{\xi \in [\xi_{2m-1},\xi_{2m}]} = -\frac{2}{\sqrt{|\gamma|}}\sum_{j=2m}^{2n+1}(-1)^{j+1}\xi_j \,. 
\label{zhat4}
\end{equation}
Comparing \eqref{zhat4} to \eqref{M2chargef} and using \eqref{M5charges}, it is easy to see that the displaced $\hat z$ is equal to 
\begin{equation}
	\lim_{\rho \rightarrow 0} \left( \hat{z}+ \frac{1}{2}\frac{\nu_2}{c_2^3}\, \hat b_2^0 \right)\Big|_{\xi \in [\xi_{2m-1},\xi_{2m}]}=\frac{Q_{M2,m}}{2Q_{M5,m}}\,.
\label{zhat5}
\end{equation}

Moving now to the other scale invariant coordinate, $\hat w$, we find that it follows the opposite behavior along the boundary of $\Sigma$. It is a monotonically decreasing function that is linear in $\xi$ in the red intervals and constant in the blue ones: 
\begin{equation}
\begin{aligned}
	\lim_{\rho \rightarrow 0} \hat{w} \big|_{\xi \in [\xi_{2m},\xi_{2m+1}]} &= \sqrt{|\gamma |} \left(-\sum_{j=1}^{2m}(-1)^j\xi_j + \sum_{j=2m+1}^{2n+2}(-1)^j \xi_j \right) \,, \\
	\lim_{\rho \rightarrow 0} \hat{w} \big|_{\xi \in [\xi_{2m-1},\xi_{2m}]} &= \sqrt{|\gamma |} \left( -2\xi -\sum_{j=1}^{2m}(-1)^j\xi_j + \sum_{j=2m+1}^{2n+2}(-1)^j \xi_j \right) \,.
\end{aligned}
\label{what1}
\end{equation}

Now, in order to compare with \eqref{M2chargefp}, we have to look at the following expression:
\begin{equation}
	\lim_{\rho \rightarrow 0} \left( \hat w -\frac{1}{2}\frac{\nu_3}{c_3^3}\hat{b}_3^0 \right)\Big|_{\xi \in [\xi_{2m},\xi_{2m+1}]} = -2\sqrt{|\gamma |}\sum_{j=1}^{2m}(-1)^j \xi_j \,. 
\label{what2}
\end{equation}
Using as before \eqref{M5charges} we find
\begin{equation}
	\lim_{\rho \rightarrow 0} \left( \hat w -\frac{1}{2}\frac{\nu_3}{c_3^3}\hat{b}_3^0 \right)\Big|_{\xi \in [\xi_{2m},\xi_{2m+1}]} = \frac{Q'_{M2,m}}{2Q'_{M5',m}} \,. 
\label{what3}
\end{equation}

Note that \eqref{zhat5} and \eqref{what3} also hold for the solution with an odd number of $G$-flips (the second choice of $G$ in \eqref{Gforms1}). 

%%%%%%%%%%%%%%%%%%%%%%%%%%%%%%%%%%%%%
\section{M2-M5 probes}
\label{sec:probes}
%%%%%%%%%%%%%%%%%%%%%%%%%%%%%%%%%%%%%

As we  mentioned in the Introduction, AdS$_3 \times$ S$^3 \times$ S$^3 \times \Sigma$ solutions with the same charges but different values of $\gamma$ represent near-horizon limits of the same system of intersecting M2 and M5 branes in flat space. We have shown in \cite{Bena:2024dre} that $\gamma=1$ solutions correspond to a scaling limit of a series of M2-M5 spikes, and that the AdS$_3$ radius, $\mu$, is the coordinate along the spike world-volume.  

The purpose of this Section is to add a probe M2-M5 spike to a bubbling geometry with negative $\gamma$, and to show that its location coincides exactly to the location of a region of the geometry containing a small bubble with the same M2 and M5 charges. This will establish, as we will discuss in detail in Section \ref{sec:Geometric}, that bubbling geometries with negative $\gamma$ come from the geometric transition of M2-M5 spikes. 

To find the action of this spike one may be tempted to work with the M5-brane action \cite{Bandos:1997ui}, but using this action is quite subtle and un-illustrative. It is much simpler to reduce both the background and the probe action along an isometry direction to Type-IIA String Theory, and to evaluate the action of the probe using the D4-brane Born-Infeld action. We begin with the M-theory background specified by the metric and fluxes in \eqref{DHokerAnsatz} and use the Poincaré AdS$_3$ metric:
\begin{equation}
\begin{aligned}
ds_{11}^2 &~=~ f_1^2\left(\frac{d\mu^2}{\mu^2}+\mu^2\left(-dt^2+dy^2\right)\right) ~+~ f_2^2 \, ds_{S^3}^2 ~+~ f_3^2 \, ds_{{S'}^3}^2 ~+~ f_4^2|dw|^2  \,, \\
C^{(3)}  &~=~  b_1\, \hat e^{012} ~+~ b_2\, \hat e^{345} ~+~ b_3\, \hat e^{678}  \,.
\end{aligned}
 \label{DHokerAnsatz2}
\end{equation}
Note that the overall factor $e^{2A}$ has been absorbed into the functions $f_i$'s, which now take the form:
\begin{equation}
\begin{aligned}
	f_1^6&=\frac{h^2 W_+W_-}{64(G\overline G-1)^2}\,, \hspace{30pt} f_2^6=\frac{h^2(G\overline G-1)W_-}{W_+^2}\,, \\
	f_3^6&=\frac{h^2(G\overline G-1)W_+}{W_-^2}\,, \hspace{20pt} f_4^6=\frac{|\partial_w h|^6}{h^4}(G\overline G-1)W_+W_- \,.
\end{aligned}
\end{equation} 
Next, we perform a dimensional reduction of the eleven-dimensional configuration along the $y$-direction. Applying the standard relations connecting type IIA and 11-dimensional supergravity backgrounds:
\begin{align}
\label{IIA-M}
	ds_{11}^2&=e^{-\frac{2\phi}{3}}ds_{10}^2+e^{\frac{4\phi}{3}}(dy+C_1)^2 \,, \\
	C_3'&=C_3+B_2 \wedge dy \,,
\end{align}
we find the IIA solution:
\begin{equation}
\begin{aligned}
	ds_{10}^2&=-\mu^3f_1^3 dt^2+\frac{f_1^3}{\mu}d\mu^2+\mu f_1 f_2^2 ds_{S^3}^2+\mu f_1 f_3^2 ds_{{S'3}^3}^2+\mu f_1 f_4^2|dw|^2 \,, \\
	C_3&=b_2\, \hat e^{345} ~+~ b_3\, \hat e^{678} \,, \hspace{15pt} B_2=-\mu b_1 dt\wedge d\mu \,, \hspace{15pt} e^{2\phi}=\mu^3 f_1^3\,.
\end{aligned}
\end{equation}

Using the map between M2-M5 solutions and AdS$_3$ coordinates \eqref{variables3}, we can see that the M2-M5 spikes sourcing this solution are extended along the $AdS_3\times S^3$ part of the metric, and the world-volume M2 charges are along $AdS_3$. We reduce along $y$, and this will transform the M2-M5 spike into a D4-F1 spike. The F1 charge of the D4 brane is carried by the brane world-volume electric field $F_{t\mu}$. This is nothing but a higher-dimensional generalization of the Callan-Maldacena spike \cite{Callan:1997kz}, placed in a non-trivial background.

The D4-brane world-volume will be parametrized by the coordinates $(\eta_0,\eta_1, \eta_2,\eta_3, \eta_4)$, where we identify $\eta_0=t$ and $\eta_1=\mu$, while $(\eta_2,\eta_3,\eta_4)$ are identified with the $S^3$ coordinates. Furthermore, we will allow for a general embedding in which $\xi$ and $\rho$ are functions of $\eta_1$. Therefore, the metric induced on the D4-brane takes the form:
\begin{equation}
	d\tilde{s}_5^2=-\eta_1^3 f_1^3 d\eta_0^2+\left(\frac{f_1^3}{\eta_1}+\eta_1 f_1 f_4^2 \left( \left(\frac{\partial \xi}{\partial \eta_1} \right)^2+\left(\frac{\partial \rho}{\partial \eta_1} \right)^2\right)\right)d\eta_1^2+\eta_1 f_1 f_2^2 d\tilde{s}_{S^3}^2
\end{equation}
and the induced NS-NS and RR gauge potentials are 
\begin{equation}
	\tilde{B}_2=-\eta_1 b_1 d\eta_0\wedge d\eta_1 \,, \quad \quad \tilde{C}_3=b_2 \hat{\tilde{e}}^{345}\,,
\end{equation}
where a tilde denotes a pullback to the D4-brane world-volume. 

The F1 charge of the spike is carried by a world-volume 2-form field:
\begin{equation}
	F_2~=~ \mathcal{F}d\eta_0\wedge d\eta_1~=~ (\partial_0 A_1 - \partial_1 A_0) \,d\eta_0 \wedge d\eta_1\,.
	\label{F2form}
\end{equation}
The DBI and WZ actions can now be straightforwardly computed: 
\begin{equation}
\begin{aligned}
	S_{DBI}&=-T_4\int\,d^5\eta \, e^{-\phi}\sqrt{-\det \left( \tilde{G}_{\alpha \beta}+F_{\alpha \beta}+\tilde{B}_{\alpha \beta}\right)} \\
&=-T_4 \int \,d^5\eta \, f_2^3\sqrt{\eta_1^2 f_1^6-(\mathcal{F}-\eta_1 b_1)^2+\eta_1^4 f_1^4 f_4^2 \big((\partial_1 \xi)^2+(\partial_1 \rho)^2\big)} \,, \\
	S_{WZ}&=-T_4 \int\, e^{\tilde{B}_2+\tilde{F}_2}\wedge \oplus_n \tilde{C}_n=-T_4 \int  d^5\eta 
	\, \left(\mathcal{F}-\eta_1 b_1\right)b_2 -T_4 \int \, \tilde{C}_5\, .
	\label{Action1}
\end{aligned}
\end{equation}
To determine the expression for $\tilde{C}_5$, we make use of the following relations:
\begin{align}
\label{conventions}
    F_p&=dC_{p-1} \hspace{88pt} \text{for } p<3 \,, \nonumber \\
    F_p&=dC_{p-1}+H_3\wedge C_{p-3} \hspace{25pt} \text{for } p \geq 3\,,  \\
    F_6&=\star F_4 \,, \hspace{20pt} F_8=\star F_2 \,.  \nonumber
\end{align}
Using these and taking into account that $C_1=0$ we obtain:
\begin{equation}
\begin{aligned}
	dC_5=&\star \left( db_2 \wedge \hat e^{345}\right)+\mu b_3 db_1 \wedge dt \wedge d\mu \wedge \hat e^{678} \, \\
	&+\star \left( db_3 \wedge \hat e^{678} \right)+\mu b_2 db_1 \wedge dt \wedge d\mu \wedge \hat e^{345} \,,
\end{aligned}
\label{dC5a}
\end{equation}
where $\star$ denotes the Hodge star operator in ten dimensions. The first line in the above expression corresponds to a component of $C_5$ along the Riemann surface, $\Sigma$, and $dt\wedge d\mu \wedge \hat e^{678}$, while the second produces a component along $\Sigma$ and $dt \wedge d\mu \wedge \hat e^{345}$. Since only the pullback of $C_5$ onto the D4-brane probe world-volume is relevant for our purposes, we retain only the contributions from the second line, which give:
\begin{equation}
dC_5|_{\rm D4}	-\mu \frac{f_1^3f_2^3}{f_3^3}\left(\partial_{\xi}b_3\,d\rho-\partial_{\rho}b_3\, d\xi \right)\wedge dt \wedge d\mu \wedge d\Omega_3 + \mu b_2 \left(\partial_{\xi}b_1\,d\xi+\partial_{\rho}b_1\,d\rho \right)\wedge dt \wedge d\mu \wedge d\Omega_3 \,.
\label{dC5b}
\end{equation}
This expression is rather cumbersome to integrate, but as we will see, we will eventually only need $dC_5|_{\rm D4}$.

The momentum conjugate to $\mathcal{F}$ encodes the number of fundamental strings (or equivalently M2-branes) that generate the spike terminating on the D4 (or M5) branes:
\begin{equation}
\label{Cmomentum}
	\Pi=\frac{\partial \mathcal{L}}{\partial {(\partial_0 A_1)}}=\left(-b_2+\frac{f_2^3\left(\mathcal{F}-\eta_1 b_1 \right)}{\sqrt{\eta_1^2 f_1^6-(\mathcal{F}-\eta_1 b_1)^2+\eta_1^4 f_1^4 f_4^2 \big((\partial_1 \xi)^2+(\partial_1 \rho)^2\big)}}\right) \,.
\end{equation} 
Solving for $\mathcal{F}$, we find
\begin{equation}
	\mathcal{F}~=~ \eta_1 b_1 \pm \frac{f_1^2\sqrt{\eta_1^2f_1^2+f_4^2\big((\partial_1 \xi)^2+(\partial_1 \rho)^2\big)\eta_1^4}}{\sqrt{f_2^6+(b_2+\Pi)^2}}|b_2+\Pi| \,.
\label{F}
\end{equation}
The Hamiltonian density can now be straightforwardly derived: 
\begin{equation}
\begin{aligned}
	\mathcal{H}& = -\eta_1 b_1 b_2 +\eta_1 f_1^2 \sqrt{\left(f_1^2+f_4^2 \big((\partial_1 \xi)^2+(\partial_1 \rho)^2 \big)\eta_1^2 \right)\left(f_2^6+(b_2+\Pi)^2 \right)} \pm \eta_1 b_1 |b_2+\Pi| +\tilde{C}_5\,,
\end{aligned}
\label{finalH}
\end{equation}
where we used \eqref{F} to express $\mathcal{F}$ in terms of $\Pi$. 

This Hamiltonian gives the potential felt by the probe spike when moving in the Riemann-surface. It is not hard to see that the probe will solve the equations of motion on the $\rho=0$ line, at some value of $\xi$. In a given background the location of this minimum depends on the value of $\Pi$. 

The expression in \eqref{finalH} admits two different forms, depending on the choice of solution for $\mathcal{F}$ in \eqref{F} and on the sign of $b_2+\Pi$. This determines whether the M5 brane carries M2-brane or anti-M2-brane charge. Knowing in hindsight that we are looking for probes that respect the supersymmetry of the background, we select the minus solution in \eqref{F} and assume that $b_2+\Pi<0$, which leads to:
\begin{equation}
	\mathcal{H}=\eta_1 \Pi \, b_1 + \eta_1 f_1^2 \sqrt{\left(f_1^2+f_4^2 \big((\partial_1 \xi)^2+(\partial_1 \rho)^2 \big)\eta_1^2 \right)\left(f_2^6+(b_2+\Pi)^2 \right)} +\tilde{C}_5 \,.
\end{equation}

Since our focus is on probe branes located at a fixed point on $\partial \Sigma$, we take the $\rho \rightarrow 0$ limit of $\delta \mathcal{H}/\delta \xi$ and set $\partial_1 \xi$ to zero. Additionally, we rewrite $b_i$ in terms of $\hat b_i$ and, for negative $\gamma$, we will choose the following values for the signs $\nu_i$: $\nu_1=-1,\nu_2=-1,\nu_3=1$ . As for $\tilde{C}_5$, let us write the part that concerns us as:
\begin{equation}
\label{genC5}
	C_5=g_1(\xi,\rho)dt\wedge d\mu \wedge d\Omega_3+g_2(\mu)d\xi \wedge dt \wedge d\Omega_3 + g_3(\mu) d\rho \wedge dt \wedge d\Omega_3 + \ldots 
\end{equation}
where the $\ldots$ denote the components of $C_5$ that do not pull back on the brane world-volume. The corresponding $dC_5$ is: 
\begin{equation}
\label{dC5 gen}
	dC_5=\left(\partial_{\xi}g_1+\partial_{\mu}g_2 \right)d\xi \wedge dt \wedge d\mu \wedge d\Omega_3 + \left(\partial_{\rho}g_1+\partial_{\mu}g_3 \right) d\rho \wedge dt \wedge d\mu \wedge d\Omega_3 + \ldots
\end{equation}
Its world-volume pullback, $\tilde{C}_5$, is then:
\begin{equation}
\label{pullC5}
	\tilde{C}_5=\left(g_1-g_2 \partial_{\mu} \xi -g_3 \partial_{\mu}\rho \right) dt \wedge d\mu \wedge d\Omega_3 \,. 
\end{equation}
and $\delta \tilde{C}_5/ \delta \xi$ by:
\begin{equation}
\label{delta C5}
	\frac{\delta \tilde{C}_5}{\delta \xi}= \frac{\partial \tilde{C}_5}{\partial \xi}-\partial_{\mu}\left( \frac{\partial \tilde{C}_5}{\partial (\partial_{\mu}\xi)} \right)=\partial_{\xi}g_1+\partial_{\mu}g_2 \,. 
\end{equation}
We see, therefore, that we only need the component of $dC_5$ along $d\xi$\,.
Retaining, thus, only the $d\xi$ contributions in \eqref{dC5b} and using the fact that $c_1 c_2 c_3 f_1 f_2 f_3=  \tilde \sigma h$, with $\tilde \sigma=-1$, we obtain:
\begin{equation}
	\frac{\delta \mathcal{H}}{\delta \xi}= \partial_{\xi} \left(f_1^3\sqrt{f_2^6+\left(\Pi-\frac{1}{(-\gamma)^{3/2}}\hat b_2\right)^2}\right)+\frac{\gamma \sqrt{-\gamma}}{(1+\gamma)^3}\left(\Pi-\frac{1}{(-\gamma)^{3/2}}\hat b_2\right)\, \partial_{\xi}\hat b_1 +\frac{\gamma^3}{(1+\gamma)^3}\frac{h^3}{f_3^6}\partial_{\rho}\hat b_3  \,,
\label{delH} 
\end{equation}
where we have divided by the overall factor of $\eta_1$. 

For our computation, we will use the $b_2$ gauge of Section \ref{sec:M2chg-spikes}, and without loss of generality we will consider probes only on the left of $\xi_1$. Then, for both the 3-flip and 4-flip solutions, we find that an M5-M2 probe with a given $\Pi$ wants to sit at the following location $\xi_0$: 
\begin{equation}
\label{x0}
	\xi_0=-\frac{1}{4}\sqrt{-\gamma}\,\Pi+\xi_1-\xi_2+\xi_3 \,.
\end{equation}

We can now show that if one replaces this M2-M5 probe with a tiny bubble (corresponding to a tiny blue interval) located at exactly the same location, the ratio of the M2 and M5 charges of the probe will be exactly as that of the bubble. As we will explain in more detail in the next Section, this establishes that the bubbles of the bubbling solution come from the back-reaction of M2-M5 spikes.

\begin{figure}[htbp]
    \centering
    \includegraphics[width=.8 \textwidth]{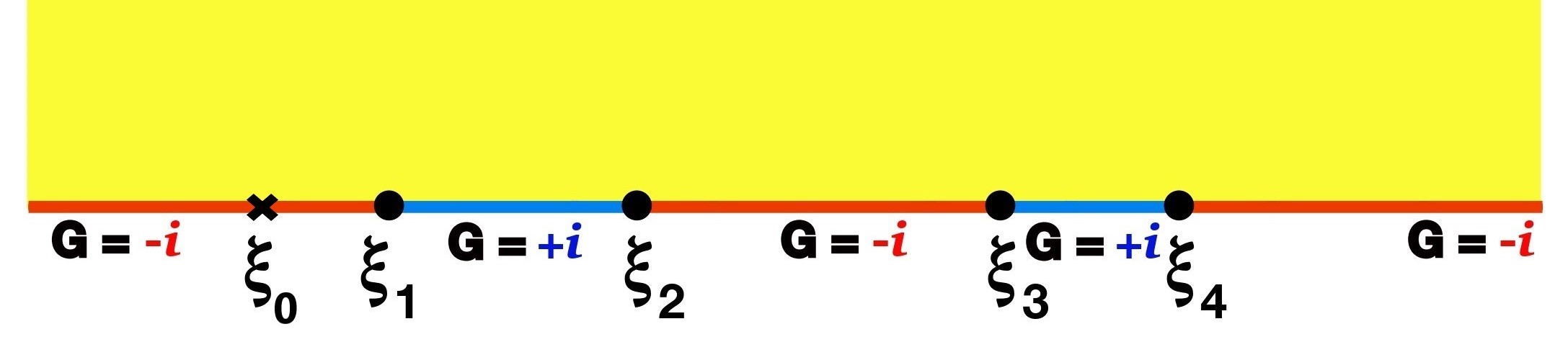}
    \includegraphics[width=.8 \textwidth]{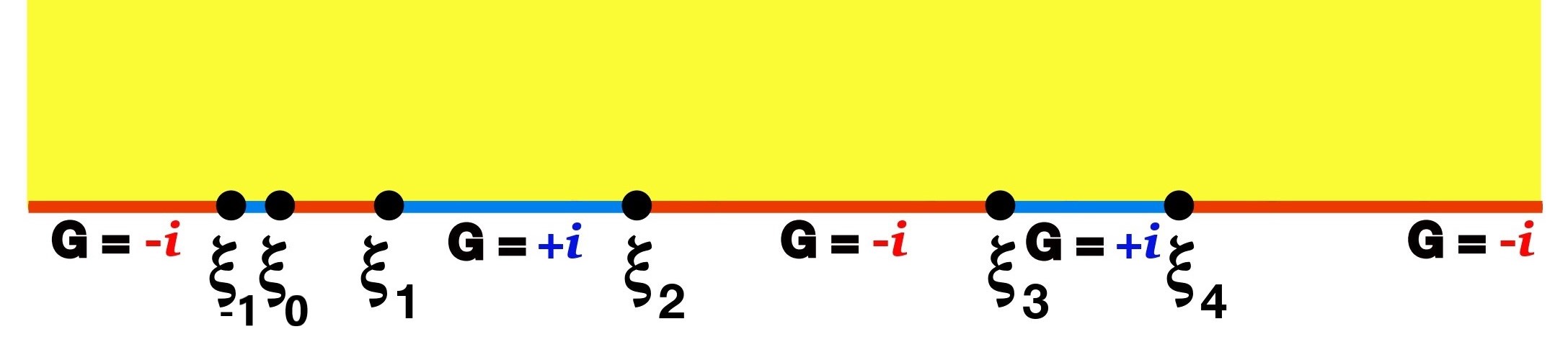}
    \caption{An M2-M5 probe at $\xi_0$ (denoted by $\times$ in the top panel) has exactly the same M2-M5 charge ratio, $\displaystyle \frac{Q_{M2}}{Q_{M5}}$, as a tiny blue interval between $\xi_{-1}$ and $\xi_0$ in the bottom panel.}
    \label{fig:probing}
\end{figure}

For a 3- or 4-flip solution, the M2-charge of the $[\xi_1,\xi_2]$ interval is given by \eqref{M2chargef}:
\begin{equation}
	\frac{Q_{M2}}{Q_{M5}}=\frac{4}{\sqrt{-\gamma}}(-\xi_2+\xi_3) \,.
\end{equation}
If we now add to this solution a tiny blue interval $[\xi_{-1},\xi_0]$, located at the minimum of the M2-M5 probe potential, the M2-charge of this tiny interval will be:
\begin{equation}
\label{Picomp}
	\frac{Q_{M2}}{Q_{M5}}=\frac{4}{\sqrt{-\gamma}}(-\xi_0+\xi_1-\xi_2+\xi_3) \,,
\end{equation} 
where $Q_{M5}$ is the M5 flux on the tiny bubble.

Given that $\Pi \sim Q_{M2}/Q_{M5}$, we see that the $\xi_0$ that gives the position of the tiny bubble in  \eqref{Picomp} is exactly the same as the $\xi_0$ that gives the minimum of the potential felt by an M2-M5 probe \eqref{x0} with the same $\frac{Q_{M2}}{Q_{M5}}$. Hence, as the blue interval becomes smaller and smaller, its location is approximated better and better by a probe M2-M5 spike with the same M2-M5 charge ratio. Note that, as long as $\xi_0-\xi_{-1}$ remains small, this matching is independent of the value of $\xi_{-1}$. This is expected: the location of the M2-M5 probe depends only on the ratio  $\frac{Q_{M2}}{Q_{M5}}$, and is not changed if one modifies $Q_{M5}$ keeping this ratio fixed. Similarly, the location of the tiny blue bubble depends on $\xi_0$ (which controls $\frac{Q_{M2}}{Q_{M5}}$ for this bubble \eqref{Picomp}) but is independent of the value of $\xi_{-1}$ (which controls the M5 flux on the bubble) as long as $\xi_0$ remains fixed.\footnote{Note that in order to compare the location of the tiny bubble with the location of the probe, one needs to use the same  $\hat{b}^0_2$ gauge of \eqref{b20 gauge} for both of them, because the Page charge is gauge dependent.}

Exactly the same phenomenon was observed in the M2-M5 LLM solutions \cite{Lin:2004nb}, where the location of tiny bubbles and probe branes was shown to coincide \cite{Massai:2014wba}, confirming that these bubbling solutions come from the geometric transitions of M2 branes polarized into M5 branes \cite{Bena:2000zb,Bena:2004jw}.

This phenomenon is illustrated in Figure \ref{fig:probing}.

%%%%%%%%%%%%%%%%%%%%%%%%%%%%%%%%%%%%%
\section{The geometric transition of the mohawk}
\label{sec:Geometric}
%%%%%%%%%%%%%%%%%%%%%%%%%%%%%%%%%%%%%

From the scaling behavior  (\ref{scaling0}) it is natural to assume that different values of $\gamma$ should amount to different choices of how we zoom in on a solution sourced by a certain brane configuration.  However, there  is a major qualitative difference in going from $\gamma >0$ to $\gamma < 0$, and it is not just the fact that the scaling of  $(u,z)$  flips between  ``zooming in''  and  ``zooming out.'' (Remember we are taking $-1 < \gamma < 1$.)  For $\gamma >0$, the solution can only have singular sources corresponding to M5 and M5' branes. Whereas for $\gamma < 0$, the solution cannot  have singular  M5 and M5' sources, but is sourced by smooth fluxes threading  homological cycles.   Moreover, the transition at  $\gamma =0$ is not smooth as it involves a  decompactifieation of $S^3$ to $\IR^3$ \cite{Bachas:2013vza}.  

The purpose of this section is to explain how various bubbling solutions emerge from  geometric transitions generated by the back-reaction of M2-M5 spikes. We will also explain why this geometric transition is not visible in $\gamma>0$ solutions. To do this, it is useful to review the standard lore of geometric transitions.

%%%%%%%%%%%%%%%%%%%
\subsection{Geometric transitions of generic wrapped branes}
\label{ss:geomtrans}
%%%%%%%%%%%%%%%%%%%

One of the first examples of a geometric transition in string theory was analyzed in \cite{Gopakumar:1998ki}, but, since then, there have been a wealth of examples in many settings (see, for example, \cite{Lin:2004nb,Klebanov:2000hb,Vafa:2000wi,Bena:2005va,Bena:2007kg}).  In particular, the transitions we see here are essentially higher-codimension analogues of the transitions of M2 branes polarized into M5 branes  \cite{Bena:2000zb,Bena:2004jw} to LLM M2-M5 bubbling solutions \cite{Lin:2004nb}, or of the transition of zero-area black rings to microstate geometries \cite{Bena:2005va}.

In supergravity,  the back-reaction of branes  leads to a warp factor that diverges as one approaches the location of the branes. The metric along the brane world-volume is multiplied by an inverse power of this warp factor\footnote{This factor is $Z^{-1/2}$ for D-branes,  $Z^{-2/3}$ for M2 branes and  $Z^{-1/3}$ for M5 branes.}.  If the branes wrap a compact cycle, $\cW$, then this causes the size of  $\cW$ to shrink to zero at the brane locus.    At the same time, this  warp factor {\it expands} the directions transverse to the brane locus.  Indeed, around the branes there is always a cycle,  a Gaussian surface, $\cG$,  threaded by the magnetic fluxes sourced by the branes. The integrals of these fluxes over $\cG$ gives the   number of branes in the source.  Before the branes back-react this surface can be shrunk all the way down to the $\delta$-function source  of the branes.   When the branes back-react, the transverse expansion created by the warp factor blows up this infinitesimal Gaussian surface, creating a finite-sized, non-contractible cycle, $\cG$.

A naive expectation, based  on the behavior of ordinary matter, might be that  the collapse of a singular source should create an even stronger singularity.  However, a collapsing brane, like a collapsing string, condenses into its massless sector, whose excitations then spread out in the transverse directions. Thus, the original singular brane source is actually removed from the space-time to be replaced by quantized flux through the now non-contractible Gaussian cycle.  Throughout this process, the charge measured on the Gaussian surface remains constant with the $\delta$-function source replaced by smooth, quantized flux on $\cG$.

%%%%%%%%%%%%%%%%%%%%%%%%%%%%%%%%
\begin{figure}[h]
    \centering
    \includegraphics[width=.8 \textwidth]{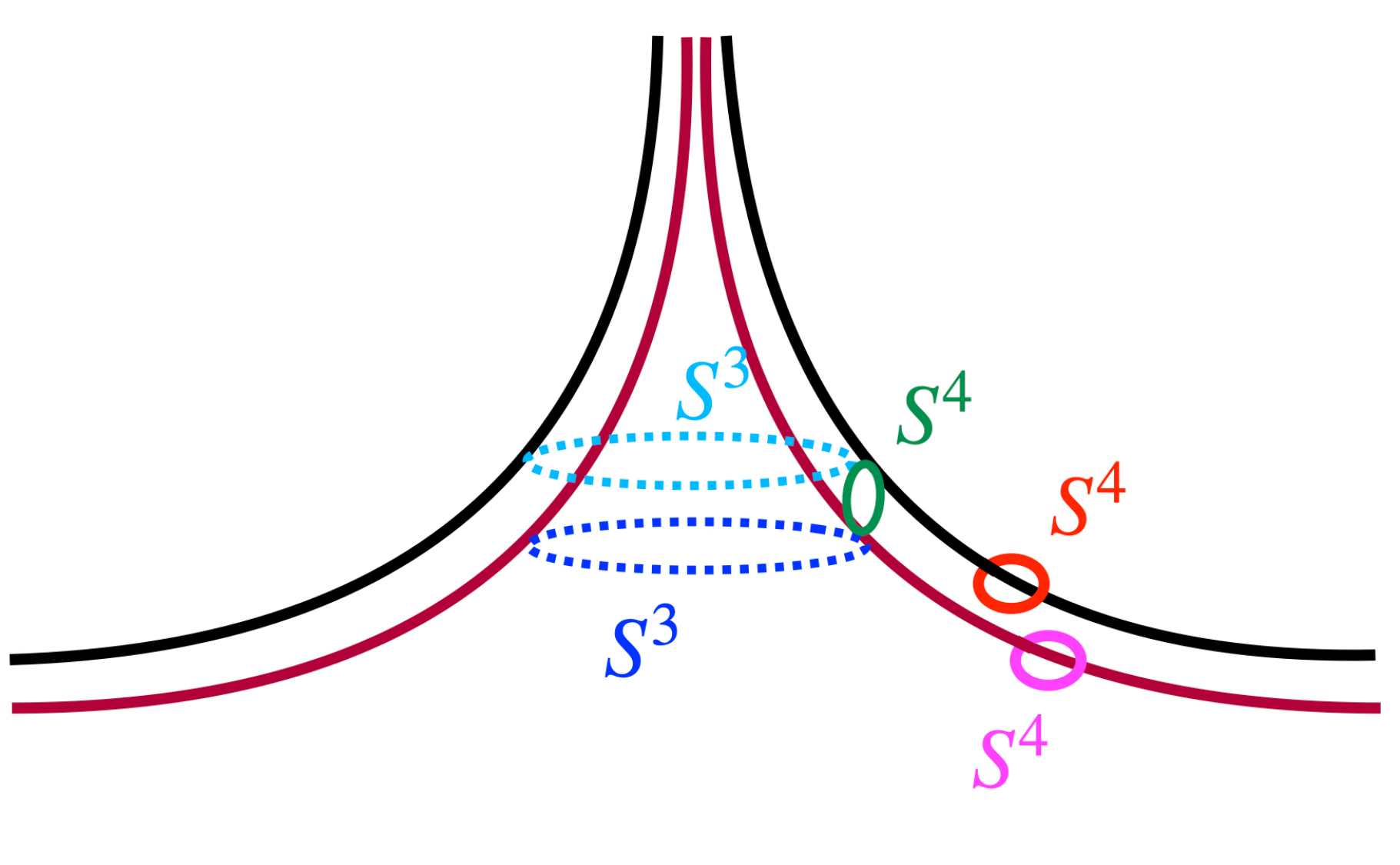}  
    \caption{The geometric transition of two M2-M5 spikes in even-flip solutions: The Gaussian four-cycles surrounding the M5 branes, in magenta and in red, corresponding to the A and C cycles in Figure \ref{fig:Topology1}, now become large. Also, because the $S^3$ inside the world-volume of the M5 branes  (dotted dark and light blue) collapses at each M5 location there will be a new topologically-non-trivial four-cycle, corresponding to the B cycle in Figure \ref{fig:Topology1}, shown in dark green.}  
    \label{fig:Topology3}
\end{figure}
%%%%%%%%%%%%%%%%%%%%%%%%%%%%%%%%

One can also  reverse this perspective by either taking a limit of the supergravity solution in which the quantized fluxes become small, and hence the Gaussian cycle becomes small, or by zooming out and considering the solution on scales much larger that the cycles.  Either way, in this limit the  cycle becomes indistinguishable from a brane.  This reversal of the geometric transition was evident in Section \ref{sec:probes}: the location of a probe M2-M5 spike coincides with the location of the corresponding Gaussian bubble, in the small-bubble limit.  

It is also possible for the collapsing, compact cycle, $\cW$, to create additional homological cycles that are distinct from, but necessarily  intersect, the homological cycles, $\cG$, created by the Gaussian surfaces.  At a generic point in a space-time, the cycle, $\cW$,  is  finite, but pinches off (usually smoothly, or with an orbifold singularity) at the original brane sources (once the geometric transition is incorporated). Usually, one such pinch-off   does not create topology:  the archetype is a sphere collapsing to zero size at the ``center of space,'' just as with radial coordinates at the origin of $\IR^{n+1}$, with $\cW = S^n$.  However, two such pinch-offs can create a non-trivial cycle, $\cC$, described by fibering $\cW$ over an interval (or higher-dimensional surface)  that runs between two pinch-offs.  The archetype  is an $\cW = S^n$, fibered over an interval between two pinch-off points, creating a new cycle $ \cC=S^{n+1}$ (or an orbifold thereof). 

Indeed, if there are multiple brane sources wrapping the same compact cycle, this cycle will shrink to zero size at every brane location, creating a collection of non-trivial, intersecting topological $\cG$ and $\cC$ cycles.  The $\cG$-cycles will be threaded by  fluxes that exactly match the charges of the brane sources  that were present before the back-reaction.   

If there are several species of branes localized on $\cW$, then the fluxes threading the intersecting cycles, $\cC$ and $\cG$, can both play an essential role in the geometric transition  through   Chern-Simons interactions.   In particular, if there is a singular object wrapping $\cW$, that has two distinct brane charges, both charges will be preserved by the geometric transition. Specifically, when there is a magnetic charge, $Q$, and an electric charge, $\tilde Q$, the magnetic charge $Q$ will be replaced by a flux on $\cG$ while  $\tilde Q$ will be generated by the Chern-Simons interaction between the magnetic flux on $\cG$ and new magnetic fluxes, $Q'$, on the cycles, $\cC$, created by the transition.

As we will discuss, this is precisely how the M2 brane charges arise in the system considered here:  The original, singular sources of M2 and M5 charges are replaced, after the geometric transition, by M5 fluxes on Gaussian cycles, $\cG$, and M5' fluxes on new cycles, $\cC$,  with  the M2 charges emerging from the Chern-Simons interaction of M5 and M5' fluxes.

This situation, can lead to a very interesting physical ambiguity.  One started with  $Q$ magnetic branes and $\tilde Q$ electric branes.  After the transition one is left with  magnetic fluxes $Q$ and $Q'$.  In some circumstances one can decrease the charge $Q$, while increasing  $ Q'$ so as to keep $\tilde Q$  fixed. One can then take a limit in which one obtains  a singular source with $\tilde Q$ electric branes and $ Q'$ magnetic branes that are quite distinct from the original branes.  More importantly, distinctly different brane configurations can, after their geometric transitions, arrive at exactly the same geometry.   This phenomenon was encountered in the  solutions \cite{Bena:2004jw, Lin:2004nb} describing M2-M5 polarization, in giant graviton solutions \cite{Bena:2004qv}, and in Polchinski-Strassler duals of ${\cal N}=1^*$ vacua \cite{Polchinski:2000uf}.  Here we will see, once again, how it connects M2-M5 solutions to M2-M5' solutions.

%%%%%%%%%%%%%%%%%%%%%%%%%%%%%%%%
\begin{figure}[h]
    \centering
    \includegraphics[width=.8 \textwidth]{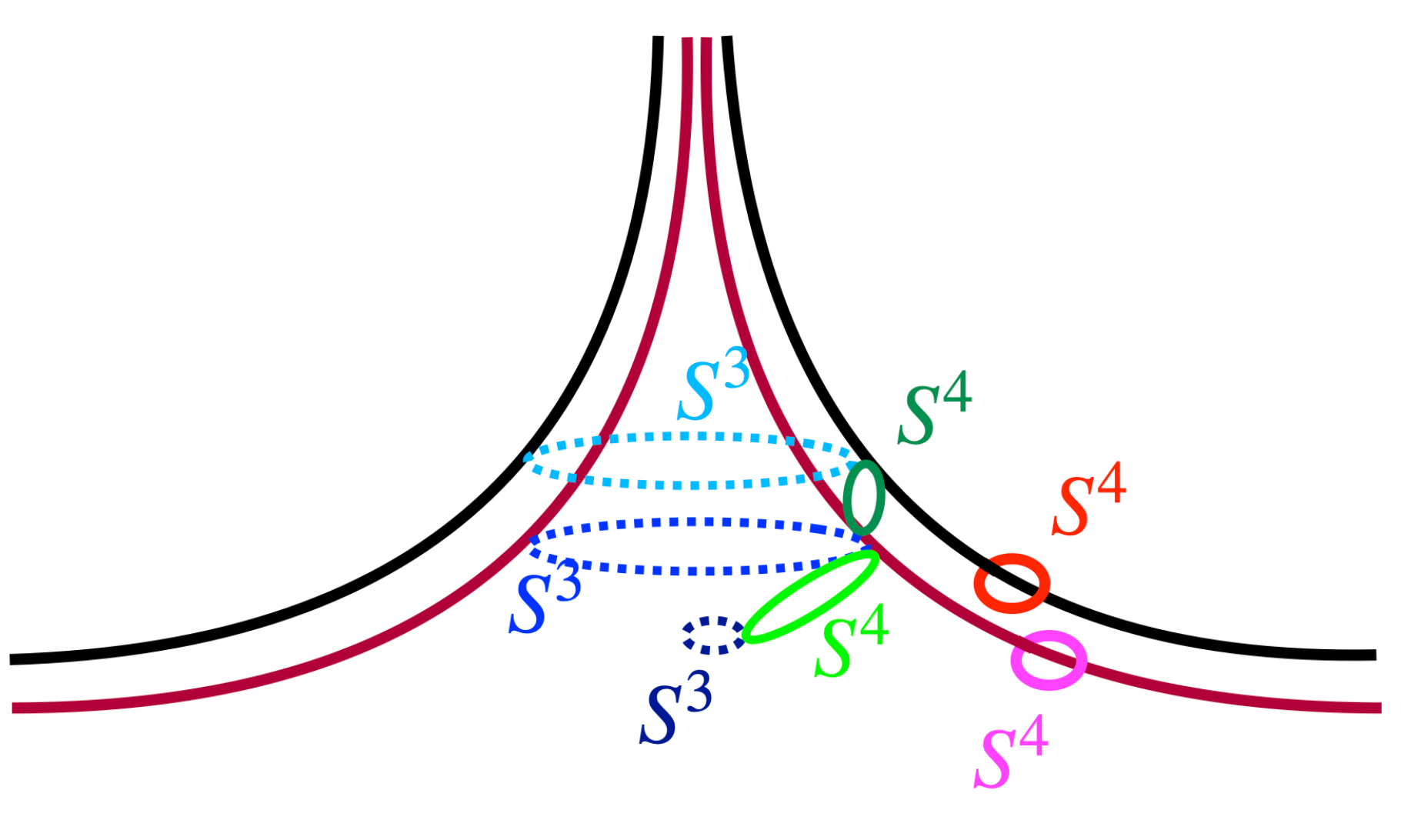}  
    \caption{The geometric transition of two M2-M5 spikes in odd-flip solutions: The Gaussian four-cycles surrounding the M5 branes, are in magenta  and red. Also, because the $S^3$ inside the world-volume of the M5 branes collapses at each M5 location  (dotted blue and dotted light blue)  and at the center of space (dotted dark blue), there will be two new topologically-non-trivial four-cycles, denoted by B and D in Fig. \ref{fig:Topology2} and shown here in dark green and green.}
    \label{fig:Topology4}
\end{figure}
%%%%%%%%%%%%%%%%%%%%%%%%%%%%%%%%

%%%%%%%%%%%%%%%%%%%
\subsection{Geometric transitions of the M2-M5 system}
\label{ss:geomtransM2M5}
%%%%%%%%%%%%%%%%%%%

Near the surface of the spike,  our supergravity solutions are  themelia \cite{Bena:2022fzf}: 
if one zooms in near any point of the spike we obtain a 16-supercharge bound state of M5 and M2 branes.  The M2 branes are smeared on the $S^3$ at a constant value of $u$, while the M5 branes are wrapping that $S^3$. The M5 branes alone would shrink this $S^3$, but the M2 branes alone would blow it up, since it is transverse to their world-volume. However, when the two branes form a bound state, it is the M5 branes that always win, regardless of the amount of their M2 charge\footnote{One can see this in equation (D.32) of \cite{Bena:2023rzm}.}. Hence, the back-reaction of the M2-M5 spikes shrinks  the $S^3$ inside the M5 world-volume to zero size, regardless of the value of $u$. The M5 branes wrapping this $S^3$ undergo a geometric transition and disappear from spacetime. At the same time, their Gaussian surfaces become large and non-contractible, and the flux wrapping these surfaces stays the same. If there are multiple M5 branes, this $S^3$ shrinks at every M5 location, and this creates new topologically non-trivial four-cycles. 

Since we are imposing spherical ($SO(4) \times SO(4)$) symmetry, there is a well-defined ``center of space" at $(u=0,v=0)$.  If the scaling of the solution is done in such a way as to keep the  center of space inside the solution, the $S^3$ inside the M5 branes will also shrink at this point. This will create another topologically non-trivial four-cycle.

Thus, the back-reaction of multiple M5 spikes will give rise to a bubbling solution, containing two types of four-cycles: 

%%%%%%%%%%
\begin{itemize}
%%%%%%%%%%
\item[(a)] The Gaussian four-cycles, $\cG$,  that surround each M5-M2 spike, and on which one integrates $F_4$ to calculate the M5 charge of the spike. 

\item[(b)] The four-cycles, $\cC$, that come from the contraction of the $S^3$ along the M5 world-volume at each spike location and, if present,  at the center of space.
%%%%%%%%%%
\end{itemize}
%%%%%%%%%%

\medskip

To illustrate these cycles and how the geometric transition happens, we can focus on a  mohawk made of two M2-M5 spikes. The scaling limit of this mohawk, can give rise to {\em two} types of bubbling $AdS_3$ solutions:

%%%%%%%%%%
\begin{itemize}
%%%%%%%%%%
\item[1.]  If the scaling limit does not contain the point $(u=0,v=0)$, that is, it does not contain the $z$-axis at the center of space, the solution will have M5 asymptotics (containing an $AdS_7$, or deformed versions thereof). This is precisely the  even-flip solution.  The cycles that result from the geometric transition are depicted in Figs.~\ref{fig:Topology1} and \ref{fig:Topology3}. The Gaussian cycles, $\cG$, are the (magenta) A and C cycles, which are now large and topologically non-trivial. Furthermore, the $S^3$ that shrinks at the location of each M5 brane creates a new (green) non-trivial four-cycle, $\cC$, denoted by B. Thus,  the three-bubble even-flip solution, depicted in Fig.~\ref{fig:Topology1}, corresponds to a scaling limit of two back-reacted M2-M5 spikes that preserves the M5 asymptotics. In the limit when the flux on either the A or the C cycles becomes small, the blue regions, $(\xi_1, \xi_2)$ or $(\xi_3, \xi_4)$, shrink, and the small bubbles looks again like singular sources. 

\item[2.]  An odd-flip solution necessarily has an additional flip at infinity, and this creates an M2 asymptotic region (containing an $AdS_4$ up to a $\ZZ_2$ orbifold) \cite{Bachas:2013vza}.  The ``partner'' of this flip at infinity is the point $\xi_5$ that is connected to the flip at infinity by the right-most  blue interval in Fig.~\ref{fig:Topology2}.  On this blue interval the $S^3$ in the M5 brane pinches off, and so we can take this interval as defining the $z$-axis at the center of space $(u=0, v=0)$.     (Note that  $\rho =0$ along this  blue interval  and so both $u$ and $v$ vanish, while the coordinate $\xi$ sweeps out the $z$-axis (or $u^2 z$-axis) as depicted in Fig.~\ref{fig:steepness}.)  The cycles that result from the geometric transition are depicted in Figs.~\ref{fig:Topology2} and  \ref{fig:Topology4}. The Gaussian cycles, $\cG$,  are, once again, the (magenta) A and C cycles, and the cycle B is as described above. The new cycle, D,  is shown in green and is created by  the $S^3$ that pinches off at the ``center of space'' and at the  right-most  M5.

%%%%%%%%%%
\end{itemize}
%%%%%%%%%%

%%%%%%%%%%%%%%%%%%%%%%%%%%%%%%%%
\begin{figure}[htbp]
    \centering
    \includegraphics[width=.8 \textwidth]{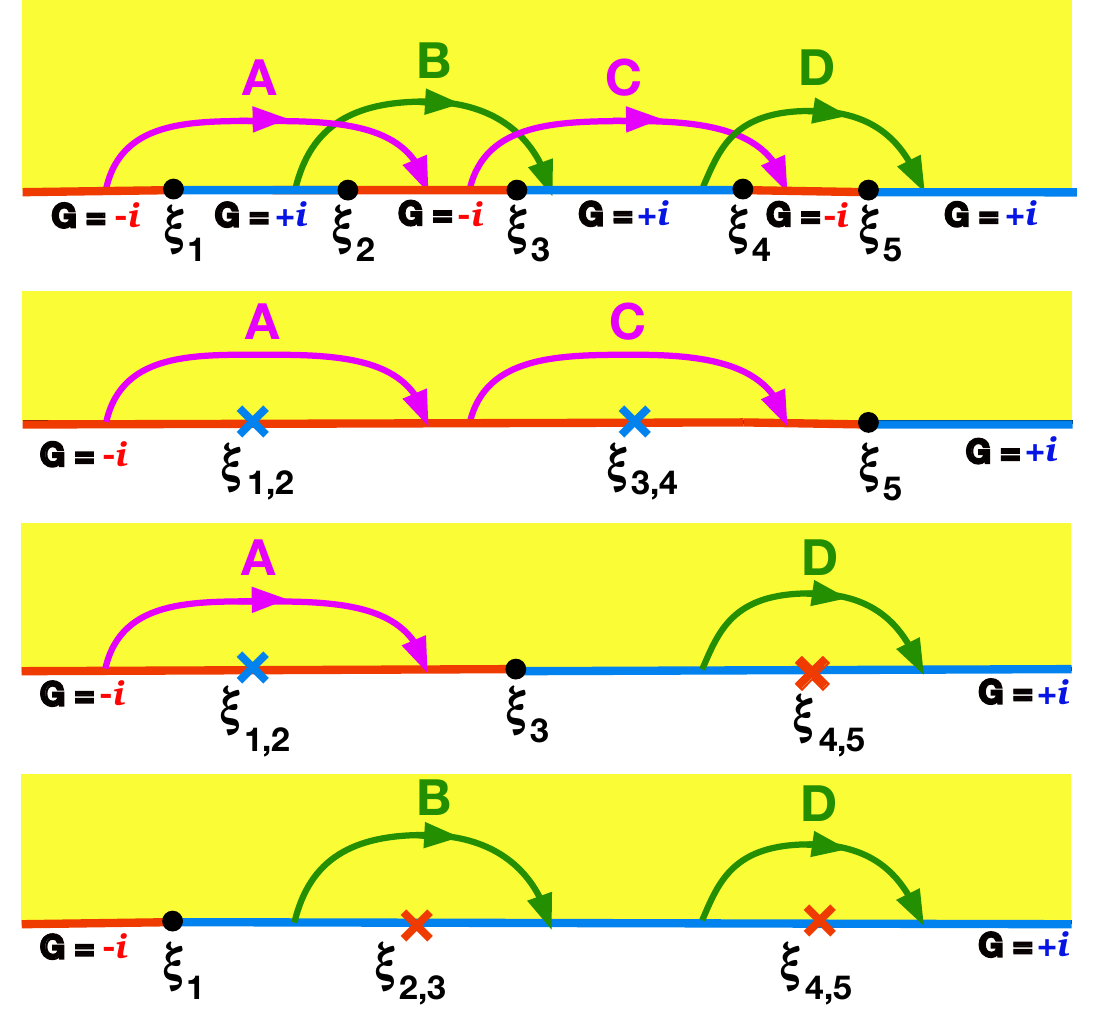}  
    \caption{The three possible M2-M5-M5' spike configurations that give rise to the same four-bubble solution (top panel). When, the two blue intervals become small, this solution matches that corresponding to two M2-M5 spikes (second panel). When one blue and one red non-adjacent intervals become small, the solution can be seen as sourced by an M2-M5 and by an M2-M5' spike (third panel). When the two red intervals become small the solution matches that corresponding to two M2-M5' spikes (bottom panel).}
    \label{fig:TopologyDegen}
\end{figure}
%%%%%%%%%%%%%%%%%%%%%%%%%%%%%%%%

One important point here is that the 2n-bubble, odd-flip solution and a (2n-1)-bubble even flip solutions correspond to exactly the  same M2-M5 spikes. The only difference is the section of that solution that is captured by taking the AdS limit.  The odd-flip solution  has a region with M2 asymptotics and has a ``center of space'', while the even-flip solution only captures  the region near the M5 branes, and the asymptotic region is dominated by those (deformed) M5 branes. A similar feature was found in $\gamma=1$ solutions \cite{Bena:2024dre} and was hinted at in \cite{Bachas:2013vza}:  the no-flip solutions can be obtained from solutions with a flip by taking a limit that merges the two flips at infinity.

The other important point is that the (magenta) A and C cycles that surround {\it compact} blue intervals in Figures \ref{fig:Topology1} and  \ref{fig:Topology2}, and that emerge from the geometric transition of the M2-M5 spikes, are now on the same footing as the cycles   ($B$ and $D$)  defined by the red intervals that are threaded by M5' fluxes.    
The geometric transition thus puts the M5 and M5' sources on a completely democratic footing and the M2 charges now emerge from the M5 and M5' fluxes via the Chern-Simons interaction.  This is directly  reflected in  equations \eqref{totalM2} and \eqref{totalM2p} for the M2 charge.

Indeed, if we now consider a mohawk made out  M2-M5' spikes, the role of the four-cycles is reversed. The Gaussian cycles, $\cG'$, surrounding the M2-M5' spikes are those that contain the $S^3$, and so are the   $\cC$-cycles of the solution resulting from the M2-M5 transition. Similarly, the cycles, $\cC'$,  that come from the collapse of the three-spheres inside the M5' world-volume were the Gaussian cycles, $\cG$, of the M2-M5 transition. A more general mohawk, made of both M2-M5 and M2-M5' spikes will in general give rise to a multi-bubble solution.

We therefore see that the bubbling solution corresponding to $N_2$ M2 branes terminating on $k$ M5 branes and the bubbling solution corresponding to $N_2$ M2 branes terminating on $N_2/k$ M5' branes are the same: both will have  cycles carrying  $F_4$ fluxes associated with M5 branes, as well as cycles carrying  $F_4$ fluxes associated with M5' brane.  The naive, non-back-reacted intuition suggests that  M2 branes can have two different boundary conditions where they end on either M5 branes or on M5' branes but this is a perturbative artifact:  brane back-reaction gives rise to s single family of bubbling solutions. Once the geometric transition happens, the two boundary conditions are  one and the same. For $\gamma<0 $ solutions this equivalence was observed in Section 7 of \cite{Bachas:2013vza}, and, as we noted earlier,  this equivalence has appeared in many other contexts.

This equivalence is even more striking for odd-flip solutions with higher numbers of bubbles, as illustrated in Figure \ref{fig:TopologyDegen}. A bubbling solution with two bubbles with M5' flux (B and D) and two bubbles with M5 flux (A and C), can come from either: \\
a) the geometric transition of two M2-M5' spikes, whose location becomes precise when the red intervals degenerate \\
b) the geometric transition of two M2-M5 spikes, whose location becomes precise when the blue intervals degenerate \\
c) the geometric transition of one M2-M5 spike and one M2-M5' spike, whose location becomes precise when the blue $(\xi_1,\xi_2)$ and the red $(\xi_4,\xi_5)$ intervals degenerate.

It is important to emphasize that the geometric transition to a bubbling solution will be a feature of the full M2-M5 mohawk solution, and not only of its scaling limits \eqref{scaling0} that give AdS$_3 \times$ S$^3 \times$ S$^3 \times \Sigma$ solutions. Since these scaling limits are controlled by $\gamma$, one can also ask why is this geometric transition not visible in $\gamma>0$ solutions, which cannot be sourced by smooth topology and fluxes, but only by singular sources that carry the M5 charges. Indeed, the $\gamma>0$ solutions describing M2-M5 mohawks \cite{Bena:2024dre} look naively as if all the Gaussian cycles had collapsed into singular sources.  This happens because of the scaling limit \eqref{scaling0} taken to obtain  $\gamma>0$ solutions. The $\gamma>0$ scaling limits can be rewritten as $u \rightarrow 0, z \rightarrow \infty$ keeping $zu^2$ fixed, and this zooming-out effectively collapses the fluxed cycles to singular sources. 

A similar example where scaling limits can collapse a bubble and undo a geometric transition, resulting in a singular geometry, is the Klebanov-Strassler solution \cite{Klebanov:2000hb}. This smooth solution comes from the geometric transition of a collection of D3 and D5 branes on the conifold. There is also a singular solution corresponding to these branes, which does not include the geometric transition, found by Klebanov and Tseytlin \cite{Klebanov:2000nc}, but this singular solution does not capture correctly the infrared physics of this system. Hence, one can think about the Klebanov-Strassler solution as resolving the singularity of the Klebanov-Tseytlin solution. However, if one scales the Klebanov-Strassler solution is a certain way (for example by taking the large-radius limit) one can see that this reduces this solution into the singular Klebanov-Tseytlin limit, thus effectively undoing the geometric transition. But this does not imply that the singular solution is physical. The correct solution is always the one with a geometric transition. Similarly here, the singular $\gamma >0$ solutions should not be thought of as physical solutions describing the full backreacted M2-M5-M5' brane system, bur rather as scaling limits of these solutions that collapse their bubbling structure.

Hence, $\gamma >0$ solutions represent a scaling limit of the full bubbling solution where certain bubbles are collapsed, and are replaced by singularities carrying M5 charges.  It is interesting to observe that even if the bubbles are invisible, $\gamma >0$ solutions still retain the memory of the equivalence described above, between the $k$-M5-brane single-spike solution and the $N_2/k$-M5'-brane single-spike solution. There is a single solution, but for different values of $k$ the scaling limit collapses either the M5 or the M5' Gaussian surfaces. And this is confirmed by the supergravity construction: $\gamma=1$ solutions describing a single M2-M5 spike only exist when the number of M5 branes is smaller than $\sqrt{N_2}$ \cite{Harvey:2025ttz}\footnote{We thank Kristan Jensen for interesting discussions on this point.}.

%%%%%%%%%%%%%%%%%%%%%%%%%%%%%%%%%%%%%
\section{Final comments}
\label{sec:Conclusions}
%%%%%%%%%%%%%%%%%%%%%%%%%%%%%%%%%%%%%

We have shown that all AdS$_3 \times$ S$^3 \times$ S$^3$ solutions warped over a simple Riemann surface come from the scaling near-horizon limit of a system of M2, M5 and M5' branes. These branes form spikes, and the back-reaction of these spikes gives rise to a bubbling geometry. As we discussed in the previous section, the scaling limits that give rise to negative-$\gamma$ solutions preserve this bubbling structure, but the positive-$\gamma$ limits collapse the bubbles into singular brane sources. 

Besides the bubbling structure, the other major $\gamma$-dependent feature is,  of course, the superalgebra. Our analysis in Section \ref{ss:metric} is based on the brane system and is universal for all  values of  $\gamma$.  However, it only determines the Poincar\'e supersymetries and it is insensitive to the superconformal structure that emerges in the scaling limit.  This suggests a simple conclusion:  the Poincar\'e supersymetries are indeed universal and  it is only the superconformal completion to $D(2,1; \gamma) \oplus D(2,1; \gamma)$  that depends   on how one takes the scaling limit.

This raises a further intriguing question. Here we are considering \nBPS{4} solutions, but ultimately we would like to add momentum as well as ``gluing fluxes'' to these solutions, \cite{Bena:2022wpl,Bena:2022fzf,Bena:2024qed}  to construct \nBPS{8} horizonless solutions with black-hole asymptotics.  Such solutions will involve a smaller superalgebra that could perhaps be common to the $D(2,1; \gamma) \oplus D(2,1; \gamma)$ for all values of $\gamma$, and it is possible that adding momentum will be insensitive to $\gamma$.  On the other hand, based on our experience with superstrata \cite{Bena:2015bea}, it is possible that the construction of solutions with generic momentum waves may only be possible on non-singular geometries and thus require $\gamma < 0$.  This will be resolved in future work.

The AdS$_3 \times$ S$^3 \times$ S$^3$ solutions depend on the two-Riemann surface coordinates, and describe a one-parameter family of scaling limits of the M2-M5-M5' system, parametrized by $\gamma$. The full M2-M5-M5' solution is much more complicated, and is governed by a master function of three variables obeying a non-linear Monge-Amp\`ere equation \cite{Lunin:2007mj, Bena:2023rzm}. This equation is in general impossible to solve analytically. It would be interesting to use the existence of the one-parameter family of scaling limits to try to find new ways to solve the Monge-Amp\`ere equation and perhaps even find the full M2-M5-M5' solution.

There is another intriguing question about the brane realization of more general AdS$_3 \times$ S$^3 \times$ S$^3$ solutions that are warped over non-trivial Riemann surfaces, like the Janus solution of  \cite{DHoker:2009lky} (reviewed in section 8.2 of \cite{Bachas:2013vza}).   Here we have worked entirely with the non-compact Poincar\'e half-plane.  A non-trivial Riemann-surface (with genus $\ge 2$) can be described by a polygonal patch in this half-plane with a gluing prescription for the edges of the polygon. Such a gluing prescription implies gluing of the three (projectively scaled)  radial coordinates, $(u,v,z)$, and hence of the M2-M5-M5' system in the near-brane region.   The question becomes whether this gluing prescription survives all the way out to infinity on the branes, or whether it can be localized in the near-brane region while the asymptotic region is simply that of flat branes.  It would be very interesting to resolve this issue and see whether there is a single intersecting brane system that asymptotes to flat branes at infinity but whose near-brane limit gives the full Janus solution, or possibly other classes of brane plumbing based on more complicated Riemann surfaces. If, however, the gluing prescription of the near-brane region extends to infinity it would imply that more complicated plumbing of AdS$_3$ solutions do not emerge from a limit of flat-brane plumbing.

Going beyond this paper, there are a huge range of supergravity solutions that have  the form   AdS$\times$ S$ \times$ S $ \times \Sigma$, where $\Sigma$ is either a Riemann surface or a sphere.  In particular, the D3-D5-NS5 system in IIB supergravity appears to be akin to the  M2-M5-M5' system studied here.   Indeed, starting from the brane system in Section \ref{ss:metric} and, instead of writing $\vec u, \vec v \in \IR^4$ in terms of a radial variable and an $S^3$, one can separate off $u_4$ and $v_4$, and write each remaining $\IR^3$ in terms of   a radial variable and an $S^2$.  Reducing to IIA on  $u_4$ and then T-dualizing on  $v_4$ leads to the D3-D5-NS5 system.  The underlying geometry is $AdS_4 \times S^2 \times S^2 \times \Sigma$ and it  captures  the back-reaction of D3 branes sandwiched between D5 and NS5 branes. It was shown  in \cite{Assel:2011xz} that these solutions are dual to families of (2+1)-dimensional conformal field theories.

There are many parallels between these two brane systems, and this will be explored in detail in an upcoming paper \cite{D3D5NS5}. Both give rise to $AdS \times S \times S$ solutions warped over a Riemann surface, and both capture solutions that come the back-reaction of spikes in which semi-infinite M2/D3 branes end on M5/D5 branes or on M5'/NS5 branes, or both.  On a technical level, both solutions involve two functions on the Riemann surface that must obey linear equations, whose sources determine the branes and fluxes, and whose boundary conditions are fixed by regularity and the asymptotics at infinity.

There are, however, some surprising differences.  At the technical level, the two functions governing the D3-D5-NS5 system are much simpler and more intuitive than those defined by (\ref{harmonic}) and  (\ref{Geqn}), and, when combined with regularity conditions, lead to rather different families of solutions.  In particular, in the D3-D5-NS5 system there are solutions in which the {\it warped}  Riemann surface is a compact space in which the are no semi-infinite D3 branes. These compact solutions are dual to 2+1 dimensional conformal field theories that one finds in the infrared of a system of D3 branes sandwiched between D5 and NS5 branes.  There seem to be no such solutions for the M2-M5-M5' system.  The possibilities were carefully analyzed in \cite{Bachas:2013vza}, and there it was shown that for the warped Riemann surface to be compact the function $h$ must be smooth and bounded and, since it is harmonic, it must therefore be constant. This means the $\Sigma$ must be a flat torus, and the solution collapses to that of  \cite{deBoer:1999gea}.   It would be interesting to see if  some possibility has been missed in \cite{Bachas:2013vza}, or understand at a more fundamental level, why systems of intersecting branes can lead to conformal field theories in some dimensions and not in others.  

\medskip 

\noindent{\bf Acknowledgments:} We would like to thank Costas Bachas, Antoine Bourget, Soumangsu Chakraborty, Eric D'Hoker and Kristan Jensen for interesting and insightful discussions.
The work of IB and NPW was supported in part by the ERC Grant 787320 - QBH Structure. The work of DT is supported by the Israel Science Foundation
(grant No. 1417/21), by the German Research Foundation through a German-Israeli Project Cooperation (DIP) grant “Holography and the Swampland”, by Carole and Marcus Weinstein through the BGU Presidential Faculty Recruitment Fund, by the ISF Center of Excellence
for theoretical high energy physics, by the VATAT Research Hub in the Field of quantum computing and by the ERC starting Grant dSHologQI (project number 101117338). The work of NPW was also supported in part by the DOE grant DE-SC0011687.

%%%%%%%%%%%%%%%%%%%%%%%%%%%%%%%%%%%%%%%%%%%%%%%%%%%%%
\newpage

\begin{adjustwidth}{-1mm}{-1mm} % to adjust the L and R margins 

\bibliographystyle{utphys}      

\bibliography{references}       % calls file "microstates.bib"

\providecommand{\href}[2]{#2}\begingroup\raggedright\begin{thebibliography}{10}

\bibitem{Bachas:2013vza}
C.~Bachas, E.~D'Hoker, J.~Estes, and D.~Krym, ``{M-theory Solutions Invariant
  under $D(2,1;\gamma) \oplus D(2,1;\gamma)$},''
  \href{http://dx.doi.org/10.1002/prop.201300039}{{\em Fortsch. Phys.}
  {\bfseries 62} (2014) 207--254},
  \href{http://arxiv.org/abs/1312.5477}{{\ttfamily arXiv:1312.5477 [hep-th]}}.

\bibitem{Okazaki:2025rjl}
T.~Okazaki and D.~J. Smith, ``{$SL(2,\mathbb{Z})$ dualities of boundary
  conditions in Abelian M2-brane SCFTs},''
  \href{http://arxiv.org/abs/2507.22644}{{\ttfamily arXiv:2507.22644
  [hep-th]}}.

\bibitem{Harvey:2025ttz}
W.~Harvey, K.~Jensen, and T.~Uzu, ``{Comparing top-down and bottom-up
  holographic defects and boundaries},''
  \href{http://dx.doi.org/10.1007/JHEP08(2025)167}{{\em JHEP} {\bfseries 08}
  (2025) 167}, \href{http://arxiv.org/abs/2504.13244}{{\ttfamily
  arXiv:2504.13244 [hep-th]}}.

\bibitem{Harvey:2023pdv}
W.~Harvey, K.~Jensen, and T.~Uzu, ``{Smeared end-of-the-world branes},''
  \href{http://dx.doi.org/10.1103/PhysRevD.111.066017}{{\em Phys. Rev. D}
  {\bfseries 111} no.~6, (2025) 066017},
  \href{http://arxiv.org/abs/2311.13643}{{\ttfamily arXiv:2311.13643
  [hep-th]}}.

\bibitem{Drukker:2022pxk}
N.~Drukker, Z.~Kong, and G.~Sakkas, ``{Broken Global Symmetries and Defect
  Conformal Manifolds},''
  \href{http://dx.doi.org/10.1103/PhysRevLett.129.201603}{{\em Phys. Rev.
  Lett.} {\bfseries 129} no.~20, (2022) 201603},
  \href{http://arxiv.org/abs/2203.17157}{{\ttfamily arXiv:2203.17157
  [hep-th]}}.

\bibitem{Witten:2024yod}
E.~Witten, ``{Instantons and the large ${\mathcal{N}} = 4$ algebra},''
  \href{http://dx.doi.org/10.1088/1751-8121/ada64d}{{\em J. Phys. A} {\bfseries
  58} no.~3, (2025) 035403}, \href{http://arxiv.org/abs/2407.20964}{{\ttfamily
  arXiv:2407.20964 [hep-th]}}.

\bibitem{Callan:1997kz}
C.~G. Callan and J.~M. Maldacena, ``{Brane death and dynamics from the
  Born-Infeld action},''
  \href{http://dx.doi.org/10.1016/S0550-3213(97)00700-1}{{\em Nucl. Phys. B}
  {\bfseries 513} (1998) 198--212},
  \href{http://arxiv.org/abs/hep-th/9708147}{{\ttfamily arXiv:hep-th/9708147}}.

\bibitem{Constable:1999ac}
N.~R. Constable, R.~C. Myers, and O.~Tafjord, ``{The Noncommutative bion
  core},'' \href{http://dx.doi.org/10.1103/PhysRevD.61.106009}{{\em Phys. Rev.
  D} {\bfseries 61} (2000) 106009},
  \href{http://arxiv.org/abs/hep-th/9911136}{{\ttfamily arXiv:hep-th/9911136}}.

\bibitem{Bena:2023rzm}
I.~Bena, A.~Houppe, D.~Toulikas, and N.~P. Warner, ``{Maze topiary in
  supergravity},'' \href{http://dx.doi.org/10.1007/JHEP03(2025)120}{{\em JHEP}
  {\bfseries 03} (2025) 120}, \href{http://arxiv.org/abs/2312.02286}{{\ttfamily
  arXiv:2312.02286 [hep-th]}}.

\bibitem{Lunin:2007mj}
O.~Lunin, ``{Strings ending on branes from supergravity},''
  \href{http://dx.doi.org/10.1088/1126-6708/2007/09/093}{{\em JHEP} {\bfseries
  09} (2007) 093}, \href{http://arxiv.org/abs/0706.3396}{{\ttfamily
  arXiv:0706.3396 [hep-th]}}.

\bibitem{Bena:2024dre}
I.~Bena, S.~Chakraborty, D.~Toulikas, and N.~P. Warner, ``{The M2-M5 Mohawk},''
  \href{http://dx.doi.org/10.1007/JHEP07(2025)127}{{\em JHEP} {\bfseries 07}
  (2025) 127}, \href{http://arxiv.org/abs/2407.01665}{{\ttfamily
  arXiv:2407.01665 [hep-th]}}.

\bibitem{Bena:2008wt}
I.~Bena, N.~Bobev, and N.~P. Warner, ``{Spectral Flow, and the Spectrum of
  Multi-Center Solutions},''
  \href{http://dx.doi.org/10.1103/PhysRevD.77.125025}{{\em Phys. Rev.}
  {\bfseries D77} (2008) 125025},
\href{http://arxiv.org/abs/0803.1203}{{\ttfamily arXiv:0803.1203 [hep-th]}}.
%%CITATION = 0803.1203;%%.

\bibitem{Giusto:2012yz}
S.~Giusto, O.~Lunin, S.~D. Mathur, and D.~Turton, ``{D1-D5-P microstates at the
  cap},'' \href{http://dx.doi.org/10.1007/JHEP02(2013)050}{{\em JHEP}
  {\bfseries 1302} (2013) 050},
\href{http://arxiv.org/abs/1211.0306}{{\ttfamily arXiv:1211.0306 [hep-th]}}.
%%CITATION = ARXIV:1211.0306;%%.

\bibitem{Lunin:2007ab}
O.~Lunin, ``{1/2-BPS states in M theory and defects in the dual CFTs},''
  \href{http://dx.doi.org/10.1088/1126-6708/2007/10/014}{{\em JHEP} {\bfseries
  10} (2007) 014}, \href{http://arxiv.org/abs/0704.3442}{{\ttfamily
  arXiv:0704.3442 [hep-th]}}.

\bibitem{DHoker:2008lup}
E.~D'Hoker, J.~Estes, M.~Gutperle, and D.~Krym, ``{Exact Half-BPS Flux
  Solutions in M-theory. I: Local Solutions},''
  \href{http://dx.doi.org/10.1088/1126-6708/2008/08/028}{{\em JHEP} {\bfseries
  08} (2008) 028}, \href{http://arxiv.org/abs/0806.0605}{{\ttfamily
  arXiv:0806.0605 [hep-th]}}.

\bibitem{DHoker:2008rje}
E.~D'Hoker, J.~Estes, M.~Gutperle, and D.~Krym, ``{Exact Half-BPS Flux
  Solutions in M-theory II: Global solutions asymptotic to AdS(7) x S**4},''
  \href{http://dx.doi.org/10.1088/1126-6708/2008/12/044}{{\em JHEP} {\bfseries
  12} (2008) 044}, \href{http://arxiv.org/abs/0810.4647}{{\ttfamily
  arXiv:0810.4647 [hep-th]}}.

\bibitem{DHoker:2008wvd}
E.~D'Hoker, J.~Estes, M.~Gutperle, D.~Krym, and P.~Sorba, ``{Half-BPS
  supergravity solutions and superalgebras},''
  \href{http://dx.doi.org/10.1088/1126-6708/2008/12/047}{{\em JHEP} {\bfseries
  12} (2008) 047}, \href{http://arxiv.org/abs/0810.1484}{{\ttfamily
  arXiv:0810.1484 [hep-th]}}.

\bibitem{DHoker:2009lky}
E.~D'Hoker, J.~Estes, M.~Gutperle, and D.~Krym, ``{Janus solutions in
  M-theory},'' \href{http://dx.doi.org/10.1088/1126-6708/2009/06/018}{{\em
  JHEP} {\bfseries 06} (2009) 018},
  \href{http://arxiv.org/abs/0904.3313}{{\ttfamily arXiv:0904.3313 [hep-th]}}.

\bibitem{DHoker:2009wlx}
E.~D'Hoker, J.~Estes, M.~Gutperle, and D.~Krym, ``{Exact Half-BPS Flux
  Solutions in M-theory III: Existence and rigidity of global solutions
  asymptotic to AdS(4) x S**7},''
  \href{http://dx.doi.org/10.1088/1126-6708/2009/09/067}{{\em JHEP} {\bfseries
  09} (2009) 067}, \href{http://arxiv.org/abs/0906.0596}{{\ttfamily
  arXiv:0906.0596 [hep-th]}}.

\bibitem{Bobev:2013yra}
N.~Bobev, K.~Pilch, and N.~P. Warner, ``{Supersymmetric Janus Solutions in Four
  Dimensions},'' \href{http://dx.doi.org/10.1007/JHEP06(2014)058}{{\em JHEP}
  {\bfseries 06} (2014) 058}, \href{http://arxiv.org/abs/1311.4883}{{\ttfamily
  arXiv:1311.4883 [hep-th]}}.

\bibitem{Lunin:2008tf}
O.~Lunin, ``{Brane webs and 1/4-BPS geometries},''
  \href{http://dx.doi.org/10.1088/1126-6708/2008/09/028}{{\em JHEP} {\bfseries
  0809} (2008) 028},
\href{http://arxiv.org/abs/0802.0735}{{\ttfamily arXiv:0802.0735 [hep-th]}}.
%%CITATION = ARXIV:0802.0735;%%.

\bibitem{deBoer:1999gea}
J.~de~Boer, A.~Pasquinucci, and K.~Skenderis, ``{AdS / CFT dualities involving
  large 2-D N=4 superconformal symmetry},''
  \href{http://dx.doi.org/10.4310/ATMP.1999.v3.n3.a5}{{\em Adv. Theor. Math.
  Phys.} {\bfseries 3} (1999) 577--614},
\href{http://arxiv.org/abs/hep-th/9904073}{{\ttfamily arXiv:hep-th/9904073
  [hep-th]}}.
%%CITATION = HEP-TH/9904073;%%.

\bibitem{Gibbons:2013tqa}
G.~Gibbons and N.~Warner, ``{Global structure of five-dimensional fuzzballs},''
  \href{http://dx.doi.org/10.1088/0264-9381/31/2/025016}{{\em
  Class.Quant.Grav.} {\bfseries 31} (2014) 025016},
\href{http://arxiv.org/abs/1305.0957}{{\ttfamily arXiv:1305.0957 [hep-th]}}.
%%CITATION = ARXIV:1305.0957;%%.

\bibitem{Bandos:1997ui}
I.~A. Bandos, K.~Lechner, A.~Nurmagambetov, P.~Pasti, D.~P. Sorokin, and
  M.~Tonin, ``{Covariant action for the superfive-brane of M theory},''
  \href{http://dx.doi.org/10.1103/PhysRevLett.78.4332}{{\em Phys. Rev. Lett.}
  {\bfseries 78} (1997) 4332--4334},
  \href{http://arxiv.org/abs/hep-th/9701149}{{\ttfamily arXiv:hep-th/9701149}}.

\bibitem{Lin:2004nb}
H.~Lin, O.~Lunin, and J.~M. Maldacena, ``{Bubbling AdS space and 1/2 BPS
  geometries},'' {\em JHEP} {\bfseries 10} (2004) 025,
\href{http://arxiv.org/abs/hep-th/0409174}{{\ttfamily arXiv:hep-th/0409174}}.
%%CITATION = HEP-TH/0409174;%%.

\bibitem{Massai:2014wba}
S.~Massai, G.~Pasini, and A.~Puhm, ``{Metastability in Bubbling AdS Space},''
\href{http://arxiv.org/abs/1407.6007}{{\ttfamily arXiv:1407.6007 [hep-th]}}.
%%CITATION = ARXIV:1407.6007;%%.

\bibitem{Bena:2000zb}
I.~Bena, ``{The M theory dual of a three-dimensional theory with reduced
  supersymmetry},'' \href{http://dx.doi.org/10.1103/PhysRevD.62.126006}{{\em
  Phys. Rev. D} {\bfseries 62} (2000) 126006},
  \href{http://arxiv.org/abs/hep-th/0004142}{{\ttfamily arXiv:hep-th/0004142}}.

\bibitem{Bena:2004jw}
I.~Bena and N.~P. Warner, ``{A Harmonic family of dielectric flow solutions
  with maximal supersymmetry},''
  \href{http://dx.doi.org/10.1088/1126-6708/2004/12/021}{{\em JHEP} {\bfseries
  0412} (2004) 021},
\href{http://arxiv.org/abs/hep-th/0406145}{{\ttfamily arXiv:hep-th/0406145
  [hep-th]}}.
%%CITATION = HEP-TH/0406145;%%.

\bibitem{Gopakumar:1998ki}
R.~Gopakumar and C.~Vafa, ``{On the gauge theory / geometry correspondence},''
  \href{http://dx.doi.org/10.4310/ATMP.1999.v3.n5.a5}{{\em Adv. Theor. Math.
  Phys.} {\bfseries 3} (1999) 1415--1443},
  \href{http://arxiv.org/abs/hep-th/9811131}{{\ttfamily arXiv:hep-th/9811131}}.

\bibitem{Klebanov:2000hb}
I.~R. Klebanov and M.~J. Strassler, ``{Supergravity and a confining gauge
  theory: Duality cascades and chi SB resolution of naked singularities},''
  \href{http://dx.doi.org/10.1088/1126-6708/2000/08/052}{{\em JHEP} {\bfseries
  0008} (2000) 052},
\href{http://arxiv.org/abs/hep-th/0007191}{{\ttfamily arXiv:hep-th/0007191
  [hep-th]}}.
%%CITATION = HEP-TH/0007191;%%.

\bibitem{Vafa:2000wi}
C.~Vafa, ``{Superstrings and topological strings at large N},''
  \href{http://dx.doi.org/10.1063/1.1376161}{{\em J. Math. Phys.} {\bfseries
  42} (2001) 2798--2817}, \href{http://arxiv.org/abs/hep-th/0008142}{{\ttfamily
  arXiv:hep-th/0008142}}.

\bibitem{Bena:2005va}
I.~Bena and N.~P. Warner, ``{Bubbling supertubes and foaming black holes},''
  \href{http://dx.doi.org/10.1103/PhysRevD.74.066001}{{\em Phys. Rev.}
  {\bfseries D74} (2006) 066001},
\href{http://arxiv.org/abs/hep-th/0505166}{{\ttfamily arXiv:hep-th/0505166}}.
%%CITATION = HEP-TH/0505166;%%.

\bibitem{Bena:2007kg}
I.~Bena and N.~P. Warner, ``{Black holes, black rings and their microstates},''
  \href{http://dx.doi.org/10.1007/978-3-540-79523-0}{{\em Lect. Notes Phys.}
  {\bfseries 755} (2008) 1--92},
\href{http://arxiv.org/abs/hep-th/0701216}{{\ttfamily arXiv:hep-th/0701216}}.
%%CITATION = HEP-TH/0701216;%%.

\bibitem{Bena:2004qv}
I.~Bena and D.~J. Smith, ``{Towards the solution to the giant graviton
  puzzle},'' \href{http://dx.doi.org/10.1103/PhysRevD.71.025005}{{\em Phys.
  Rev. D} {\bfseries 71} (2005) 025005},
  \href{http://arxiv.org/abs/hep-th/0401173}{{\ttfamily arXiv:hep-th/0401173}}.

\bibitem{Polchinski:2000uf}
J.~Polchinski and M.~J. Strassler, ``{The String dual of a confining
  four-dimensional gauge theory},''
\href{http://arxiv.org/abs/hep-th/0003136}{{\ttfamily arXiv:hep-th/0003136
  [hep-th]}}.
%%CITATION = HEP-TH/0003136;%%.

\bibitem{Bena:2022fzf}
I.~Bena, N.~{\v{C}}eplak, S.~D. Hampton, A.~Houppe, D.~Toulikas, and N.~P.
  Warner, ``{Themelia: the irreducible microstructure of black holes},''
  \href{http://arxiv.org/abs/2212.06158}{{\ttfamily arXiv:2212.06158
  [hep-th]}}.

\bibitem{Klebanov:2000nc}
I.~R. Klebanov and A.~A. Tseytlin, ``{Gravity duals of supersymmetric SU(N) x
  SU(N+M) gauge theories},''
  \href{http://dx.doi.org/10.1016/S0550-3213(00)00206-6}{{\em Nucl. Phys. B}
  {\bfseries 578} (2000) 123--138},
  \href{http://arxiv.org/abs/hep-th/0002159}{{\ttfamily arXiv:hep-th/0002159}}.

\bibitem{Bena:2022wpl}
I.~Bena, S.~D. Hampton, A.~Houppe, Y.~Li, and D.~Toulikas, ``{The (amazing)
  super-maze},'' \href{http://dx.doi.org/10.1007/JHEP03(2023)237}{{\em JHEP}
  {\bfseries 03} (2023) 237}, \href{http://arxiv.org/abs/2211.14326}{{\ttfamily
  arXiv:2211.14326 [hep-th]}}.

\bibitem{Bena:2024qed}
I.~Bena, R.~Dulac, A.~Houppe, D.~Toulikas, and N.~P. Warner, ``{Waves on
  mazes},'' \href{http://dx.doi.org/10.1007/JHEP02(2025)105}{{\em JHEP}
  {\bfseries 02} (2025) 105}, \href{http://arxiv.org/abs/2404.14477}{{\ttfamily
  arXiv:2404.14477 [hep-th]}}.

\bibitem{Bena:2015bea}
I.~Bena, S.~Giusto, R.~Russo, M.~Shigemori, and N.~P. Warner, ``{Habemus
  Superstratum! A constructive proof of the existence of superstrata},''
  \href{http://dx.doi.org/10.1007/JHEP05(2015)110}{{\em JHEP} {\bfseries 05}
  (2015) 110},
\href{http://arxiv.org/abs/1503.01463}{{\ttfamily arXiv:1503.01463 [hep-th]}}.
%%CITATION = ARXIV:1503.01463;%%.

\bibitem{Assel:2011xz}
B.~Assel, C.~Bachas, J.~Estes, and J.~Gomis, ``{Holographic Duals of D=3 N=4
  Superconformal Field Theories},''
  \href{http://dx.doi.org/10.1007/JHEP08(2011)087}{{\em JHEP} {\bfseries 08}
  (2011) 087}, \href{http://arxiv.org/abs/1106.4253}{{\ttfamily arXiv:1106.4253
  [hep-th]}}.

\bibitem{D3D5NS5}
I.~Bena, A.~Bourget, R.~Dulac, D.~Toulikas, and N.~P. Warner $\!\!$, {\it to
  appear}.

\end{thebibliography}\endgroup

\end{adjustwidth}
%%%%%%%%%%%%%%%%%%%%%%%%%%%%%%%%%%%%%%%%%%%%%%%%%%%%%

%%%%%%%%%%%%%%%%%%%%%%%%%%%%%%%%%%%%%

\end{document}